\documentclass[apj]{emulateapj}

\usepackage{apjfonts}

\newcommand{\tnm}{\tablenotemark}
\newcommand{\tnt}{\tablenotetext}

\newcommand{\ergs}{ergs s$^{-1}$}
\newcommand{\flux}{ergs cm$^{-2}$ s$^{-1}$}
\newcommand{\intens}{ergs cm$^{-2}$ s$^{-1}$ deg$^{-2}$}
\newcommand{\fluxhz}{ergs cm$^{-2}$ s$^{-1}$ Hz$^{-1}$}

\newcommand{\cdens}{cm$^{-2}$}

\newcommand{\chandra}{{\it Chandra}}

\newcommand{\im}{\item}

\newcommand{\iiu}{$L_{R}/L_{\rm 4.5\mu m}$}
\newcommand{\liiu}{$\log{(L_{R}/L_{\rm 4.5\mu m})}$}

\newcommand{\lumtwo}{$L_{\rm 2\mu m}$}

\newcommand{\average}[1]{\ensuremath{\langle#1\rangle} }

\newcommand{\lsun}{$L_{\sun}$}

\newcommand{\W}{\hphantom{0}}

\newcommand{\nh}{$N_{\rm H}$}

\newcommand{\luv}{$L_{2500\; \rm \AA}$}
\newcommand{\lr}{$L_{R}$}
\newcommand{\lbol}{$L_{\rm bol}$}

\newcommand{\lirac}{$L_{\rm 4.5\mu m}$}
\newcommand{\lx}{$L_{\rm X}$}

\newcommand{\fhard}{$F_{\rm 2-7\; keV}$}

\newcommand{\bootes}{Bo\"{o}tes}
\newcommand{\spitzer}{{\it Spitzer}}

\newcommand{\cstar}{${\tt CLASS\_STAR}$}

\begin{document}
\slugcomment{Accepted for publication in The Astrophysical Journal}

\title{A large population of mid-infrared selected,
obscured \\ active galaxies in the Bo\"{o}tes field}
\shorttitle{OBSCURED INFRARED AGNS}
\shortauthors{HICKOX ET AL.}
\author{R.~C. Hickox\altaffilmark{1}}
\author{C. Jones\altaffilmark{1}}
\author{W.~R. Forman\altaffilmark{1}}
\author{S.~S. Murray\altaffilmark{1}}
\author{M. Brodwin\altaffilmark{2,3}}
\author{M.~J.~I. Brown\altaffilmark{4}}
\author{P.~R. Eisenhardt\altaffilmark{2}}
\author{D. Stern\altaffilmark{2}}
\author{C.~S. Kochanek\altaffilmark{5}}
\author{D. Eisenstein\altaffilmark{6}}
\author{R.~J. Cool\altaffilmark{6}}
\author{B.~T. Jannuzi\altaffilmark{3}}
\author{A. Dey\altaffilmark{3}}
\author{K. Brand\altaffilmark{3,7}}
\author{V. Gorjian\altaffilmark{2}}
\author{N. Caldwell\altaffilmark{1}}

\altaffiltext{1}{Harvard-Smithsonian Center for Astrophysics, 60 Garden Street,
 Cambridge, MA 02138; rhickox@cfa.harvard.edu.}
\altaffiltext{2}{Jet Propulsion Laboratory, California Institute of Technology, Pasadena, CA 91109.}
\altaffiltext{3}{National Optical Astronomy Observatory, Tucson, AZ
 85726-6732.}
\altaffiltext{4}{School of Physics, Monash
University, Clayton 3800, Victoria, Australia.}
\altaffiltext{5}{Department of Astronomy, The Ohio State University, 140 West 18th Avenue, Columbus, OH 43210.}
\altaffiltext{6}{Steward Observatory, 933 North Cherry Avenue, Tucson, AZ 85721.}
\altaffiltext{7}{Space Telescope Science Institute, 3700 San Martin
 Drive, Baltimore, MD 21218.}


\begin{abstract}
We identify a population of 640 obscured and 839 unobscured AGNs at redshifts $0.7<z\lesssim3$ using
multiwavelength observations of the 9 deg$^2$ NOAO Deep Wide-Field
Survey (NDWFS) region in \bootes.  We select AGNs on the basis of
\spitzer\ IRAC colors obtained by the IRAC
Shallow Survey.  Redshifts are obtained from optical spectroscopy or
photometric redshift estimators.  We classify the IR-selected AGNs as
IRAGN 1 (unobscured) and IRAGN 2 (obscured) using a simple criterion
based on the observed optical to mid-IR color, with a selection
boundary of $R-[4.5]=6.1$, where $R$ and [4.5] are the Vega magnitudes
in the $R$ and IRAC 4.5 $\mu$m bands, respectively.  We verify this
selection using X-ray stacking analyses with data from the \chandra\
X\bootes\ survey, as well as optical photometry from NDWFS and
spectroscopy from  MMT/AGES.  We show that (1) these sources
are indeed AGNs, and (2) the optical/IR color selection separates
obscured sources (with average $N_{\rm H}\sim3\times10^{22}$ \cdens\
obtained from X-ray hardness ratios, and optical colors and
morphologies typical of galaxies) and unobscured sources (with no
X-ray absorption, and quasar colors and morphologies), with a
reliability of $\gtrsim80\%$.  The observed numbers of IRAGNs are
comparable to predictions from previous X-ray, optical, and IR
luminosity functions, for the given redshifts and IRAC flux limits.
We observe a bimodal distribution in $R-[4.5]$ color, suggesting that
luminous IR-selected AGNs have either low or significant dust
extinction, which may have implications for models of AGN obscuration.

\end{abstract}


\section{Introduction}
\label{intro}

In unified models of active galactic nuclei (AGNs), a significant
number of objects are expected to be obscured by a torus of gas and
dust that surrounds the central engine and blocks the optical emission
along some lines of sight \citep[see reviews by ][]{urry95, anto93}.
In addition, some models of merger-driven quasar activity predict a
prolonged phase in which the central engine is entirely obscured,
followed by a ``blowout'' of the absorbing material and a relatively
short unobscured phase \citep[e.g.,][]{silk98, spri05,hopk06apjs}.
While some obscured AGNs have been identified, the existence of a
large absorbed population ($N_{\rm H}>10^{22}$
\cdens) has been invoked to explain the slope of the cosmic X-ray
background (CXB) at $E>2$ keV, which is believed to be integrated
emission from active galaxies \citep[e.g.,][]{sett89, coma95, bran05}. 

\subsection{Obscured AGNs in the optical, X-ray, and radio}

There are three well-established methods for identifying obscured
AGNs.  The first is the existence of narrow, high-excitation emission
lines in the optical spectrum, along with the absence of a power-law
continuum and broad emission lines that are characteristic of
unobscured sources.  The lack of broad lines and continuum is
attributed to dust that obscures the broad-line region around
the central engine, but leaves visible the larger narrow-line region
\citep{urry95, anto93}.

In the standard nomenclature, AGNs with a power-law optical continuum
and broad emission lines are referred to as type 1 objects, and those
with only narrow lines as type 2 \citep{seyf43,khac74}.  In the Seyfert
galaxies, the optical luminosity of the nucleus is comparable to that
of the host galaxy, while in the quasars, the nuclear luminosity
dominates that of the host galaxy.  Many type 2 Seyfert galaxies are
known, and the ratio in number density between type 2 and type 1
Seyferts in the local Universe has been estimated to be $\gtrsim$3:1
\citep[e.g.,][]{oste88, maio95}, although there is evidence from the
X-rays \citep[e.g.,][]{ueda03, barg05, gill07} and optical
\citep[e.g.,][]{lawr91,hao_l05b}, that the ratio of type 2 to type 1
AGNs decreases with increasing luminosity, and may also change with
redshift \citep{lafr05, ball06b}.  While type 2 quasars have been
challenging to detect in the optical, $\approx$300 type 2 quasars at
redshifts $0.3<z<0.83$ have recently been identified in the Sloan
Digital Sky Survey \citep[SDSS,][]{zaka03,zaka04,zaka05}.

X-ray observations also can identify obscured AGNs, by the presence of
absorption in the spectrum due to intervening neutral gas that
preferentially absorbs soft X-rays
\citep[e.g.,][]{awak91,cacc04,guai05,alex05}.  X-ray detection of
obscuration is complementary to that in the optical, because it is
caused by absorbing neutral gas rather than dust.  Some authors have
classified X-ray AGNs similarly to optical AGNs, based on the absence
(type 1) or presence (type 2) of X-ray absorption
\citep[e.g.,][]{ster02, ueda03, zhen04}.  Typically, an X-ray AGN is defined
to be absorbed (type 2) if its spectrum implies a neutral hydrogen
column density $N_{\rm H}\gtrsim10^{22}$ \cdens\
\citep[e.g.,][]{ueda03}.

Finally, radio observations can detect the population of obscured AGNs
that are radio-loud.  Such radio galaxies were some of the first
objects detected at high redshifts \citep[for a review see][]{mcca93},
and have been identified out to $z=5.19$ \citep{vanbreu99}.
Radio-loud AGNs make up $\sim$10\% of the total AGN population, and
many are known to be obscured \citep[e.g.,][]{webs95}, however they
may represent a different mode of accretion from the radio-quiet AGNs
\citep{best05}.  For this study we concentrate on an
infrared-selected sample that is mostly radio-quiet.

Identification of obscured AGNs from their optical and X-ray
properties is complicated by the fact that these two classifications
do not always agree.  Some type 2 optical AGNs, which show no broad
emission lines, also show no absorption in their X-ray spectra and so
would be classified as type 1 X-ray AGNs \citep[e.g.,][]{mate05}.
Conversely, some type 1 optical AGNs show X-ray absorption
\citep[e.g.,][]{matt02}.  The observations of these anomalous objects
are quite robust and are not simply due to measurement errors.  We
do not expect a perfect correlation between dust extinction and gas
attenuation, but geometric or physical explanations for these
observed properties are not yet clear.

However, for $\gtrsim$70\%--80\% of AGNs, the optical and X-ray
classifications correspond \citep{tozz06,cacc04}, suggesting that in
most cases the absorption of X-rays emitted close to the central
engine is related to larger-scale obscuration of the broad-line
region.  In this paper, our classifications of IR-selected AGNs as
type 1 and type 2 are initially based on optical and IR colors
but are verified by measuring absorption in the X-rays.  We will
generally use the term ``obscuration'' to refer to dust extinction
observed in the ultraviolet (UV), optical, and IR, and ``absorption'' to refer to neutral
gas absorption in the X-rays.

\subsection{Obscured AGNs in the infrared}

The optical and X-ray selection techniques described above generally
require bright sources or long integrations to observe optical narrow
lines or X-ray absorption.  With the launch of the {\it Spitzer Space
Telescope}, the IR provides a new, highly sensitive window
to identify obscured AGNs, using new techniques to select AGNs based
on mid-IR colors \citep[e.g.,][hereafter S05]{lacy04, ster05}.  IR emission
is produced by the reprocessing of nuclear luminosity by surrounding
dust, and is not as strongly affected as optical or UV light by dust
extinction.  Therefore IR criteria can identify many AGNs that
are not detected in the optical or X-rays, because the optical lines
are highly extincted, or the X-ray emission is too faint to observe
without long exposures.

\defcitealias{ster05}{S05}

Recent works have identified populations of obscured AGNs among
IR-selected samples.  Using near-IR data from the Two Micron All Sky
Survey, \citet{cutr02} identified 210 red AGNs at $z<0.7$, and
\citet{wilk02} showed that most of these objects have X-ray properties
consistent with absorption of $N_{\rm H}=$(0.1--1)$\times10^{23}$
\cdens.  In the mid-infrared, \citet{lacy04} used the $\spitzer$ First
Look Survey to select $\sim$2000 candidate AGNs based on their
\spitzer\ Infrared Array Camera (IRAC) colors. Of these,
\citet{lacy04} identified 16 objects from their optical and mid-IR
properties that are likely to be luminous obscured AGNs at
$z\lesssim0.7$.  \citet{lacy07} obtained optical spectra for a sample
of 77 IR-selected AGNs and found that 47\% had broad emission
lines and 44\% had high-ionization narrow emission lines, while 9\%
had no AGN spectral signatures.  Similarly, \citet{mart06} used
the Multiband Imaging Photometer (MIPS) for \spitzer\ 24 $\mu$m and radio data
to select 21 luminous, obscured quasars at $z\sim2$, for which
follow-up optical spectroscopy showed that 10 of these objects had
narrow emission lines characteristic of type 2 optical AGNs, while the
remainder had no emission lines.  These optical spectra are consistent
with obscuration of the nucleus, although it is important to note that
such objects may still have broad lines in the rest-frame optical that
are redshifted out of the observed spectrum.  \citet{alon06} used
mid-IR colors to select 55 candidate obscured AGNs in the extremely
deep Great Observatories Origins Deep Survey (GOODS) fields.  Also
using GOODS data, \citet{dadd07comp} identified $\sim$100 AGNs based
on excess 24 \micron\ emission above that expected for star formation,
and used X-ray stacking to infer the presence of a significant
population in the sample of highly obscured ($N_{\rm H}>10^{24}$ \cdens) AGNs .  A
large IR-selected sample of obscured AGNs comes from \citet{poll06},
who used IRAC observations from the SWIRE survey in the 0.6 deg$^2$
Lockman Hole to select 120 obscured AGN candidates based on their
optical to IR spectral energy distributions (SEDs).  In the \bootes\
field, \citet{brow06} identified several hundred candidate $z>1$ type
2 quasars, by selecting 24 $\mu$m MIPS sources with faint, extended
optical counterparts.

In the X-rays, a few hundred type 2 AGNs have been found in the
extremely deep, pencil-beam Chandra Deep Fields (CDFs)
\citep[e.g.,][]{trei04, zhen04, trei05, dwel05, dwel06, tozz06}, and
some have been identified with AGN counterparts selected in the IR
\citep{alon06} or submillimeter \citep{alex05}.  However, these narrow
fields miss rarer, more luminous objects.  Wide-field surveys offer
the best opportunity to select a large sample of AGNs with moderate to
high luminosity ($10^{45}\lesssim L_{\rm bol}\lesssim 10^{47}$ \ergs), moderate
obscuration ($10^{22}<N_{\rm H}<10^{23}$ \cdens), and high redshifts
($0.7<z<3$), which is the goal of the present study.

The 9 deg$^2$ multiwavelength survey in the NOAO Deep Wide-Field
Survey region in \bootes\ is uniquely suited for identifying large
numbers of such obscured AGNs.  In this study, we develop IRAC and
optical selection criteria for finding obscured AGNs, and then use the
available multiwavelength data, principally X-rays, to confirm the
selection and to measure properties such as accretion luminosity and
absorbing column density.  To this end, we analyze a sample of 1479
IR-selected AGNs at $0.7<z\lesssim3$ for which we have spectroscopic
and/or photometric redshift estimates, and we select 640 candidate
luminous, obscured AGNs.

This paper is organized as follows.  In \S\ \ref{obs} we describe the
\bootes\ multiwavelength observations, and in \S\ \ref{sample} we
discuss the sample of IR-selected AGNs.  In \S\ \ref{anal} we develop
criteria based on optical-IR colors for selecting obscured AGNs.  In
\S\ \ref{tests} we confirm these selection criteria using the X-ray
and optical properties of these objects, and measure X-ray
luminosities and absorbing column densities.  In \S\ \ref{photoz} we
verify the photometric redshift estimates, and in \S\ \ref{caveats} we
discuss contamination and incompleteness in the IR-selected AGN
samples.  In \S\ \ref{discussion} we place the population of
IR-selected AGNs in the context of the known and expected populations
of obscured and unobscured objects, and in \S\ \ref{summary} we
summarize our results. Throughout this paper we use a cosmology with
$\Omega_{\rm m}=0.3$, $\Omega_{\Lambda}=0.7$, and $H_0=70$ km$^{-1}$
s$^{-1}$ Mpc$^{-1}$.  Unless otherwise noted, we use the Vega system
for optical and infrared magnitudes.

\section{Bo\"{o}tes data set}
\label{obs}
The 9 deg$^{2}$ survey region in \bootes\ of the NOAO Deep Wide-Field
Survey \citep[NDWFS;][]{jann99} is unique among extragalactic
multiwavelength surveys, in its wide field and uniform coverage using
space- and ground-based observatories, including the {\it Chandra
X-Ray Observatory} and \spitzer.  Extensive optical spectroscopy makes
this field especially well suited for studying the statistical properties
of a large number of AGNs (C.S. Kochanek et.~al. 2008, in preparation).

The \bootes\ field  was observed by the \spitzer\ IRAC Shallow
Survey \citep{eise04}.  Three or more 30 s exposures were taken per
position, in all four IRAC bands (3.6, 4.5, 5.8, and 8 $\mu$m), with
$5\sigma$ flux limits of 6.4, 8.8, 51, and 50 $\mu$Jy, respectively.
The sample includes $\approx$370,000 sources detected at 3.6 $\mu$m,
including $>80\%$ of the X-ray sources.  We limit our IRAC sample to
$\approx$15,500 objects that have 5$\sigma$ detections in all four
bands and at least three good exposures (for reliable rejection of cosmic
rays), which cover an area of 8.5 deg$^2$.

X-ray data are taken from the X\bootes\ survey, which is a mosaic
comprised of 126 5 ks \chandra\ ACIS-I exposures and is the largest
contiguous field observed to date with \chandra\ \citep{murr05}.  Due
to the shallow exposures and low background in the ACIS CCDs, X-ray
sources can be detected to high significance with as few as four
counts. In this field, 3293 X-ray point sources with four or more counts
are detected \citep{kent05}, of which 2960 lie within the area covered
by IRAC.  Optical identifications for the X-ray sources are presented
in \citet{brand06a}.  We also use radio data from the Very Large Array
(VLA) FIRST 20 cm radio survey \citep{beck95}, which detects 930
sources in the area covered by the IRAC, to a limiting flux of
$\approx$1 mJy.

Optical photometry in the \bootes\ field comes from the NDWFS, which
used the Mosaic-1 camera on the 4-m Mayall Telescope at Kitt Peak
National Observatory.  Deep optical imaging was performed over the
entire 9.3 deg$^{2}$ in the $B_{W}$, $R$, and $I$ bands with 50\%
completeness limits of 26.7, 25.5, and 24.9 mag, respectively
\citep{jann99}.  Optical spectroscopy in the \bootes\ field comes from
the AGN and Galaxy Evolution Survey (AGES), which uses the Hectospec
multifiber spectrograph on the MMT.  We use AGES Data Release 1 (DR 1)
and Internal Release 2 (IR 2), which consist of all the AGES spectra
taken in 2004--2005.  In AGES DR 1, targets include (1) all extended
sources with $R\le 19.2$ (2) a randomly selected sample of 20\% of all
extended sources with $19<R\le 20$, and (3) all extended sources with
$R\le 20$ and IRAC 3.6, 4.5, 5.8, and 8.0 $\mu$m magnitudes $\le$15.2,
15.2, 14.7, and 13.2, respectively.  In addition, (4) fainter sources
were observed, selected mainly from objects with counterparts of
\chandra\ X-ray sources \citep{murr05, brand06a, kent05}, radio
sources from the VLA FIRST survey, and objects selected from 24 $\mu$m observations with
MIPS (E.\ Le Floc'h et al. 2008, in preparation).  AGES IR 2 contains
$I$-selected targets with $I\le 21.5$ for point sources and $I\le
20.5$ for extended sources.  Because X-ray sources were preferentially
targeted, the survey contains a large number of spectral
identifications for distant AGNs.  Galaxy spectra are classified by
template fits into three categories: optically normal galaxies,
broad-line AGNs (BLAGNs), and narrow-line AGNs (NLAGNs).

We use the optical and IRAC photometry described in \citet{brod06},
for which optical and IRAC sources are matched using a 1\arcsec\
radius.  We then match the \chandra\ X-ray sources, AGES optical
spectra, and VLA FIRST 20 cm sources to the IRAC sources, using radii of
3.5\arcsec, 2\arcsec, and 2\arcsec, respectively.  There were 1298
matches to X-ray sources, 6450 matches to AGES spectra, and 196
matches to radio sources. There were no sources with multiple
matches (owing to the $\sim$2\arcsec\ point-spread function [PSF] of the IRAC images, no two
sources in the $5\sigma$ catalog are closer than 3\arcsec).

To estimate the number of spurious matches, we offset the positions of
the IRAC sources by 16\arcsec\ and re-perform the source matching.
This places the IRAC sources at ``random'' positions away from the
X-ray or AGES sources but retains their surface density distribution
on larger scales.  We re-perform the matching with offsets in eight
directions and derive the median number of matches from these eight
trials.  For the full sample of $\approx$15,500 IRAC sources detected
at $5\sigma$ in all four bands, we expect spurious matches to 20 X-ray
sources, 45 AGES spectra, and 3 radio sources.  In this paper we focus
on a sample of 1479 IR-selected AGNs (\S\ \ref{sample}), for which we
expect spurious matches to only 2 X-ray sources, 4 AGES spectra, and no
radio sources.  Details of the IRAC $5\sigma$ sample and matches to
the optical and X-ray catalogs are given in Table \ref{tblsample}.

To calculate luminosities and to fit models to SEDs for the objects in
our sample, we require estimates of redshift.  For all objects with
AGES spectra, which have $17.5<R<22$, we have reliable spectroscopic
redshifts with uncertainties of $\sigma_{z} < 0.001$.  However, 51\%
of our IR-selected AGNs (as defined in \S\ \ref{sample}) do not have
optical spectra, either because they were not spectroscopically
targeted, or because they are fainter than the AGES spectroscopic
limits.  For these, we use photometric redshifts from the catalog of
\citet{brod06}, who use fluxes from the four IRAC bands, as well as
$B_{W}$, $R$, and $I$ in the optical. photo-$z$'s are obtained through
a hybrid technique; for objects with strong spectral features such as
most optically normal galaxies, redshifts are estimated using template
fitting.  For objects (such as AGNs) that have more featureless SEDs, an
artificial neural net is used.  Uncertainties in the photo-$z$ are
$\sigma_{z}=0.06(1+z)$ for galaxies at $z<1$ and
$\sigma_{z}=0.12(1+z)$ for optically bright AGNs.  Photo-$z$
uncertainties increase for fainter sources due to larger photometric
errors.  In \S\ \ref{photoz} we address possible systematic errors in
the photo-$z$'s and show that there are no large biases in the
photo-$z$ estimates that would significantly affect our conclusions.
However, because of the limited accuracy of the photo-$z$'s, in this
paper we do not use them to measure precise quantities such as the
evolution of the obscured AGN fraction with luminosity or redshift.

\begin{figure}
\epsscale{1.2}
\label{figsel}
\plotone{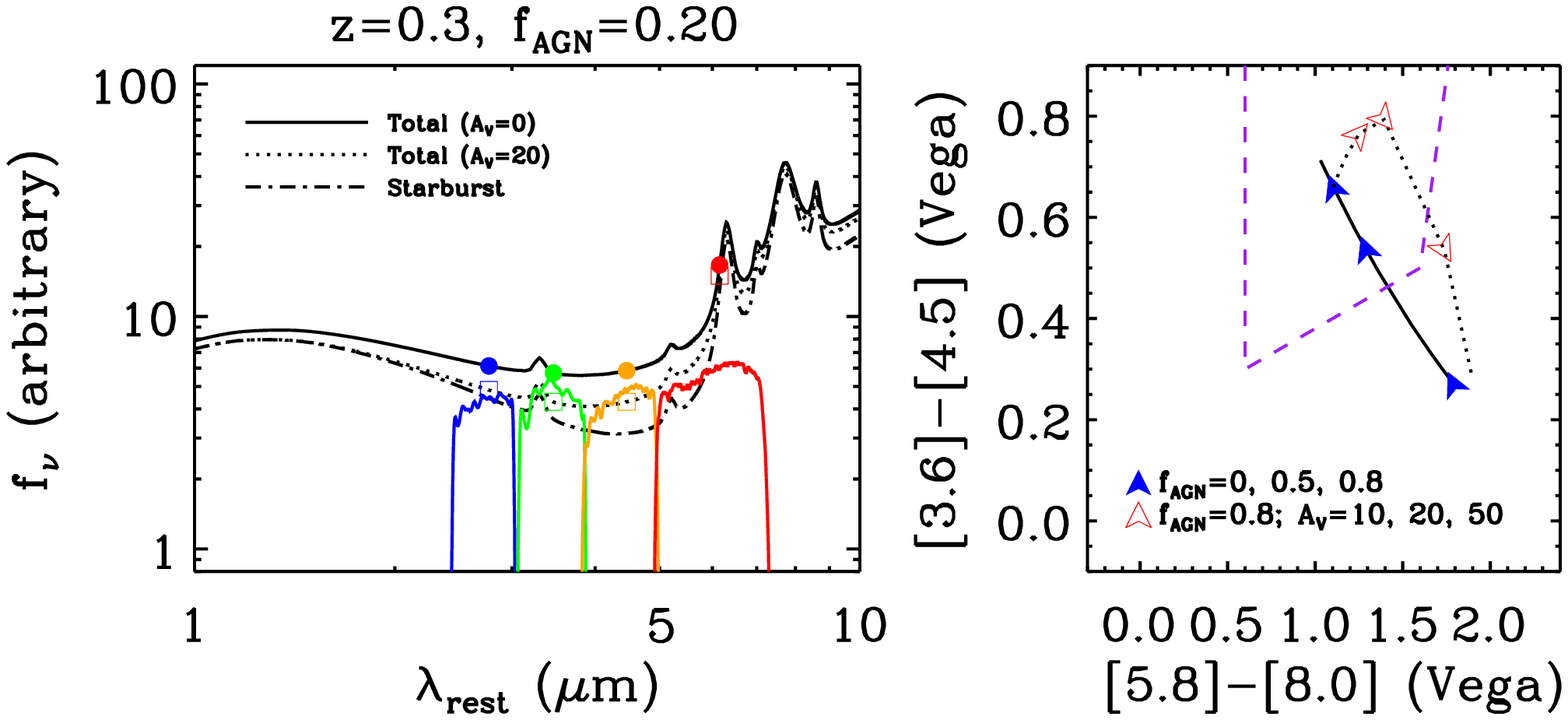}
\plotone{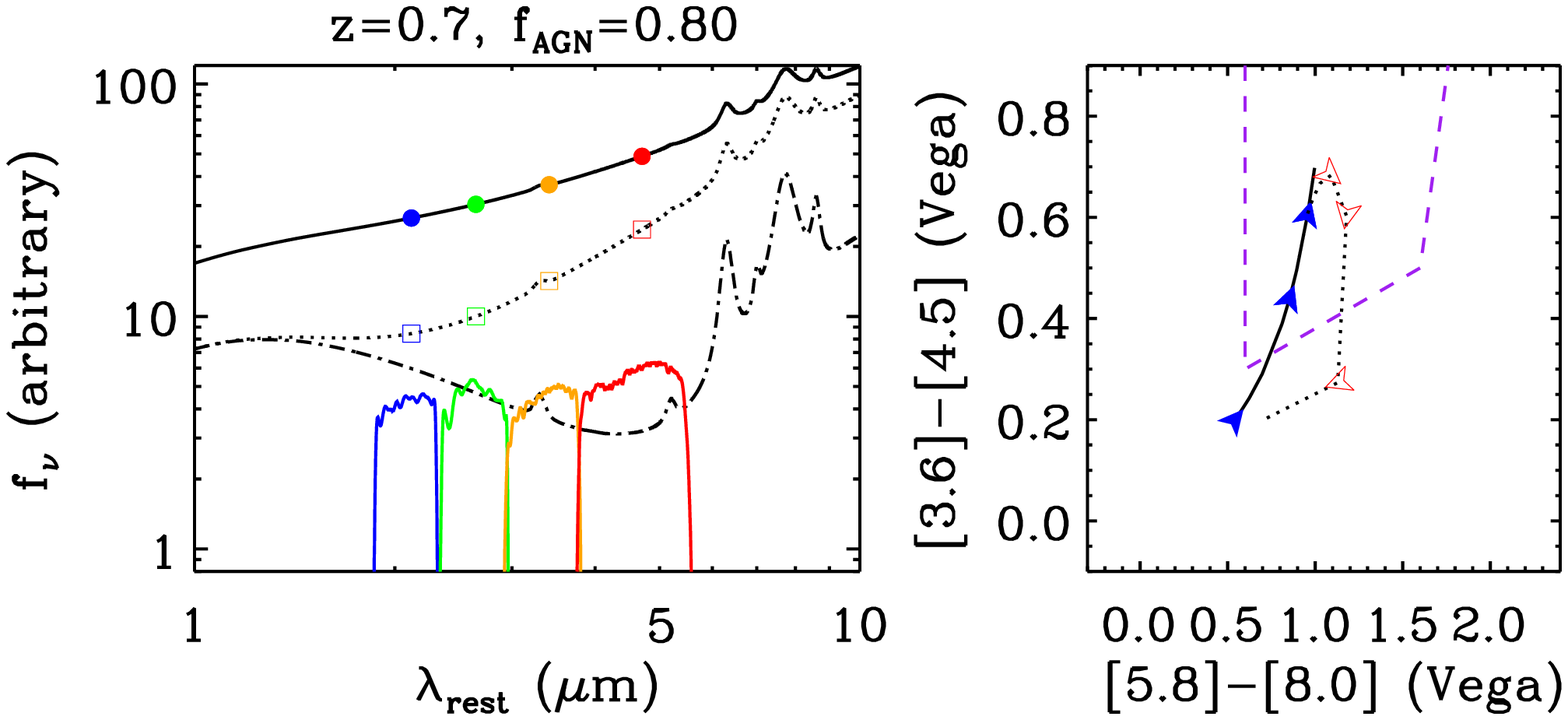}
\plotone{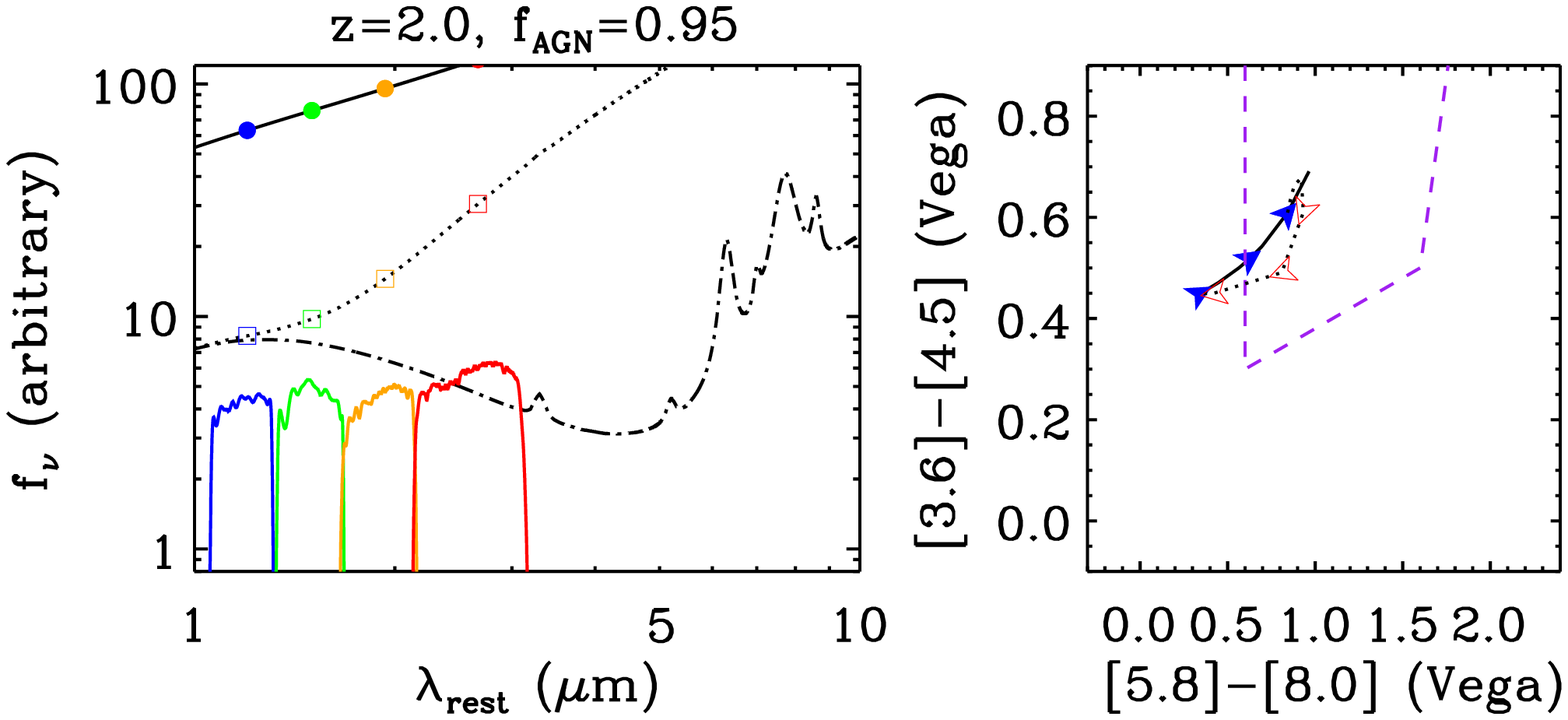}
\caption{Illustration of mid-IR AGN color selection.  On the left, the
  solid lines show the rest-frame spectrum consisting of the sum of a
  starburst \citep[][dot-dashed lines]{sieb07} plus AGN power-law
  ($\alpha_\nu=-1$) templates.  The three panels are for objects at
  $z=0.3$, 0.7, and 2, and AGN contribution to the rest-frame mid-IR
  (3--8 \micron) emission $f_{\rm AGN}=0.2$, 0.8, and 0.95,
  respectively.  Fluxes (shown by the filled circles) are determined
  by convolving the spectra with the responses of the four IRAC bands
  (shown).  The dotted spectra and open squares in the left panels
  show the same model with the AGN power law extincted for $A_V=20$,
  with a Galactic extinction curve.  The panels at right show
  corresponding locations on the color-color diagram for these models
  along with the \citetalias{ster05} AGN color selection region.  The
  solid lines in the color-color diagrams represent the colors as a
  function of increasing $f_{\rm AGN}$ from 0 to 0.95 for each
  redshift (the blue arrows show $f_{\rm AGN}=0$, 0.5 and 0.8).  Note
  how increasing the contribution of the red AGN power law brings the
  objects into the \citetalias{ster05} color selection region for all
  three redshifts, with the objects entering the selection region at
  $f_{\rm AGN}=0.3-0.5$.  The dotted line in the right panel shows the
  effects of obscuration of the nuclear component (for $A_V=0-100$),
  for the spectrum with $f_{\rm AGN}=0.8$, with the open arrows
  representing $A_V=10$, 20, and 50.  Very high dust extinction
  $A_V\sim30$--50 will move the object out of the \citetalias{ster05}
  selection region.
\label{figcol}
\vskip0.2cm
}

\end{figure}

\begin{figure}
\epsscale{1.2}
\plotone{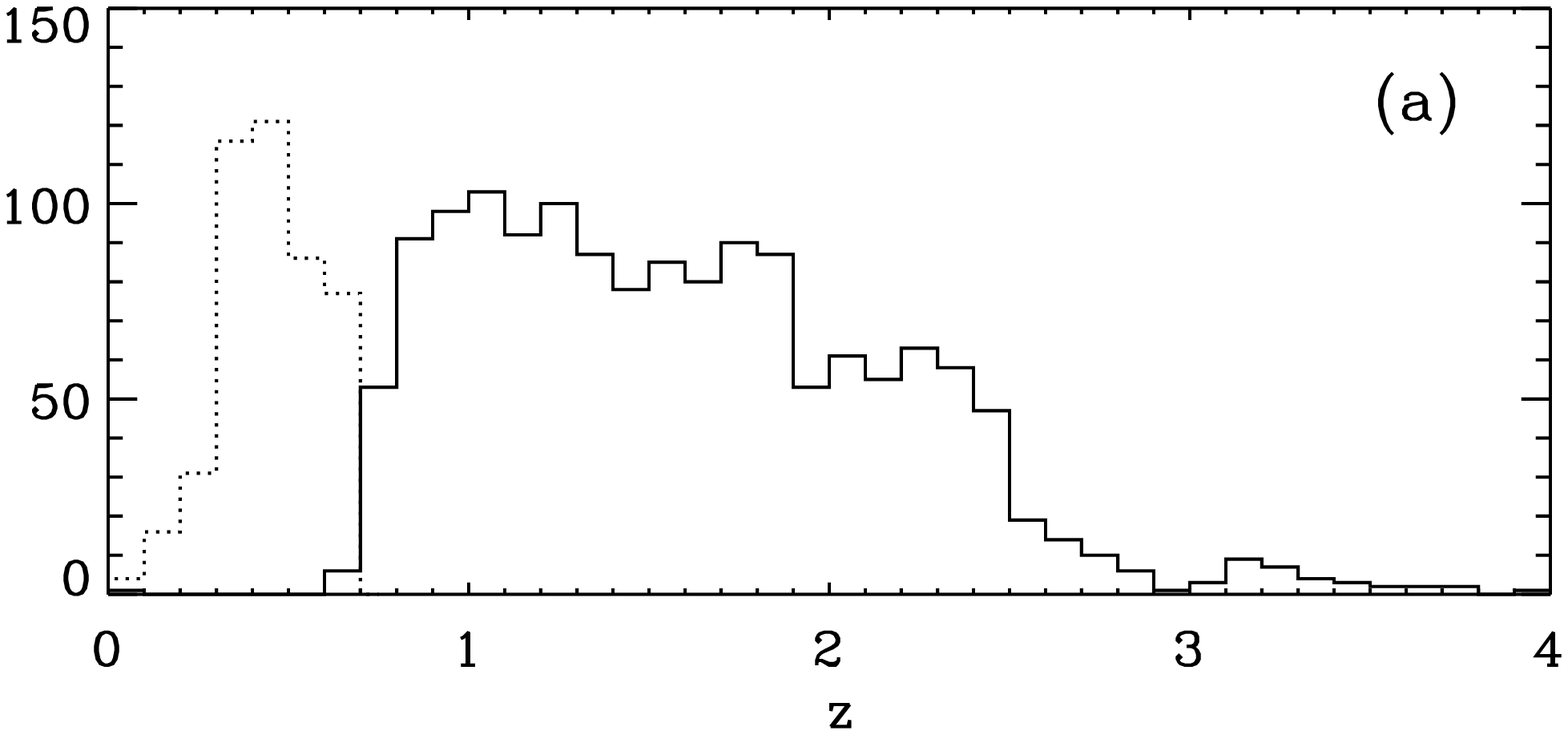}
\plotone{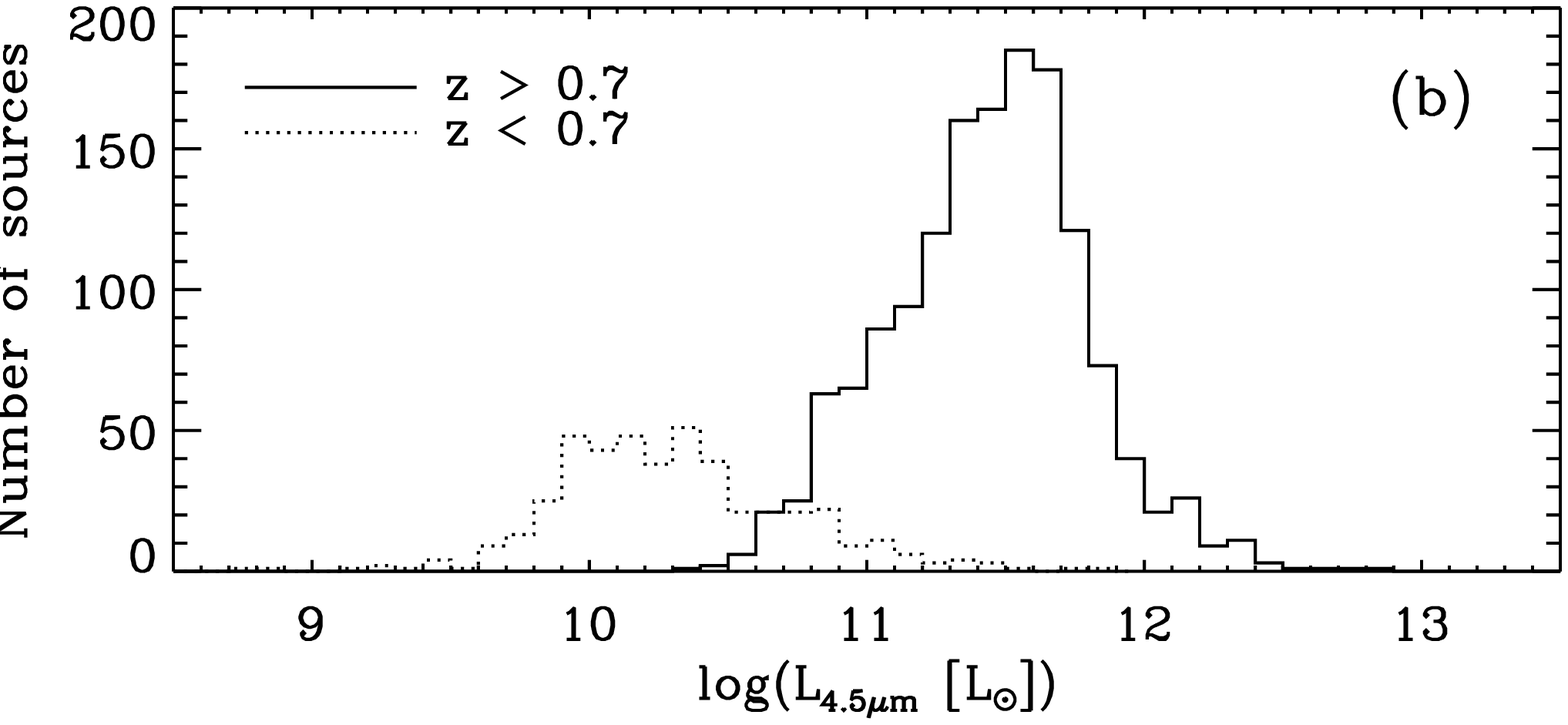}
\plotone{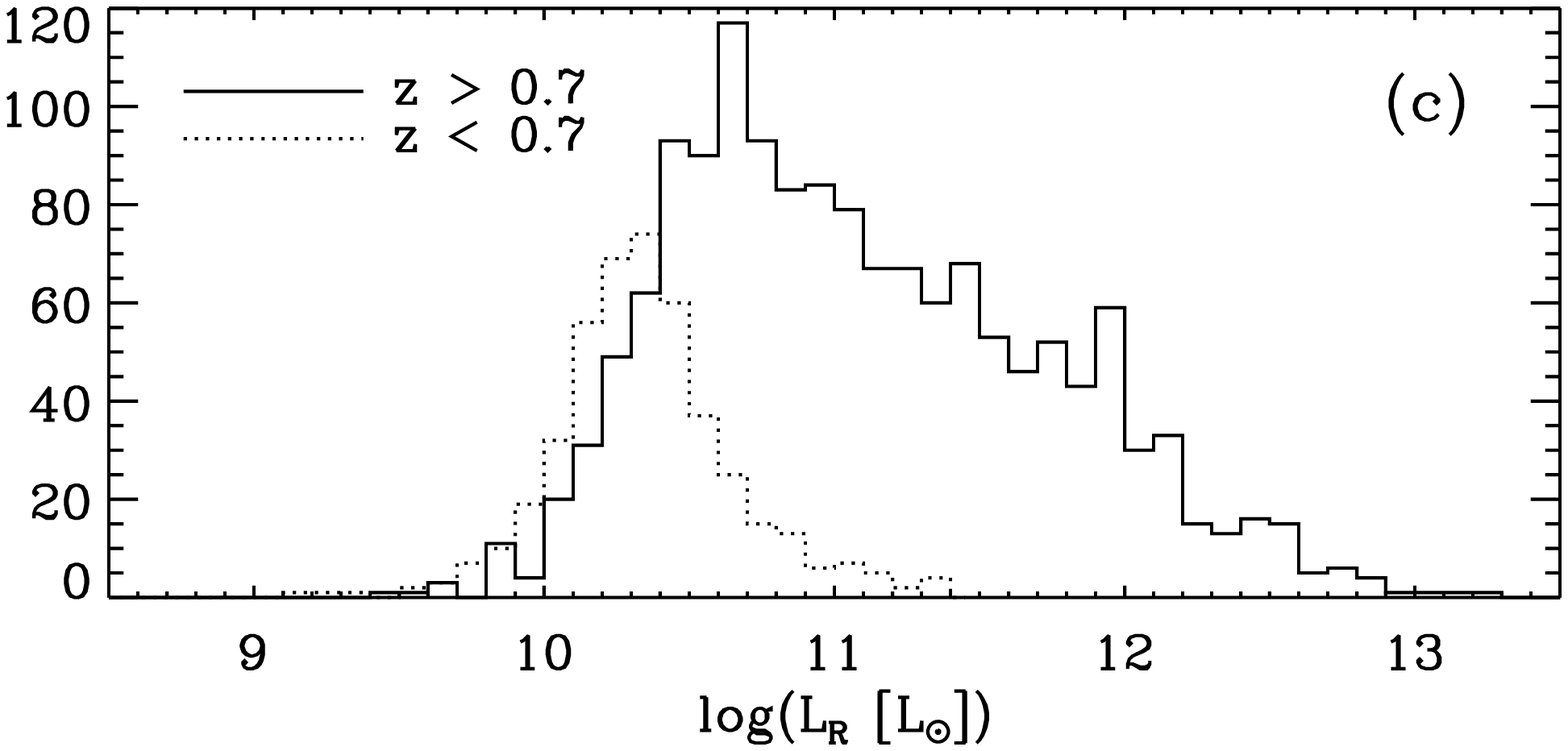}
\caption{Distributions of (a) redshift, (b) \lirac, and (c) $L_R$ for
  the 1929 infrared-selected AGNs in the sample.  \lirac\ and $L_R$
  are $\nu L_\nu$ in the observed $R$ and IRAC 4.5 $\mu$m bands,
  respectively (see \S\ \ref{lum}).  Many of the objects at $z<0.7$
  (dotted line) are not AGNs but ``normal'' galaxies; above this
  redshift these objects are typically too faint to be detected in all
  four bands of the IRAC Shallow Survey.  We restrict our analysis to
  the 1479 objects with $z>0.7$, shown by the solid line.
\label{figdist}}
\vskip0.2cm
\end{figure}

\section{Infrared-selected AGN sample}

\label{sample}
The AGN sample used in this paper is selected in the mid-IR, which is
less affected by obscuration than optical or soft X-ray emission.  In
the (rest-frame) near- to mid-IR from 1 to 10 $\mu$m, AGNs have
markedly different SEDs from normal or starburst galaxies. AGNs
typically have a roughly power law continuum in the near- to mid-IR,
$S_\nu\propto \nu^{\alpha_{\nu}}$, where $\alpha_{\nu} \simeq -1$
\citep[e.g., S05;][]{glik06}.  In contrast, normal and starburst
galaxies have bluer continua in the rest-frame mid-IR, due to the fact
that the spectrum from the stellar population of the galaxy peaks at
$\simeq$1.6 $\mu$m, and falls at longer wavelengths.  In addition,
star-forming galaxies have prominent emission features at 3--10 $\mu$m
due mainly to lines from polycyclic aromatic hydrocarbons (PAHs) in dust
\citep{puge89}.  This difference in SEDs allows us to effectively
distinguish AGN-dominated objects from normal and starburst galaxies
using observed colors in the mid-IR.

\citetalias{ster05} developed a set of IRAC color-color selection
criteria based on the IRAC Shallow Survey photometry and AGES spectra,
described in \S\ \ref{obs}.  In this paper we use those criteria
to select AGNs.  To illustrate the \citetalias{ster05} color-color
selection, we show in Fig.\ \ref{figcol} the IRAC $[3.6]-[4.5]$ and
$[5.8]-[8.0]$ colors for a two-component template spectrum consisting
of a starburst galaxy \citep{sieb07} plus AGN power law with
$\alpha_\nu=-1$.  We show these colors for three redshifts and for
various values of the fraction $f_{\rm AGN}$ of the rest-frame 3-10
$\mu$m luminosity that is emitted by the AGN.  Because the colors of
the power law AGN spectrum are constant with redshift, increasing
$f_{\rm AGN}$ moves the colors into the \citetalias{ster05} AGN
selection region, regardless of the redshift of the source.  Fig.\
\ref{figcol} also shows the effect of dust extinction of the nuclear
component, for a Galactic extinction curve \citep{pei92}.  For
$A_V\gtrsim30$--50 (depending on redshift), extinction can cause the
IRAC colors to move out of the \citetalias{ster05} selection region,
even for $f_{\rm AGN}$ as high as 0.8.

\begin{deluxetable*}{lcccc}
\tablewidth{5.2in}
\tabletypesize{\footnotesize}
\tablecaption{Matches of IRAC sources to  AGES
  spectra and X-ray sources
\label{tblsample}}
\tablehead{
\colhead{} &
\multicolumn{2}{c}{All IRAC\tnm{a}} &
\multicolumn{2}{c}{IRAGN ($z>0.7$)} \\
\colhead{AGES spectral type} & 
\colhead{All sources} &
\colhead{X-ray} &
\colhead{All sources} &
\colhead{X-ray}}
\startdata
Total &        15492  & 1298    & 1479 & 654\\
BLAGN &        941 & 592   & 697 & 457\\
NLAGN &        108 & 43   & 4 & 2\\
Galaxies &     5401 & 244   & 27 & 13\\
No spectrum &  9042 & 419   & 751 & 182

\enddata
\tnt{a}{Sources with $5\sigma$ detections in all four IRAC bands.}
\end{deluxetable*}

We stress that this color-color technique does not select all AGNs.
In the \bootes\ data \citep{gorj07}, as well as the extended Groth
strip \citep[EGS;][]{barm06}, only half of X-ray detected AGNs were identified
using the \citetalias{ster05} IRAC color-color criteria.  This is likely
due to the fact that some X-ray sources are too faint to be detected
in all four IRAC bands, while others might not have red power-law
mid-IR spectra.  Recent mid-IR spectroscopy of type 2 quasars with the
Infrared Spectrograph on \spitzer\ has shown that most luminous ($L_{\rm
X}>10^{44}$ \ergs) X-ray selected type 2 quasars have relatively
featureless mid-IR spectra \citep{stur06, weed06c}.  Still, many AGNs in
ultraluminous infrared galaxies (ULIRGs) show a variety of spectral
shapes including PAH emission and deep silicate absorption features
\citep{spoo05, buch06, bran07spec_aph}, which may
indicate deep obscuration of the nuclear IR emission.  Therefore, the
completeness of AGN color-color selection is still unclear.  The key
point for this study is that while color-color selection may miss
many AGNs, there should be little contamination in the
AGN color-color region from starburst-powered objects, particularly
for objects with $z>0.7$ (see \S\ \ref{lum}).  Sample completeness and
contamination are discussed in more detail in \S\ \ref{caveats}.

Our sample of IR-selected AGNs contains objects that have: (1)
$5\sigma$ detections in all four IRAC bands as well as the $R$ band of
the NOAO DWFS catalog, which we use to calculate optical luminosities;
(2) IRAC colors that fall in the \citetalias{ster05} AGN selection
region; and (3) spectroscopic redshifts from AGES or
photometric redshifts from the \citet{brod06} catalog, with $z_{\rm
phot}>0$.  These criteria select 1929 objects.  Only 13 additional
objects are not detected in the $R$ band but meet all the other
criteria, so this requirement has little effect on our results.
Excluding all objects with $z<0.7$ to minimize contamination by normal
galaxies (see \S\ \ref{lum}) leaves a sample of 1479 IR-selected AGNs,
of which 1469 have detections in all three NDWFS optical bands.
Details of AGES spectra and X-ray matches to the objects are shown in
Table \ref{tblsample}.

\section{Optical/IR SEDs and obscured AGN selection}
\label{anal}

In this section we calculate optical and IR luminosities for the
IR-selected AGNs, and perform template fits to the optical and IR SEDs
that provide evidence that roughly half of the sample has significant
nuclear extinction.  We then develop a simple optical-IR color
criterion for selecting obscured AGNs.  We show that the obscured AGN
candidates display absorption in their average X-ray spectra and have
the optical characteristics of normal galaxies, while the unobscured
candidates are on average X-ray unabsorbed and have optical colors
and morphologies typical of unobscured AGNs.

\begin{figure}
\epsscale{1.2}
\plotone{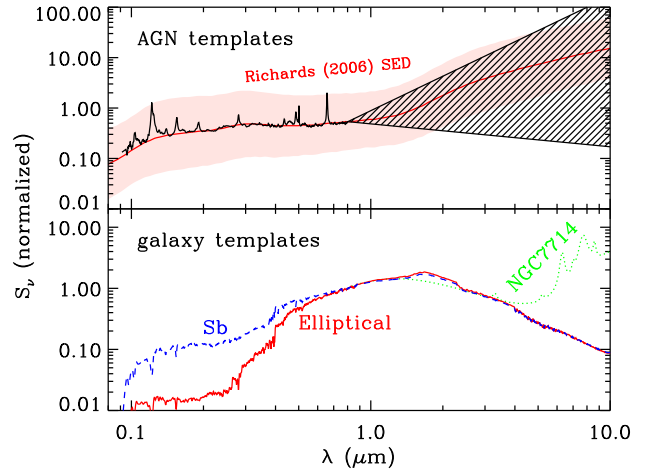}
\caption{Template spectra used for SED fits, normalized at 0.8
 \micron. The top panel shows the unabsorbed AGN template, with the
 mean SED and dispersion from \citet{rich06} for comparison.  The
 hatched region at right shows the allowed values of
 $\alpha_{\nu}$. The bottom panel shows elliptical (red solid line),
 Sb (blue dashed line) and starburst (green dotted line) galaxy
 templates.  See \S~\ref{lum} for details of the models.
\label{figtemp}
}
\vskip0.5cm
\end{figure}

\subsection{Luminosities and model fits}
\label{lum}

For each of the 1479 AGNs in our sample, we calculate the observed mid-IR
and optical luminosity densities using
\begin{equation}
L_\nu(\nu_{\rm rest})=\frac{4\pi d_{\rm L}^2}{1+z} S_\nu(\nu_{\rm obs}),
\end{equation}
where $d_{\rm L}$ is the luminosity distance for a given redshift in
our adopted cosmology \citep{hogg99}, $S_\nu$ is the flux density in \fluxhz,
and $\nu_{\rm obs}$ and $\nu_{\rm rest}$ are the observed and
rest-frame frequencies, respectively, where $\nu_{\rm
rest}=(1+z)\nu_{\rm obs}$.  Throughout the paper
we present optical and IR luminosities in terms of the bolometric
luminosity of the Sun, $L_{\sun}=3.83\times10^{33}$ \ergs.

We generally define luminosities and colors in terms of
the observed (rather than rest-frame) photometric bands; the
relationship between  rest-frame luminosity density
$L_\nu(\nu_{\rm rest})$ and the observed-frame luminosity density
$L_\nu(\nu_{\rm obs})$ is 
\begin{equation}
L_\nu(\nu_{\rm obs})=\frac{L_\nu(\nu_{\rm rest})}{(1+z)^{\alpha_\nu}},
\label{eqnrestobs}
\end{equation}
where $\alpha_\nu$ is the power-law index ($S_\nu\propto
\nu^{\alpha_{\nu}}$) for the spectrum between $\nu_{\rm obs}$ and
$\nu_{\rm rest}$.  Redshift estimates and detailed spectral shapes are
uncertain for many of the AGNs in our sample, so framing the selection
in terms of observed luminosities and colors makes our results less
subject to the details of $K$-corrections.

We define luminosities in each photometric band in terms of $\nu
L_{\nu}$, which unlike the luminosity density $L_\nu$, is not
strongly affected by corrections for redshift, at least for unobscured
quasars.  Eqn.~\ref{eqnrestobs} shows that for a typical broadband
quasar SED with IR power law with $\alpha_{\nu}=-1$, $\nu L_\nu$
remains constant with redshift.  For a typical optical continuum
(which is not exactly a power law, as described below), in the
redshift range we consider ($0.7<z<3$), the observed $\nu L_{\nu}$
varies by at most 0.25 dex.  In \S~\ref{kcor}, we estimate $K$-corrections and
show that they have no significant effect on our selection criteria.

For the mid-IR luminosity we use $L_{\rm 4.5\mu m}$, defined to be
$\nu L_{\nu}$ in the observed 4.5 $\mu$m IRAC band.  Because the
color-selected AGNs in our sample have similar IRAC SEDs, the \lirac\
is a simple and sufficiently accurate proxy for the total luminosity in the IRAC
bands.  In the optical, we use $L_{R}$, defined as $\nu L_{\nu}$
observed in the $R$ band centered on 6514 \AA.  The
distributions of the 1929 IR-selected AGNs in redshift, \lirac, and
$L_{R}$ are shown in Fig.\ \ref{figdist}.

We restrict our sample to AGNs at $z>0.7$.  At lower redshifts the
IRAC source counts are dominated by normal or star-forming galaxies
with relatively low luminosities ($L_{\rm 4.5\mu m}<10^{11}$
\lsun). Some of these objects may have red IRAC SEDs, for example, due
to heavy dust obscuration.  The model SED from \citet{sieb07} for the
heavily extincted starburst Arp 220 has IRAC colors that lie within
the \citetalias{ster05} AGN region, and less obscured sources can lie
close to  this region. Combined with photometric errors, this
results in a significant number of $z<0.7$ objects selected with the
\citetalias{ster05} criterion being either normal or starforming
galaxies.  By cutting our sample at $z>0.7$, however, we exclude most
of these ``normal'' galaxies as they are generally fainter than the
flux density limits in the 5.8 \micron, 8.0 $\mu$m, or $R$ bands
(heavily extincted starbursts, for example, are very faint in the
optical).  In addition, limiting the sample to $z>0.7$ allows for more
straightforward color selection of obscured AGNs, as shown in \S\
\ref{identify}.  Our final IR-selected AGN sample includes only the
1479 IR-selected AGNs with $z>0.7$.

\begin{figure}
\epsscale{1.2}
\plotone{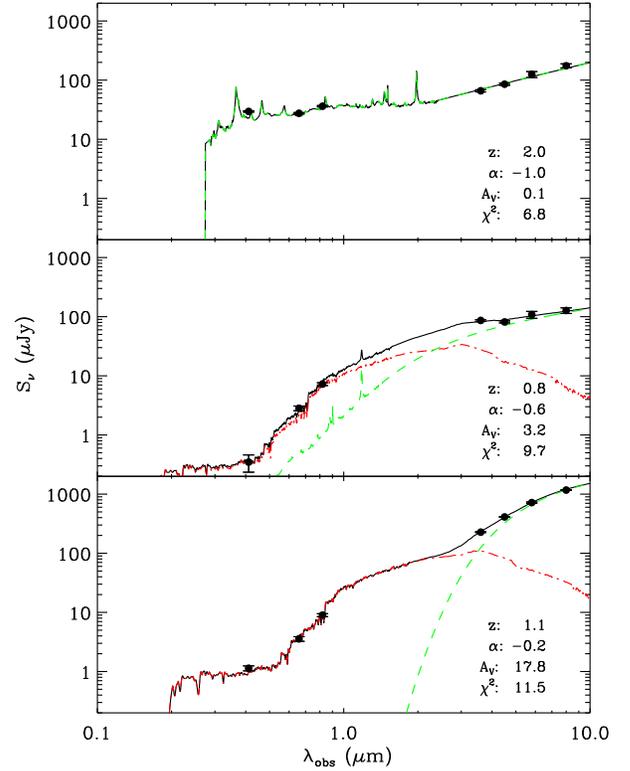}
\caption{Examples of fits to optical and IRAC photometry, for galaxy
   (red dot-dashed lines) and AGN (green dashed lines)
  spectral templates (see \S\ \ref{lum} for a description of the
  models).  These three example objects show a range
of best-fit values to the AGN power-law slope $\alpha_\nu$ (where $S_\nu
\propto \nu^{\alpha_\nu}$) and AGN template extinction $A_V$.
\label{figtempall}}
\end{figure}

\begin{figure}
\epsscale{1.2}
\plotone{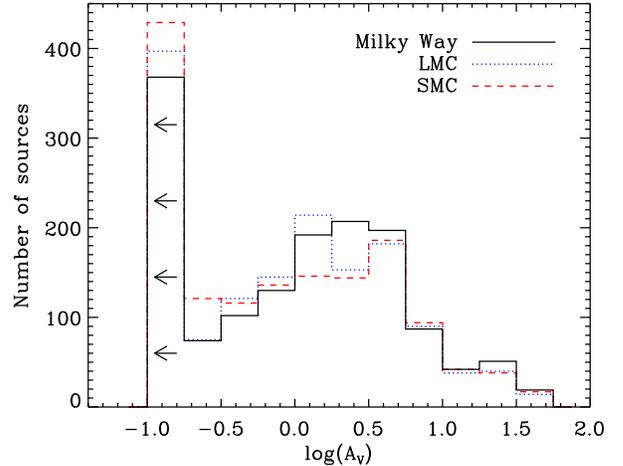}
\caption{Distribution of best-fit $A_V$ for the IR-selected AGNs.  The
  far left bin corresponds to objects with $A_V\leq0.2$.  The three
  lines are for Galactic, LMC, and SMC extinction curves.
\label{figav}}
\end{figure}
 
To model the SEDs of the IR-selected AGNs, we fit the optical and IRAC
photometry of each source with spectral templates including AGN and
host galaxy components.    For the nuclear emission in the rest-frame
optical/UV, we use the AGN template of \citet{hopk07qlf}, which
consists of the composite SED of \citet{rich06}, with optical lines
taken from the SDSS composite quasar
template \citep{vand01}.  For $\lambda > 0.8$ $\mu$m, we use a power
law component.  Our grid of models includes 14 values of the slope
from $-2.2\leq \alpha_{\nu}\leq 0.4$.  We also include extinction of
the nuclear component, with a Galactic extinction curve \citep{pei92},
and $E_{B-V}/A_V=3.1$ for 12 logarithmically spaced values of
$0<A_V<32$.  This corresponds to a total of 168 separate AGN models.

For the host galaxy emission, we use two model galaxy templates
calculated using the PEGASE population synthesis code \citep{fioc97}.
The models are chosen so that at age 13 Gyr, they correspond to
observed low-redshift ellipticals and spirals.  The models differ in
their initial specific star formation rates ($5\times10^{-3}$ vs.\
$3.5\times10^{-4}$ $M_{\sun}$ Myr$^{-1}$ per unit gas mass in
$M_{\sun}$, for elliptical and Sb, respectively), the fraction of
stellar ejecta available for new star formation (0.5 vs. 1), and
extinction (none for the elliptical galaxy, disk extinction for the
Sb).  For simplicity, we use non-evolving spectra corresponding to an
age of 3 Gyr after formation.  Assuming that massive galaxies form at
$z>6$, this age roughly corresponds to the age of such a galaxy at
$z\sim1$--2 for our adopted cosmology.  At $\lambda > 0.8$ $\mu$m, the
models include either the quiescent galaxy spectrum, or the spectrum
of the starburst galaxy NGC 7714 \citep{sieb07}.  This gives a total
of four separate host galaxy models (E, Sb, E plus starburst, and Sb
plus starburst).  The quasar and galaxy template spectra are shown in
Fig.~\ref{figtemp}.

\begin{figure*}
\epsscale{1.1}
\plottwo{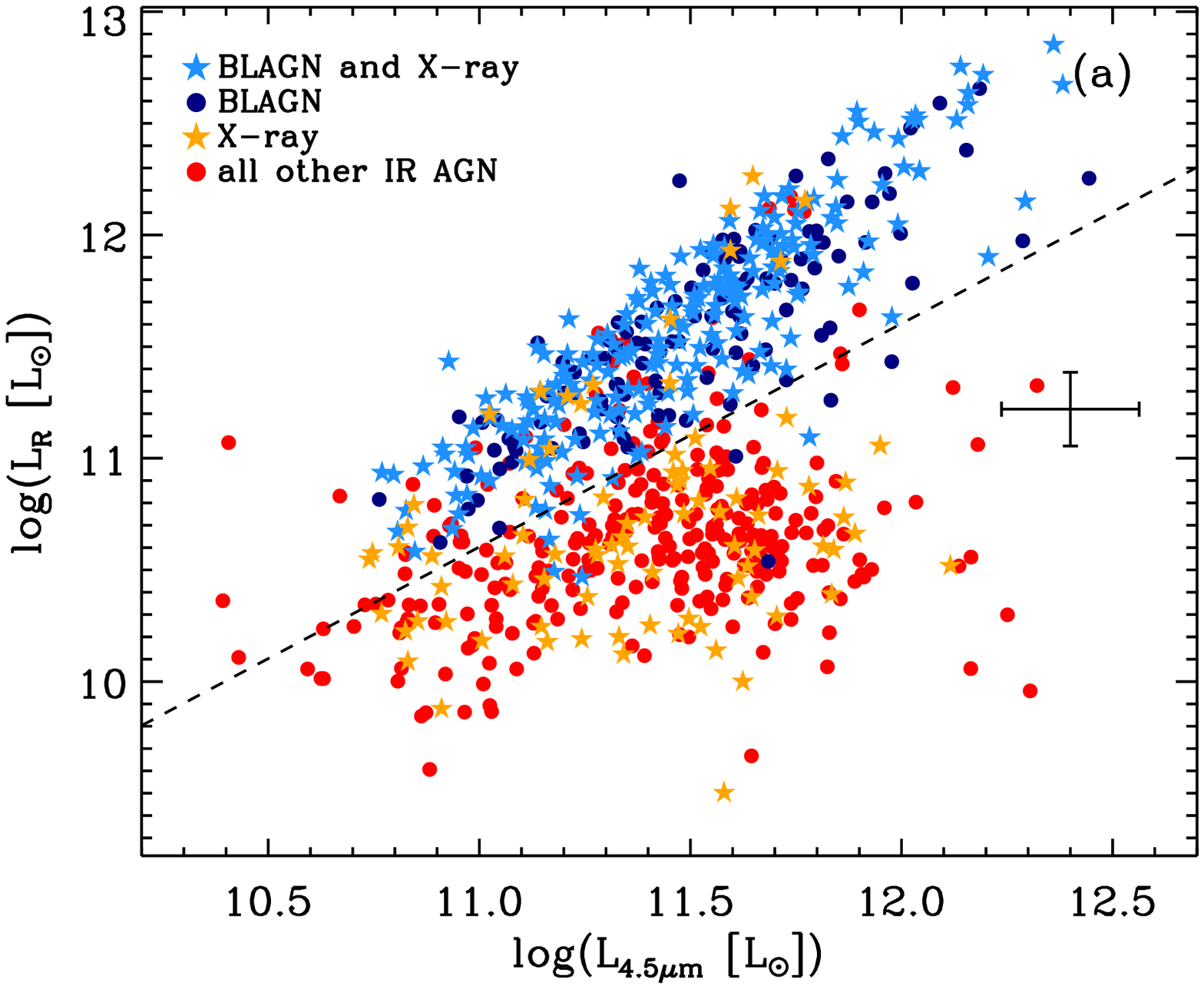}{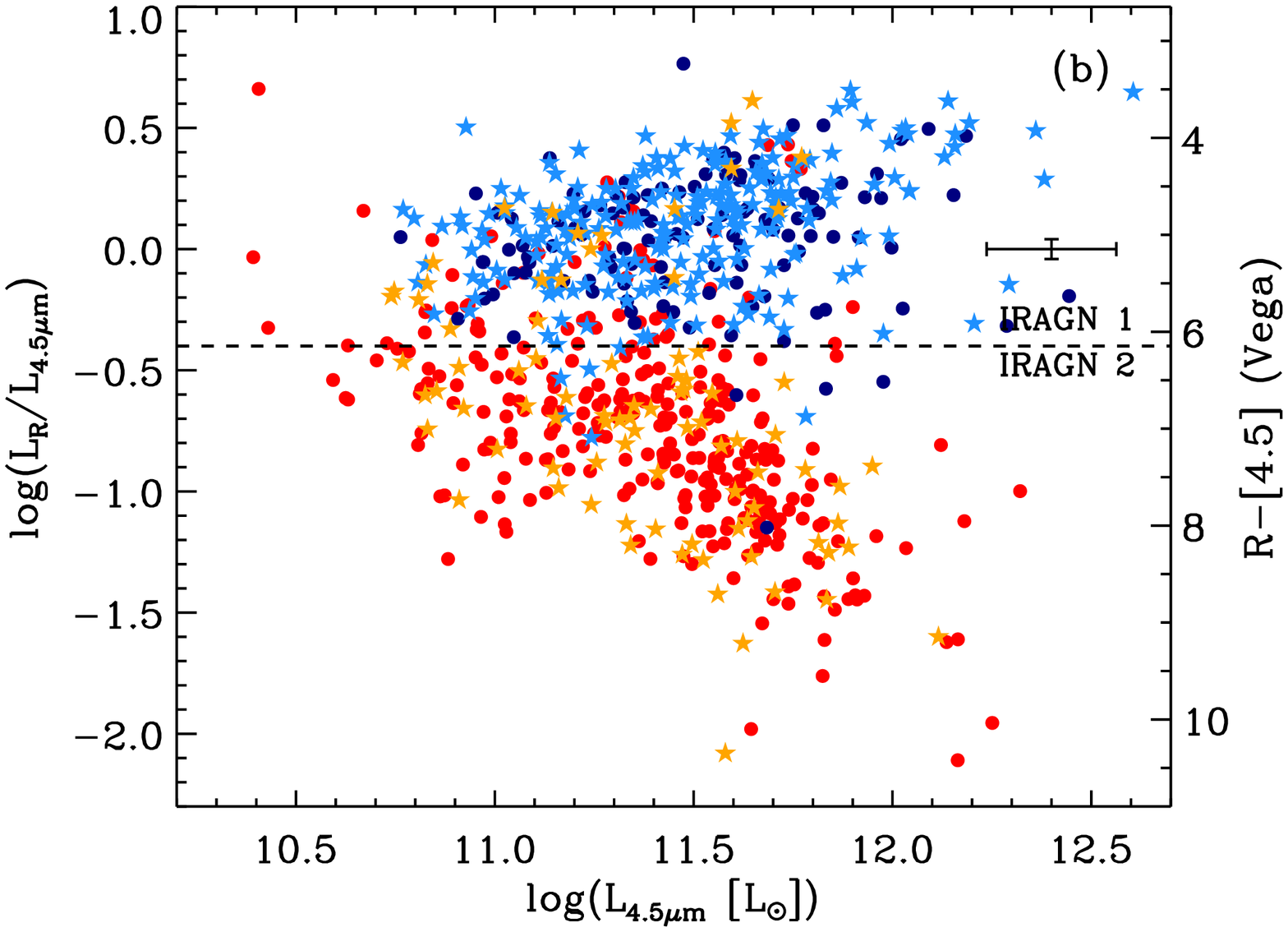}
\plottwo{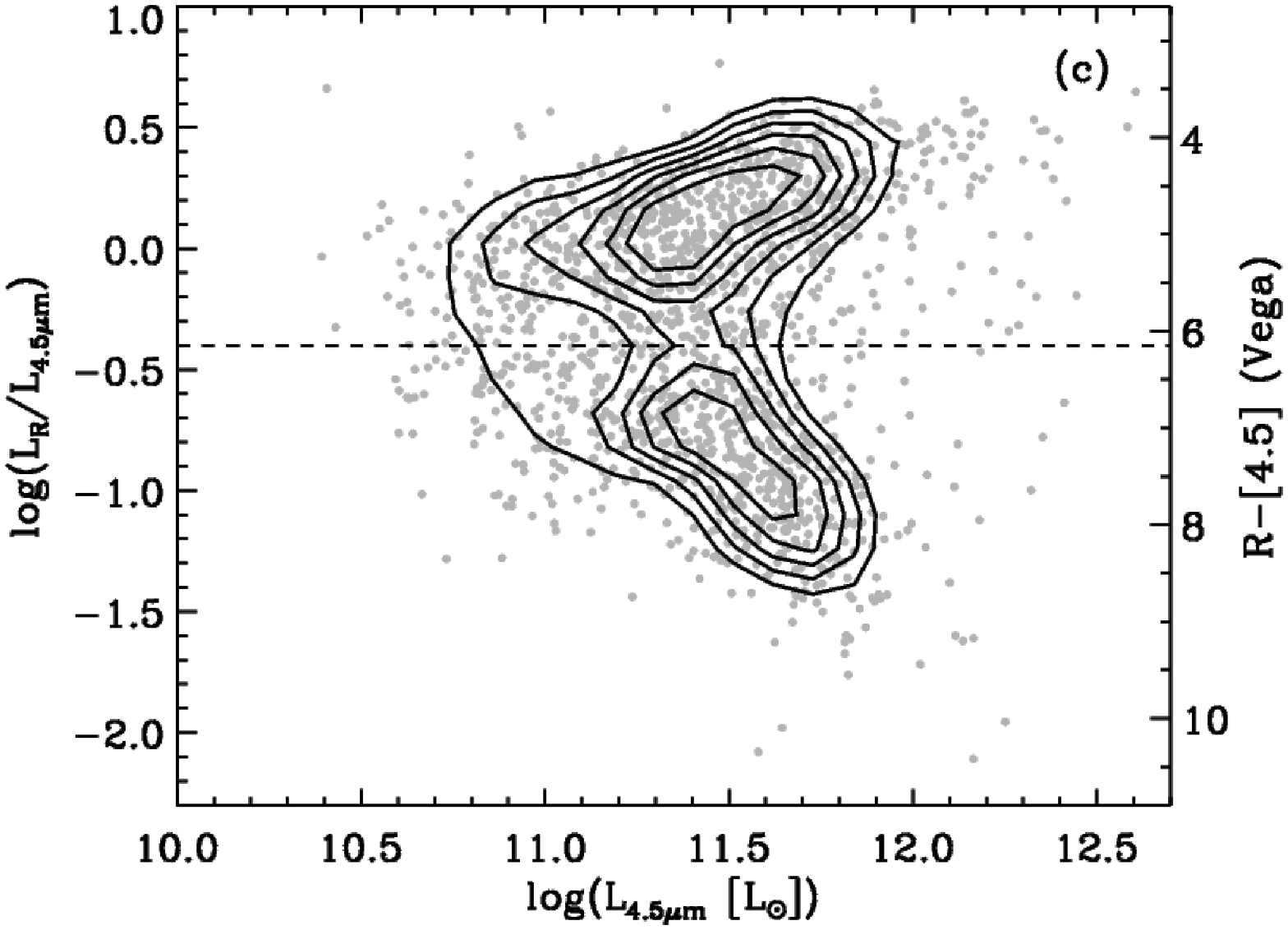}{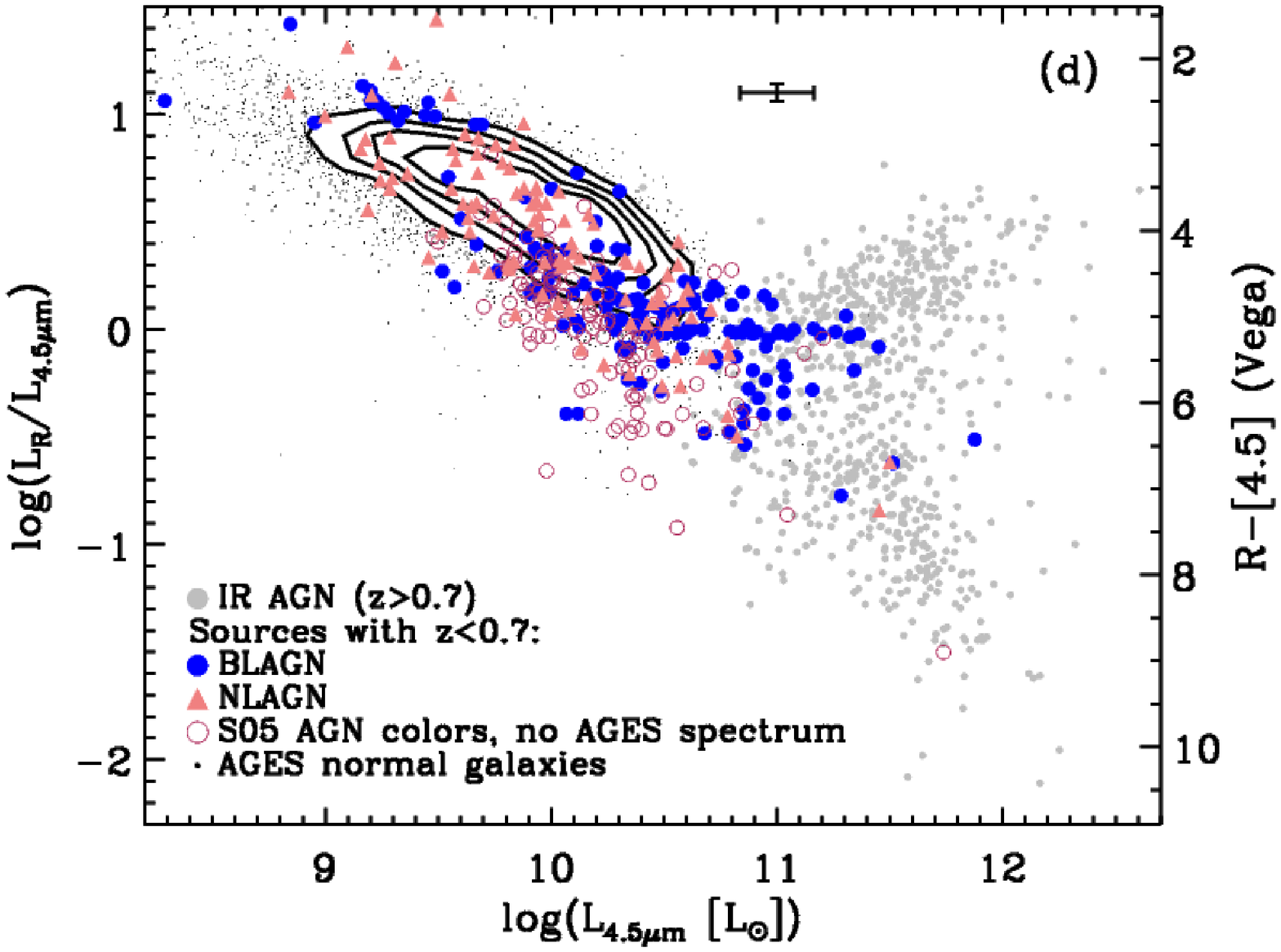}
\caption{ (a) $L_{R}$ vs.\ \lirac\ and (b) \iiu\ vs. \lirac\ for
  IR-selected AGNs.  Objects with optical spectroscopic
  classifications as BLAGNs are shown in blue, and X-ray sources are
  shown as stars.  For clarity, only one out of every two objects is
  shown.  Empirical selection criteria to separate IRAGN 1s and IRAGN
  2s are shown by the dashed line.  Error bars show the median
  uncertainties for objects lacking spectroscopic redshifts (objects
  with spectroscopic redshifts have much smaller uncertainties in
  luminosity).  Luminosity uncertainties include both redshift and
  flux uncertainties; note that the error bars in (a) are not
  independent as the luminosity errors are dominated by uncertainty in
  $z$.  (c) shows the same points as (b), but includes contours of
  source density that clearly show the bimodal distribution in \iiu.
  (d) show the same distribution as (b), but includes sources at
  $z<0.7$ with various optical spectroscopic classifications.
\label{figuvir}}
\vskip0.3cm
\end{figure*}

For all AGN and galaxy models, we account for neutral hydrogen
absorption in the intergalactic medium by setting the templates equal
to zero blueward of the Lyman limit (912 \AA), which is probed by the
shortest-wavelength ($B_W$) band only for redshifts $z>2.7$.  Additional
absorption by Ly$\alpha$ becomes significant at $z\gtrsim3$ and
depends strongly on redshift; this absorption can be as strong as 50\%
at $\lesssim4$. \citep[e.g.,][]{beck07}. However, only 42 (3\%) of the
objects in our sample lie at $z>3$, so for simplicity we ignore
redshift-dependent Ly$\alpha$ absorption in our templates.

For each IR-selected AGN in our sample, we perform $\chi^2$ fits to
the optical and IRAC photometry with the redshifted sum for each
combination of starburst template and AGN power law (for this analysis
we omit the 10 sources that do not have detections in all three NDWFS
bands).  We leave the normalizations of the AGN and galaxy components
as free parameters, and we convolve the template spectra with the
appropriate Mosaic-1 and IRAC response functions\footnotemark.  From
the template with the lowest $\chi^2$, we derive the best fit
$\alpha_\nu$ and $A_V$ for the AGN.  Example fits are shown in Fig.\
\ref{figtempall}.  From the best-fit template, we also calculate
$K$-corrected luminosities $L_{2500\; \AA}$ and $L_{2\mu m}$,
corresponding to the rest-frame $\nu L_{\nu}$ at 2500 \AA, and 2
$\mu$m, respectively.  These wavelengths are probed by the optical and
IRAC photometry for all sources with $0.7<z<2.7$. The effects of
$K$-corrections are discussed in \S~\ref{kcor}.

There is some evidence that the extinction in AGNs is best
described by curves observed for the LMC and SMC, which have greater
extinction in the UV than is observed in the Galaxy.  To check the
dependence of the fit parameters on the choice of extinction curve, we
re-perform the SED fits using LMC and SMC curves \citep{pei92}.  These
do not significantly alter the quality of the fits, although the SMC
curve gives somewhat lower $A_V$ estimates for some objects with
$A_V\sim 1$.

\footnotetext{
{\tt http://www.noao.edu/kpno/mosaic/filters/filters.html} and  {\tt http://ssc.spitzer.caltech.edu/irac/\\spectral\_response.html}.}

\subsection{Color selection of obscured AGNs}
\label{identify}

The distribution in the best-fit $A_V$ from the optical/IR SED fits is
shown in Fig.~\ref{figav}.  The AGN extinctions are
bimodal, with a large fraction of sources having $A_V\gtrsim1$,
suggesting that the IRAC selection includes many obscured AGNs.  We do
not expect moderate extinction to strongly affect the IRAC color-color
selection because the (rest-frame) near- and mid-IR emission that is
probed with IRAC suffers relatively little obscuration by gas or dust
compared to the optical, UV, or soft X-ray bands (although the near-IR
can be extincted for sufficiently large $A_V$).  As shown by the
models in the lower two panels of Fig.~\ref{figtempall}, nuclear
emission with significant extinction in the optical can still dominate
over emission from the host galaxy in the IRAC bands.  Because the
extinction curve for the mid-IR is relatively flat
\citep[e.g.,][]{pei92, inde05}, extinction does not significantly
affect the shape of the observed IRAC spectrum.  Therefore, IRAC
color-color selection can identify AGNs even for $A_V\sim30$, as shown
in Fig.~\ref{figcol}.

\begin{figure}
\epsscale{1.2}
\plotone{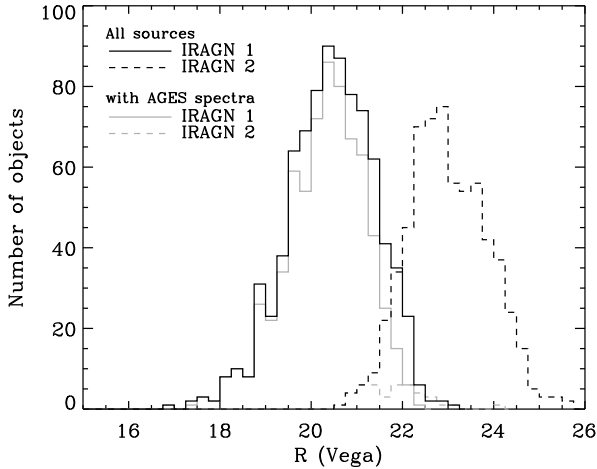}
\caption{Distribution in $R$ for the IRAGN 1 and IRAGN 2 subsets.  Gray lines show the distribution for sources with AGES spectroscopy.  
\label{figr}}
\end{figure}
 
\begin{figure}
\epsscale{1.2}
\plotone{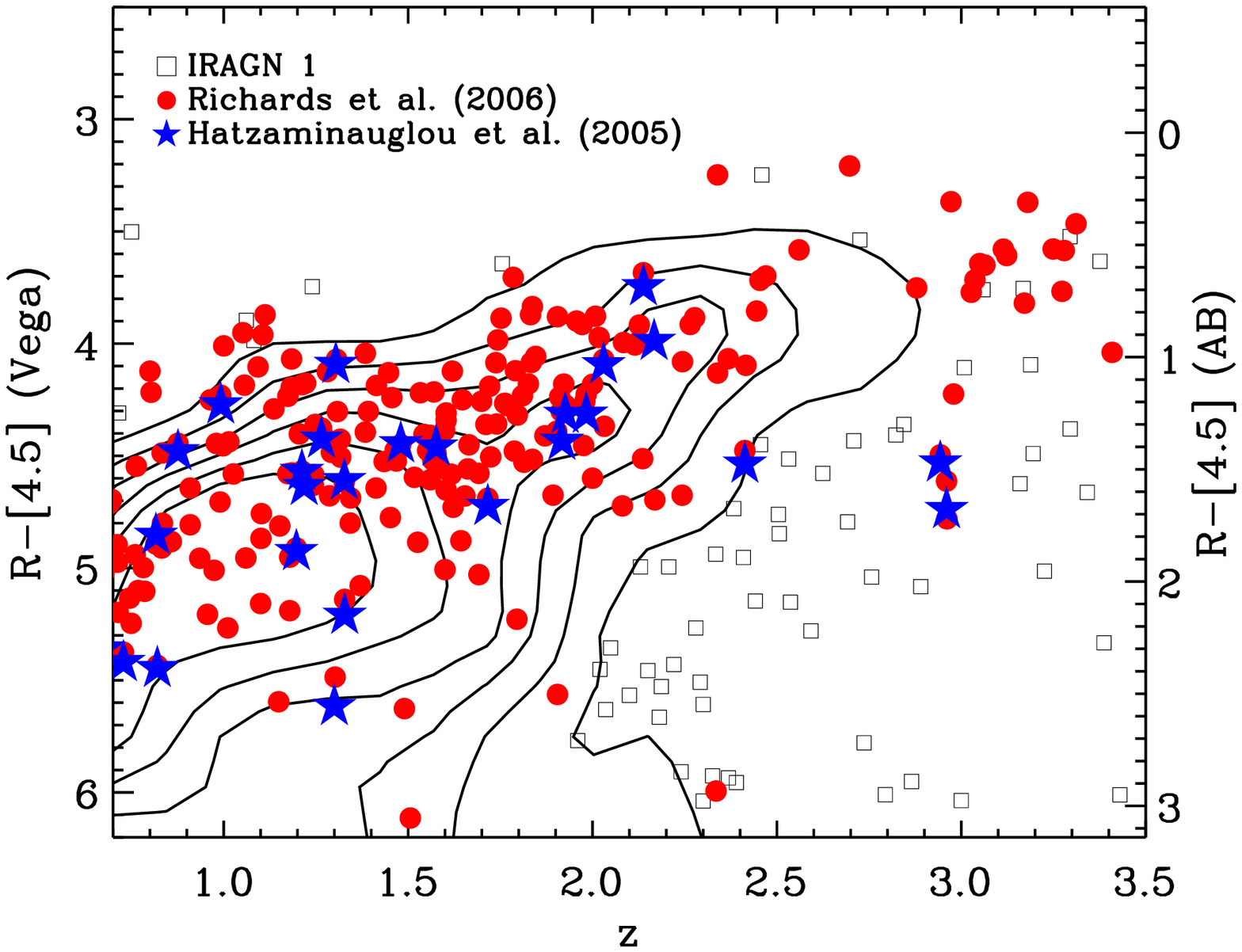}
\caption{Optical/mid-IR color ($R-[4.5]$) vs. redshift of IRAGN 1s,
compared to type 1 AGN selected from other works. Note that blue
colors are the top and red at the bottom, to correspond to the plots
in Fig.~\ref{figuvir}.  Squares and density contours are for IRAGN 1s.
Red circles and blue stars show type 1 AGN from \citet{rich06} and
\citet{hatz05}, respectively.  The figure shows that most the
color-selected IRAGN 1s show a similar distribution in color versus
redshift as other samples, although our sample includes more
moderately reddened AGNs (with $R-[4.5] > 5.5$).
\label{figcol_noabs}}
\vskip0.3cm
\end{figure}
 
\begin{figure}
\epsscale{1.2}
\plotone{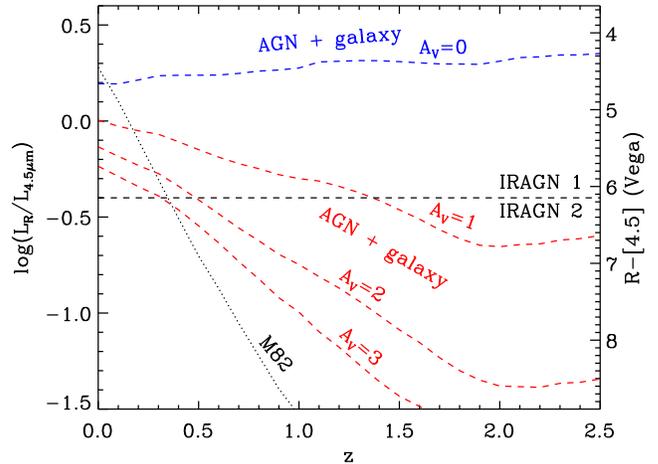}
\caption{ $L_{R}/L_{\rm 4.5 \mu m}$ vs. redshift for a template of AGN
  plus host galaxy (as described in \S\ \ref{lum}), with different
  values of the extinction on the AGN template.  Because at higher $z$
  the optical observations probe shorter wavelengths, where the
  extinction curve is steeper, $L_{R}/L_{\rm 4.5 \mu m}$ varies more
  strongly with $A_V$.  Therefore candidate obscured AGN can be more easily
  selected on the basis of optical-IR color at $z>0.7$, to which we
  restrict our sample.  We also show, for comparison, the colors of a
  template for the starburst M82 \citep{sieb07}, and the obscured AGN
  selection cut shown in Fig.~\ref{figuvir}.
\label{figtempav}}
\vskip0.3cm
\end{figure}

\begin{figure}
\epsscale{1.2}
\plotone{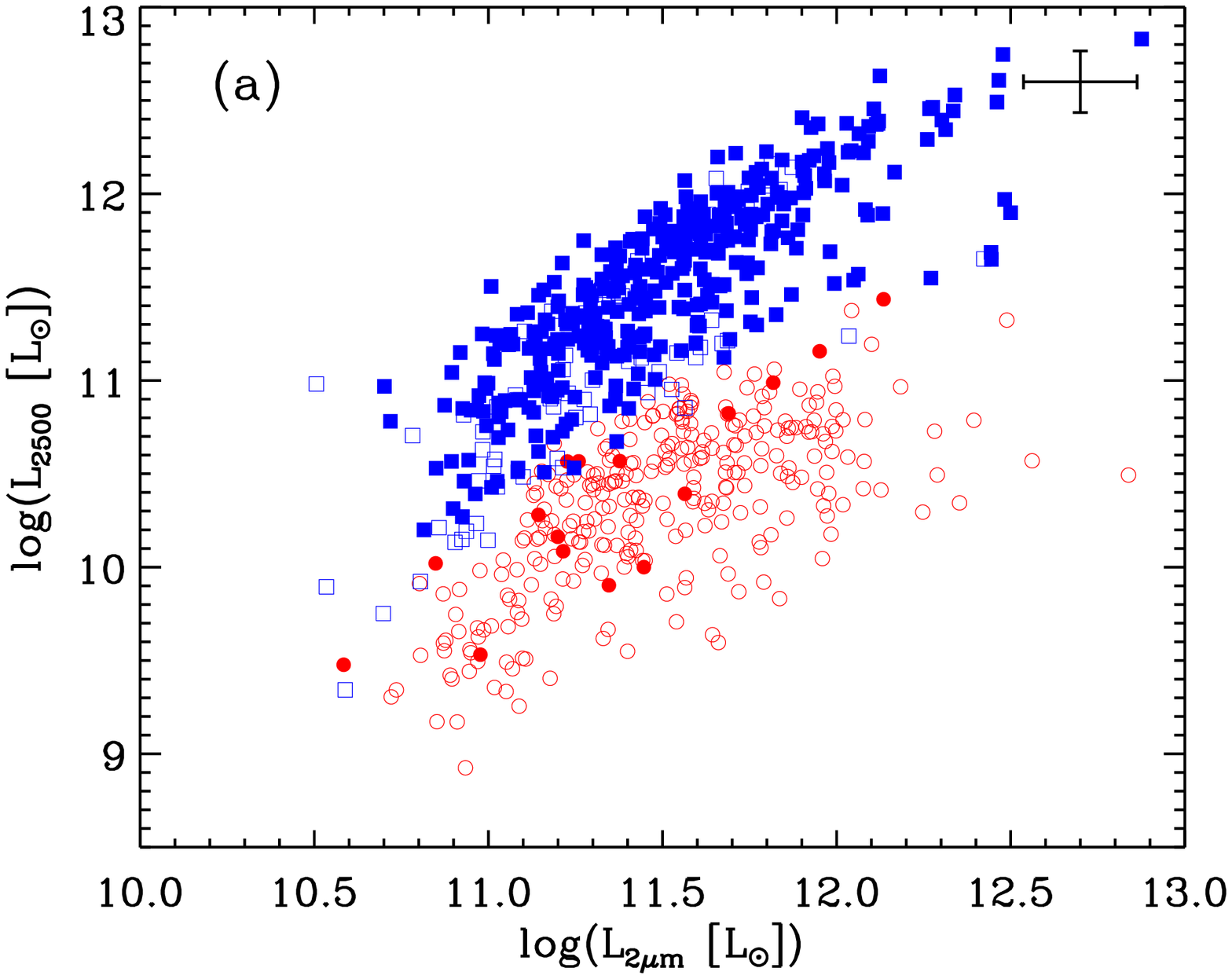}
\plotone{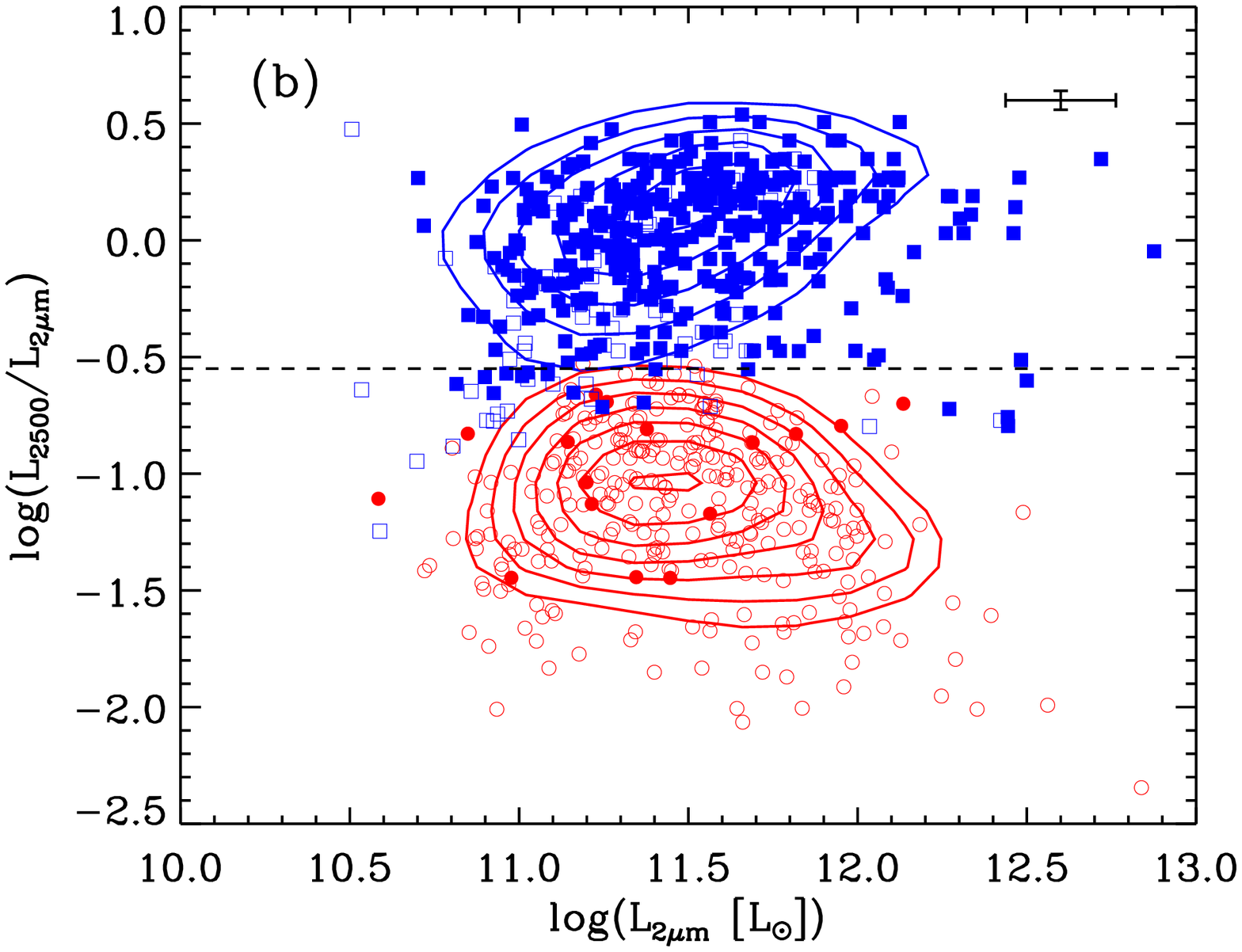}
\caption{ Similar to Fig.\ \ref{figuvir}, with $K$-corrected
  luminosities $L_{\rm 2500\; \rm \AA}$ and $L_{\rm 2 \mu m}$
  calculated from template fits (\S\ \ref{lum}).  IRAGN 1s
  (blue squares) and IRAGN 2s (red circles) are classified as defined
  in \S\ \ref{identify} and shown in Fig. \ref{figuvir}.  For clarity,
  only one out of every two objects is shown.  Filled symbols show objects
  with optical spectroscopic redshifts, while empty symbols have only
  photo-$z$'s.  Errors are as in Fig.\ \ref{figuvir}.
\label{figuvirk}}
\vskip0.5cm
\end{figure}

Since extinction by dust is much stronger in the optical than the IR,
simple optical and IR color criteria (rather than detailed SED fits)
can be used to select obscured objects.  In Fig.~\ref{figuvir} (a) we
plot \lr\ versus \lirac\ for the IR-selected AGNs.  This plot shows
two separate distributions of sources.  The first has \lr\ values that
rise along with the \lirac\ and contains nearly all (96\%) of the
IR-selected AGNs that have optical spectra of BLAGNs.  A total of 79\%
of these objects have relatively low extinction $(A_V<1)$ from the SED
fits, so we associate them with candidate unobscured AGNs and classify
them as type 1 IR-selected AGNs (IRAGN 1s).  The second population has
lower values of \iiu, and 98\% with $A_V\geq 1$, so we associate these
with candidate obscured AGNs (IRAGN 2s).  Unfortunately, spectroscopic
classification is of little help with the IRAGN 2s, since most of them
are fainter than the spectroscopic limits of AGES (Fig.\
\ref{figr}). The 4\% of the BLAGNs that lie in this ``obscured'' region
have red colors in the optical, with $R-I\sim0.7$--1.2, compared to
$R-I\sim0.3$ for a typical unreddened BLAGN.

We elucidate the distinction between the subsets by plotting the
quantity \iiu\ (or equivalently in magnitudes, $R-[4.5]$) versus
\lirac\ in Fig.\ \ref{figuvir} (b) and (c).  The contours in
Fig.~\ref{figuvir}(c) show that the distribution in \iiu\ is bimodal,
so that there are two distinct populations.  We empirically define
the boundary between these two populations to be $\log{(L_{R}/L_{\rm
4.5 \mu m})}=-0.4$, corresponding to $R-[4.5]=6.1$ (Vega) or $R-[4.5]=3.1$
(AB), as shown in all four plots in Fig.\ \ref{figuvir}.  We select
this boundary (1) to divide the region populated by AGES BLAGNs from
the region with few BLAGNs, and (2) to bisect the bimodal distribution
in \iiu\ shown in Fig.~\ref{figuvir}(c).  Because this boundary is
based in part on the AGES spectral classifications, it is possible that our
selection may be biased by the fact that AGES did not target optically
fainter sources.  However, as we show in \S\ \ref{xrayres}, X-ray
analysis independently confirms the division at
$\log{(L_{R}/L_{\rm4.5\mu m})}=-0.4$.  This criterion selects 839
IRAGN 1s and 640 IRAGN 2s.

 The IRAGN 1s have mid-IR/optical colors similar to those found for
other samples of type 1 AGNs.  Fig.~\ref{figcol_noabs} shows the
distribution in $R-[4.5]$ for the IRAGN 1s, with comparisons to
samples from \citet{rich06} and \citet{hatz05}.  Most of the IRAGN 1s
show the same trend in redshift and color as these previous samples,
although the IRAGN 1s include more moderately reddened AGNs (with
$R-[4.5]>5.5$), which make up 24\% of the total number of IRAGN
1s.

The color distribution in Fig.~\ref{figuvir} can be interpreted in
terms of how the observed $R$ and IRAC fluxes for AGNs change with
extinction and redshift.  In Fig.~\ref{figtempav} we show \iiu\
versus $z$ for a template including an elliptical host galaxy plus AGN
(with $\alpha_\nu=-1$ and various $A_V$) as described in \S\
\ref{lum}.  The model AGN has an unabsorbed, rest-frame $R$-band
luminosity 5 times that of the host galaxy.  Fig.~\ref{figtempav}
shows that extinction of the AGN component decreases the observed
\iiu. This decrease becomes larger at higher $z$ because the $R$ band
probes shorter wavelengths in the rest-frame UV, where dust extinction
is greater.  Because obscured AGNs at higher redshift tend to have higher \lirac\ (owing to the flux limits of the survey), more
luminous objects appear redder in the observed \iiu\ for the same
$A_V$.  This explains the decrease in \iiu\ with \lirac\ observed in
Figs.~\ref{figuvir}(b) and (c).  

For starburst galaxies (shown by the M82 template), \iiu\ changes even
more strongly with redshift; at low $z$, starbursts, obscured AGNs,
and unobscured AGNs can have similar values of \iiu.  However, for most
redshifts, all but the most extincted starbursts are not selected by
the \citetalias{ster05} IRAC color-color criteria; the colors for M82
and NGC 7714 fall in the \citetalias{ster05} region only at $z\gtrsim
3$.  We discuss possible contamination from these objects in
\S~\ref{contamination}.

\subsection{$K$-corrected colors}
\label{kcor}
In Fig.\ \ref{figuvirk} we include the $K$-corrections to the
IR-selected AGN luminosities and plot \luv\ versus \lumtwo.  The
$K$-corrections have negligible effect on the color classification; if
we apply an equivalent empirical boundary to separate IRAGN 1s and 2s
using the $K$-corrected luminosities [$\log{(L_{\rm 2500\; \AA}/L_{\rm
2\mu m})}=-0.55$, shown by the dashed line in Fig.~\ref{figuvirk}],
only 90 of the 1479 objects (6\%) change their
classification. Therefore, almost all the IRAGNs can be empirically
classified by their observed colors, independent of $K$-corrections,
which allows this criterion to be used for samples that do not include
accurate redshifts.

\subsection{Dependence of color selection on luminosity and redshift}
The template used in Fig.~\ref{figtempav} represents a luminous AGN
that dominates the optical emission from the host galaxy.  For
lower-luminosity AGNs, whose unobscured optical flux is smaller than
that of their hosts, extinction of the nucleus will have a
relatively small effect on \iiu.  Therefore, our selection criterion
is not applicable for samples of sources at lower luminosities and
redshifts.  In Fig.~\ref{figuvir}(d), we show \iiu\ versus \lirac\ for
subsets of objects at $z<0.7$, for comparison to the $z\geq0.7$ IRAGN
sample.  At low \lirac, optical BLAGNs and NLAGNs have \iiu\ typical of
normal galaxies, indicating that their total emission is dominated by
the hosts, and this simple color criterion cannot distinguish obscured sources.

By cutting our IRAGN sample at $z=0.7$, we include only sources with
$L_{\rm 4.5\mu m} \gtrsim 10^{11}$ \lsun\ (owing to the flux limits of
the IRAC Shallow Survey). These AGNs are luminous enough that if
unobscured, their nuclear optical luminosity is comparable to all but
the most luminous host galaxies.  Therefore, our redshift cut at
$z\geq0.7$ enables obscured AGN color selection, (1) by probing
shorter rest-frame wavelengths in the optical and (2) by selecting
luminous AGNs for which the intrinsic optical luminosity is larger
than the host.

\subsection{Are the IRAGN 2s intrinsically optically faint?}
We consider the possibility that the IRAGN 2s are not obscured, but
intrinsically faint in the observed optical band.  For example, there
exist modes of accretion that lack a luminous accretion disk and
therefore do not radiate strongly in the optical and UV; these are
known as radiatively inefficient accretion flows
\citep[e.g.,][]{nara95}.  However, in such a scenario it is difficult
to explain the observed properties of IRAGN 2s in the mid-IR.  The red
IRAC colors and high mid-IR luminosities of these objects are
characteristic of dust that has been heated to high temperatures by
high UV fluxes, and so imply some luminous UV emission from the
nucleus.  Such emission would not be present in radiatively
inefficient flows.  Therefore, we hypothesize that all the IRAGN 2s are
intrinsically luminous enough in the UV to power the observed mid-IR
emission, but they are optically faint because the nuclear emission is
obscured.

\subsection{Bolometric luminosities}
\label{bolometric}
A fundamental property of AGNs is the bolometric accretion luminosity
$L_{\rm bol}$.  For the IRAGN 2s, the nuclear optical light is
extincted, and most objects are not individually detected in X-rays,
so we cannot use optical or X-ray luminosities to estimate $L_{\rm
bol}$.  Instead, we derive $L_{\rm bol}$ by scaling from the
$K$-corrected luminosity of the AGN at 2 \micron, $L_{\rm 2 \mu
m}^{\rm AGN}$, taken from the SED fits (\S\ \ref{lum}).  $L_{\rm bol}$
is given by $L_{\rm bol}=BC_{2 \rm \mu m}L_{\rm 2 \mu m}^{\rm
AGN}$, where $BC_{2 \rm \mu m}$ is the bolometric correction. We
derive $BC_{2 \rm \mu m}$ from the luminosity-dependent quasar SED
model of \citet{hopk07qlf}, for which the correction is in the range
$BC_{2 \rm \mu m}=10$--15 for the luminosities of the sample (we note
that the luminosity-independent model of \citet{rich06} gives a
similar  $BC_{2 \rm \mu m}=12$).

\begin{figure}
\epsscale{1.2}
\plotone{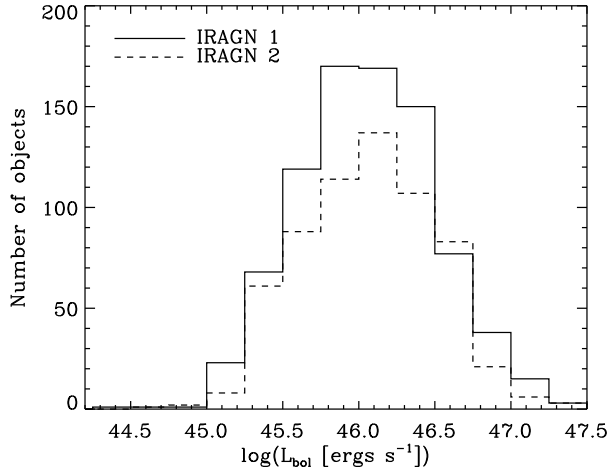}
\caption{Distribution in bolometric luminosity $L_{\rm bol}$
estimated as described in \S~\ref{bolometric}, for the
two types of IR-selected AGN. \label{figlbol}}
\vskip0.3cm
\end{figure}

The distributions in \lbol\ are shown in Fig.\ \ref{figlbol}.  The
$L_{\rm bol}$ values of (0.1--10)$\times10^{46}$ \ergs\ are similar
for IRAGN 1s and 2s and are typical of the accretion luminosities of
bright Seyferts and quasars.  We see no systematic difference between
the distributions for the two types of IRAGNs, indicating that at
these high luminosities, the fraction of obscured to unobscured
sources is relatively constant with luminosity.  However, these
results give only approximate distributions in $L_{\rm bol}$ because
of uncertainties in the photo-$z$'s for individual objects,
particularly IRAGN 2s (see \S~\ref{photoz}).  While it would be very
interesting to use this sample to study quantities such as the
evolution of the obscured AGN fraction with $z$ or $L_{\rm bol}$, to
confidently perform such measurements requires better calibration of
the IRAGN 2 redshifts.

\section{Multiwavelength tests of obscured AGN selection}

\label{tests}

Our classification of candidate AGNs as unobscured (IRAGN 1) or obscured
(IRAGN 2) is based solely on the ratio of their observed optical
to IR color.  This classification makes several predictions for the
observed emission from these sources at X-ray, optical, and infrared
wavelengths:

\begin{enumerate}

\im The average X-ray properties of the two populations should be
consistent with unabsorbed and absorbed AGNs, respectively.  Both
types should have high X-ray luminosities typical of Seyferts and
quasars.  The IRAGN 1s should have X-ray spectral shapes consistent
with unabsorbed AGNs, while $>$70\%--80\% of the IRAGN 2s should have
harder X-ray spectra due to absorption by neutral gas \citep[e.g.,][]{tozz06}.

\im For IRAGN 2s, the observed X-ray absorption should be consistent
with the extinction derived from the optical/UV colors, for a
reasonable gas-to-dust ratio.

\im The IRAGN 1s and 2s  should have optical morphologies and optical colors
characteristic of BLAGNs and galaxies, respectively.

\end{enumerate}

In \S\S\ \ref{xray}--\ref{galedd} we test each of these predictions
using the available data from \chandra, \spitzer, and optical
photometry and spectroscopy.  In each case we show that the data are
consistent with the above classification of IR-selected AGNs as
unobscured (IRAGN 1) and obscured (IRAGN 2).  

\subsection{X-ray properties}
\label{xray}

\begin{deluxetable*}{lcccccc}
\tabletypesize{\scriptsize}
\tablewidth{6.5in}
\tablecaption{Average X-ray fluxes from stacking \label{tblxray}}
\tablehead{
\colhead{} &
\colhead{Number} &
\multicolumn{2}{c}{0.5--2 keV} &
\multicolumn{2}{c}{2--7 keV} \\
\colhead{Subset} & 
\colhead{of sources\tnm{a}} &
\colhead{Counts source$^{-1}$\tnm{b}} &
\colhead{Flux source$^{-1}$\tnm{c}} &
\colhead{Counts source$^{-1}$\tnm{b}} &
\colhead{Flux source$^{-1}$\tnm{c}} &
\colhead{$HR$}}
\startdata
\multicolumn{6}{c}{\it All sources} \\
IRAGN 1s & 346 & $7.64\pm0.15$ & $ 9.90\pm 0.19$ & $2.76\pm0.09$ & $10.9\pm 0.4$\W & $-0.47\pm 0.02$  \\
IRAGN 2s & 267 & $2.48\pm0.10$ & $ 3.21\pm 0.12$ & $1.65\pm0.08$ & $ 6.50\pm 0.31$ & $-0.20\pm 0.03$  \\
Normal galaxies & 2107 & $0.49\pm0.02$ & $ 0.63\pm 0.02$ & $0.33\pm0.01$ & $ 1.30\pm 0.05$ & $-0.19\pm 0.03$  \\

\multicolumn{6}{c}{\it Non X-ray detected sources} \\
IRAGN 1s & 122 & $0.92\pm0.09$ & $ 1.20\pm 0.11$ & $0.47\pm0.06$ & $ 1.86\pm 0.25$ & $-0.32\pm 0.08$ \\
IRAGN 2s & 179 & $0.41\pm0.05$ & $ 0.53\pm 0.06$ & $0.46\pm0.05$ & $ 1.83\pm 0.20$ & \W$ 0.06\pm 0.09$ \\
Normal galaxies & 2011 & $0.12\pm0.01$ & $ 0.15\pm 0.01$ & $0.08\pm0.01$ & $ 0.33\pm 0.03$ & $-0.17\pm 0.05$

\enddata

\tnt{a}{Only sources at an angular distance $<$6\arcmin\ from the
  \chandra\ optical axis are included in the stacking analysis.}
\tnt{b}{Source counts shown are equal to 1.1 times the observed source
counts, to account for flux outside the $r_{90}$ source aperture.} 
\tnt{c}{All fluxes are in units of $10^{-15}$ \flux.} 
\end{deluxetable*}

X-ray emission is an efficient and largely unbiased way of detecting
AGN activity for objects with $N_{\rm H}\lesssim10^{24}$ \cdens\
\citep[for reviews see][]{mush04book, bran05}.  Thus, the contiguous
\chandra\ coverage of the \bootes\ field provides a useful diagnostic
for confirming our classifications of IR-selected AGNs
\citep[S05;][]{gorj07}, allowing us to estimate both the X-ray
luminosity \lx\ and the absorbing neutral hydrogen column density \nh.

The main limitation of the wide-field X\bootes\ observations is that
they are shallow, with exposures of only 5 ks yielding a 0.5--7 keV
source flux limit of $(4-8)\times10^{-15}$ \flux.  Most IR-selected
AGNs do not have firm X-ray detections, and most detected sources have
fewer than 10 counts, so we do not have X-ray spectral information for
most individual sources.  We therefore perform a stacking analysis,
which compensates for the shallowness of the X-ray observations by
averaging over the large number of IR-selected AGNs in the field.  By
summing X-ray images around the known IR positions, we determine the
average X-ray fluxes, luminosities and spectral shapes of various
subsets of these sources.

\subsubsection{X-ray stacking}
\label{xraystack}
Around the position of each object in a given sample, we extract
$40\times40$-pixel (19.7\arcsec) X-ray images in the soft band (0.5--2 keV)
and hard band (2--7 keV).  Because the \chandra\ telescope PSF varies with angle $\theta$ from the optical axis, the
aperture from which we extract source photons varies from source to
source.  We take this aperture to be the 90\% energy encircled radius
at 1.5 keV:\footnotemark
\begin{equation}
r_{90}\simeq 1\arcsec+10\arcsec(\theta/10\arcmin)^2.
\end{equation} 
We include in the stacking analysis only objects that lie within
6\arcmin\ of the optical axis of a \chandra\ pointing, for which
$r_{90}<4\farcs6$.  This excludes over half the available sources
but minimizes source confusion and maximizes signal-to-noise ratio.  Using a
model of the \chandra\ PSF (from the \chandra\ CALDB) and sources with
random positions inside the 6\arcmin\ radius, an aperture of $r_{90}$
includes 90\%--92\% of the source counts in both the 0.5--2 keV and 2--7
keV bands.  Accordingly, in our stacking analysis, we multiply the
observed source counts by 1.1 to obtain the total counts from the
source.

 \footnotetext{\chandra\ Proposer's Observatory Guide (POG), available at
 {\tt http://cxc.harvard.edu/proposer/POG}.}

Of the 126 pointings in the X\bootes\ data set, there are
eleven\footnotemark\ that have significantly higher background
intensities, due to background flares \citep[for a detailed discussion
of ACIS backgrounds see][]{hick06a}.  In our stacking analysis, we do
not include any source positions that lie within these eleven ``bad''
exposures.  The total area over which we perform the stacking, which consists
of the region covered by IRAC that lies within the central 6\arcmin\
radii of these 115 pointings, is 2.9 deg$^2$.

\footnotetext{ObsIDs 3657, 3641, 3625, 3617, 3601, 3607, 3612, 3623,
  3639, 3645, and 4228.}

An accurate measure of the stacked source flux requires subtraction of
the background, which we estimate by stacking X-ray images on random
positions around the \bootes\ field, at least 20\arcsec\ away from any
X-ray sources included in the X\bootes\ catalog \citep{kent05}.  We
performed 20 trials, stacking $\sim$30,000 positions in each trial.
As a check, we also calculate the surface brightness using the ACIS
blank-sky data sets,\footnotemark\ which are obtained using deep
exposures at high Galactic latitude and removing all detected sources.

\footnotetext{{\tt http://cxc.harvard.edu/contrib/maxim/acisbg/}}

Both estimates of the diffuse background give identical
surface brightnesses of 3.0 counts s$^{-1}$ deg$^{-2}$ in the 0.5--2 keV
band and 5.0 counts s$^{-1}$ deg$^{-2}$ in the 2--7 keV band.  We use
these values to calculate the expected background counts within a
circle of radius $r_{90}$ for each source position.  For a typical
$r_{90}=3$\arcsec\ and an exposure time of 4686 s (see below), this
corresponds to 0.03 and 0.05 background counts for each IRAC source in the
0.5--2 keV (soft) and 2--7 keV (hard) bands, respectively. 

Subtracting this background, we obtain the average X-ray flux in
counts source$^{-1}$.  We assume that all source positions have an X-ray
exposure time of 4686 s, which is the mean for all the X\bootes\
observations excluding the ``bad'' exposures.  For simplicity we
ignore variations in exposure time between pointings, as well as
variations in effective exposure time within each single ACIS-I field
of view due to mirror vignetting.  These variations are at most
$\sim$10\% and do not significantly affect our results.  We convert
count rates (in counts s$^{-1}$) to flux (in \flux) using the conversion
factors $6.0\times10^{-12}$ ergs cm$^{-2}$ count$^{-1}$ in the 0.5--2 keV band and
$1.9\times10^{-11}$  ergs cm$^{-2}$ count$^{-1}$ in the 2--7 keV band.  In addition to fluxes, we
obtain rough X-ray spectral information by calculating the hardness
ratio, defined as
\begin{equation}
{\rm HR}=\frac{H-S}{H+S},
\end{equation}
where $H$ and $S$ are the count rates in the hard and soft bands,
respectively.  Errors in count rates are calculated using the
approximation $\sigma_{X}=\sqrt{X+0.75}+1$, where $X$ is the number of
counts in a given band \citep{gehr86}.  Uncertainties in HR are
derived by propagating these count rate errors.  In the following
analysis, we use these hardness ratios and fluxes to determine the
typical absorption and X-ray luminosities from the stacking analysis.

\begin{figure}
\epsscale{1.2}
\plotone{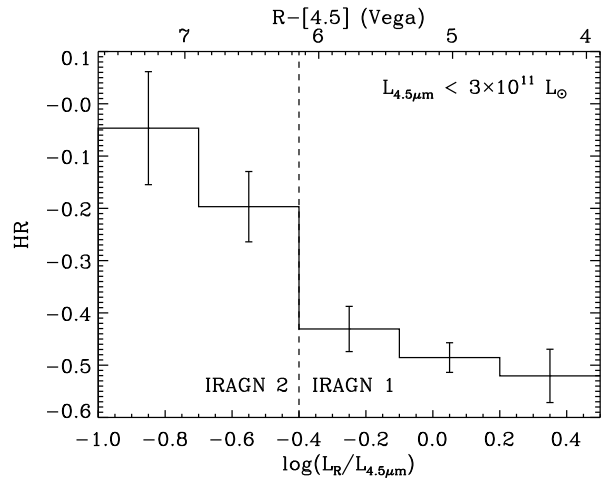}
\caption{Verification of IRAGN 2 selection using X-ray data.
  Shown is the average hardness ratio from X-ray stacking analysis, for
  objects with $L_{\rm 4.5 \mu m}<3\times10^{11}$ \lsun, in bins of \liiu. The
  dashed line shows the boundary between IRAGN 1 and 2 as defined in
  Fig.\ \ref{figuvir}.  \label{figstackxi}}
\vskip0.3cm
\end{figure}

\begin{figure}
\epsscale{1.2}
\plotone{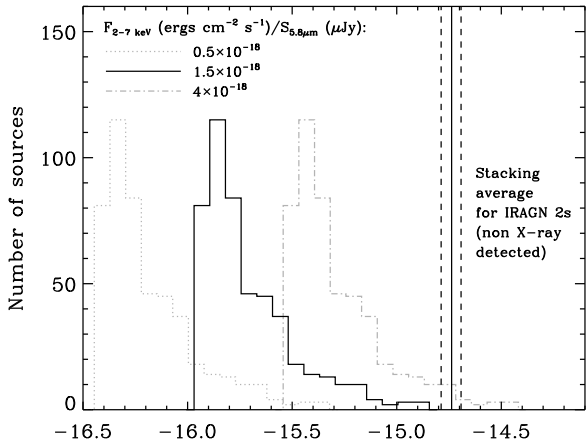}
\caption{Comparison of average 2--7 keV flux for X-ray undetected
  IRAGN 2s, versus X-ray fluxes expected for star formation.  Vertical
  lines show observed average 2--7 keV X-ray flux from stacking
  analysis of $\average{F_{\rm 2-7\;
  keV}}=(1.83\pm0.20)\times10^{-15}$ \flux\ (Table \ref{tblxray}).
  Histograms show the distribution of predicted X-ray fluxes for star
  formation, given the measured $S_{\rm 5.8 \mu m}$ for each object
  and the ratio $F_{\rm 2-7\; keV}/S_{\rm 5.8 \mu m}$ for star
  formation (see Eqn.~\ref{eqnf27}), derived from the $F_{\rm 2-10\;
  keV}/F_{\rm FIR}$ relation of \citet{rana03}, and typical starburst
  SEDs.  The three histograms represent $F_{\rm 2-7\; keV}/S_{\rm 5.8
  \mu m}=(0.5-4)\times10^{18}$ \flux\ $\mu$Jy$^{-1}$, for which the
  average values of $F_{\rm 2-7\; keV}$ are $(0.8-6)\times10^{-16}$
  \flux.  Even for the largest typical ratio of X-ray to 5.8 $\mu$m flux, the
  observed average flux is $>3$ times larger than that expected for star
  formation.   This indicates that the X-ray emission from these objects is
  dominated by nuclear accretion.  \label{figsblx}}
\vskip0.3cm
\end{figure}

\begin{deluxetable*}{cccccccc}
\tabletypesize{\scriptsize}
\tablewidth{6.3in}
\tablecaption{X-ray properties of IR-selected AGNs versus redshift \label{tblxray2}}
\tablehead{
\colhead{} &
\colhead{Number} &
\colhead{${\rm rms}(d_{\rm L})$\tnm{b}} &
\colhead{0.5--2 keV} &
\colhead{$\average{L_{\rm 0.5-2\; keV}}$} &
\colhead{2--7 keV} &
\colhead{$\average{L_{\rm 2-7\; keV}}$} &
\colhead{} \\
\colhead{$z$} & 
\colhead{of sources\tnm{a}} &
\colhead{(Gpc)}  &
\colhead{(counts src$^{-1}$)} &
\colhead{($10^{43}$ \ergs)} &
\colhead{(counts src$^{-1}$)} &
\colhead{($10^{43}$ \ergs)} &
\colhead{$HR$}}
\startdata
\multicolumn{8}{c}{\it IRAGN 1 (all sources)} \\
0.7--1.0 & 65 & \W$ 5.52\pm 0.03$ & $11.9\pm 0.4$\W & $ 5.3\pm 0.2$ & $ 4.4\pm 0.3$ & $ 6.5\pm 0.4$ & $-0.46\pm0.03$ \\
1.0--1.5 & 125 & \W$ 8.76\pm 0.03$ & $ 7.6\pm 0.3$ & $ 8.2\pm 0.3$ & $ 2.5\pm 0.2$ & $ 9.1\pm 0.5$ & $-0.50\pm0.03$ \\
1.5--2.0 & 86 & $13.1\pm 0.1$ & $ 4.4\pm 0.2$ & $10.3\pm 0.6$\W & $ 1.7\pm 0.2$ & $13.2\pm 1.2$\W & $-0.44\pm0.04$ \\
2.0--2.5 & 45 & $17.5\pm 0.1$ & $ 4.3\pm 0.3$ & $17.1\pm 1.3$\W & $ 1.9\pm 0.2$ & $25.0\pm 3.0$\W & $-0.39\pm0.06$ \\

\multicolumn{8}{c}{\it IRAGN 2 (all sources)} \\
0.7--1.0 & 31 & $ 5.6\pm 0.2$ & $ 1.7\pm 0.3$ & $ 0.8\pm 0.1$ & $ 1.9\pm 0.3$ & $ 2.8\pm 0.5$ & \W$ 0.04\pm0.11$ \\
1.0--1.5 & 69 & $ 8.6\pm 0.2$ & $ 2.3\pm 0.2$ & $ 2.5\pm 0.2$ & $ 1.8\pm 0.2$ & $ 6.3\pm 0.7$ & $-0.13\pm0.06$ \\
1.5--2.0 & 72 & $13.2\pm 0.3$\W & $ 3.1\pm 0.2$ & $ 7.2\pm 0.6$ & $ 1.7\pm 0.2$ & $13.6\pm 1.5$\W & $-0.28\pm0.06$ \\
2.0--2.5 & 76 & $18.2\pm 0.5$\W & $ 1.9\pm 0.2$ & $ 8.0\pm 0.8$ & $ 1.0\pm 0.1$ & $14.7\pm 2.0$\W & $-0.28\pm0.07$ \\

\multicolumn{8}{c}{\it IRAGN 1 (non X-ray detected sources)} \\
0.7--1.0 & 15 & $ 5.5\pm 0.1$ & $ 0.63\pm 0.28$ & $ 0.28\pm 0.12$ & $ 0.67\pm 0.29$ & $ 0.97\pm 0.41$ & \W$ 0.03\pm0.31$ \\
1.0--1.5 & 39 & $ 8.9\pm 0.1$ & $ 0.94\pm 0.18$ & $ 1.05\pm 0.20$ & $ 0.49\pm 0.14$ & $ 1.79\pm 0.51$ & $-0.32\pm0.16$ \\
1.5--2.0 & 36 & $13.0\pm 0.1$\W & $ 0.86\pm 0.18$ & $ 1.99\pm 0.43$ & $ 0.40\pm 0.14$ & $ 3.06\pm 1.04$ & $-0.37\pm0.17$ \\
2.0--2.5 & 21 & $17.4\pm 0.2$\W & $ 0.87\pm 0.26$ & $ 3.47\pm 1.02$ & $ 0.37\pm 0.19$ & $ 4.92\pm 2.47$ & $-0.40\pm0.24$ \\

\multicolumn{8}{c}{\it IRAGN 2 (non X-ray detected sources)} \\
0.7--1.0 & 19 & $ 5.6\pm 0.2$ & $ 0.18\pm 0.16$ & $ 0.08\pm 0.07$ & $ 0.32\pm 0.19$ & $ 0.49\pm 0.29$ & \W$ 0.28\pm0.49$ \\
1.0--1.5 & 49 & $ 8.7\pm 0.2$ & $ 0.36\pm 0.11$ & $ 0.38\pm 0.12$ & $ 0.52\pm 0.12$ & $ 1.82\pm 0.45$ & \W$ 0.19\pm0.19$ \\
1.5--2.0 & 41 & $13.2\pm 0.4$\W & $ 0.34\pm 0.12$ & $ 0.80\pm 0.28$ & $ 0.30\pm 0.11$ & $ 2.32\pm 0.89$ & $-0.07\pm0.26$ \\
2.0--2.5 & 53 & $18.2\pm 0.5$\W & $ 0.42\pm 0.11$ & $ 1.82\pm 0.48$ & $ 0.50\pm 0.12$ & $ 7.09\pm 1.72$ & \W$ 0.08\pm0.17$

\enddata
\tnt{a}{Only sources at an angular distance $<$6\arcmin\ from the
  \chandra\ optical axis are included in the stacking analysis.}
\tnt{b}{The root mean squared value of $d_{\rm L}$ for the objects in each redshift bin, with approximate statistical uncertainty, used for calculating $\average{L_{\rm X}}$.}
\end{deluxetable*}

\begin{figure}
\epsscale{1.2}
\plotone{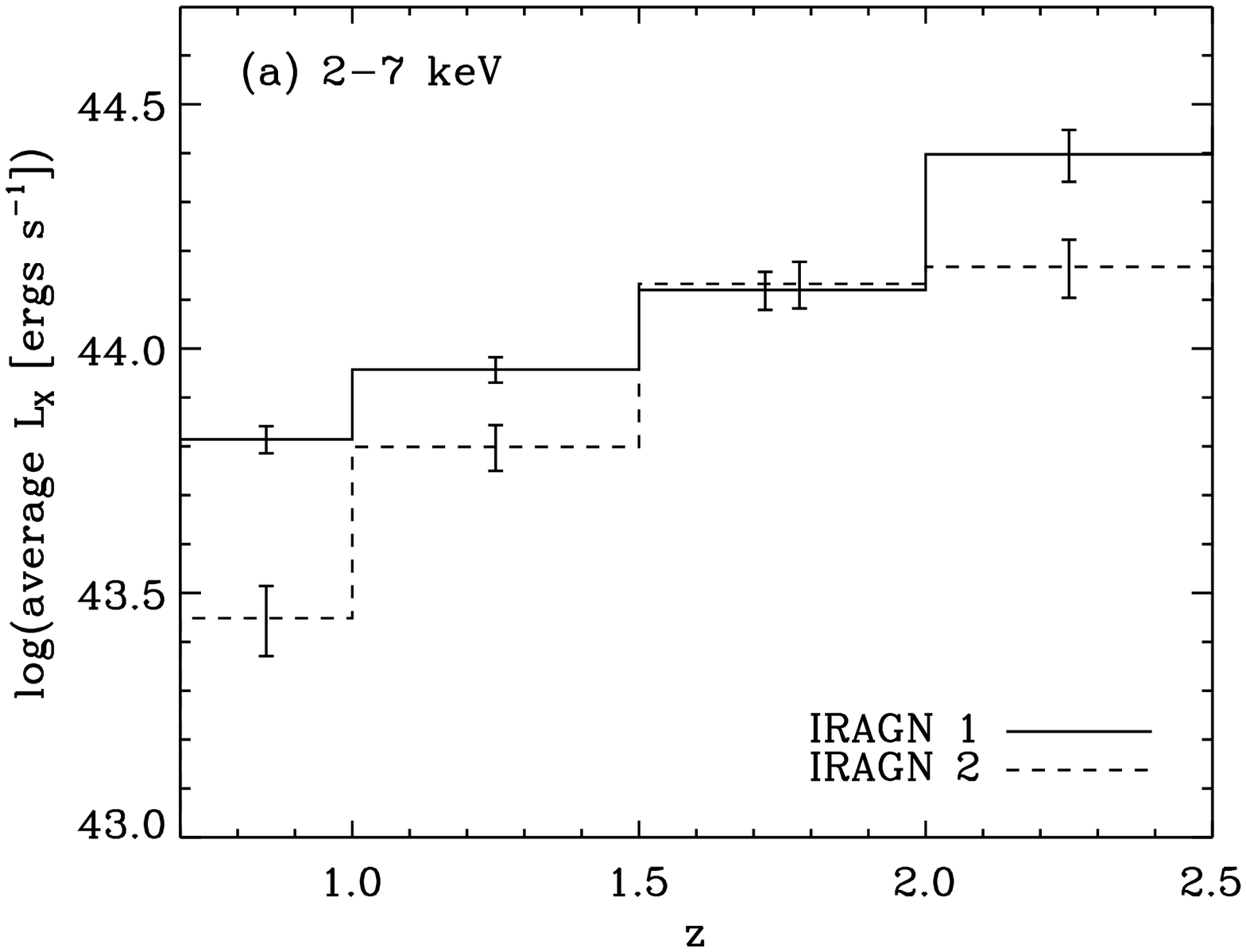}
\plotone{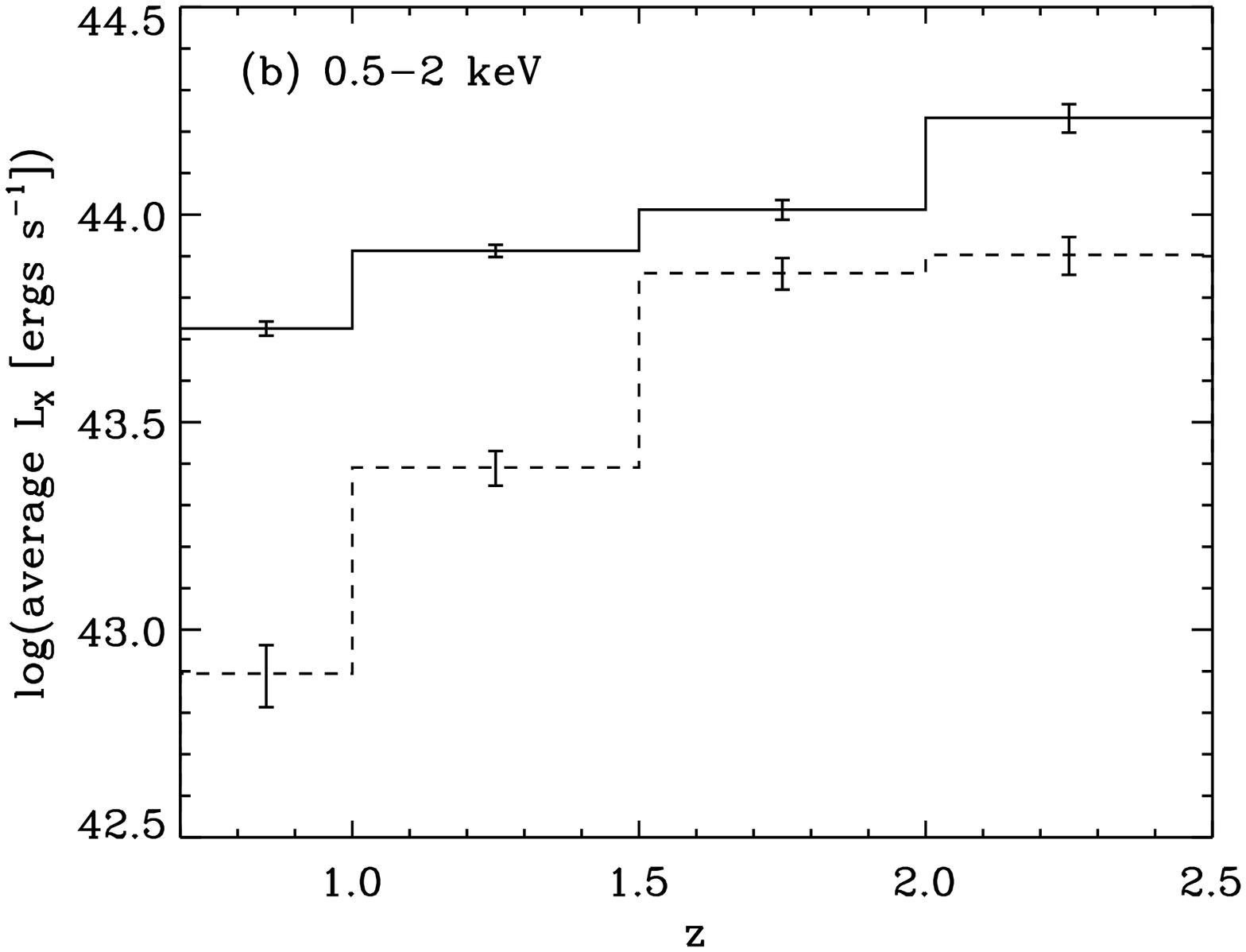}
\plotone{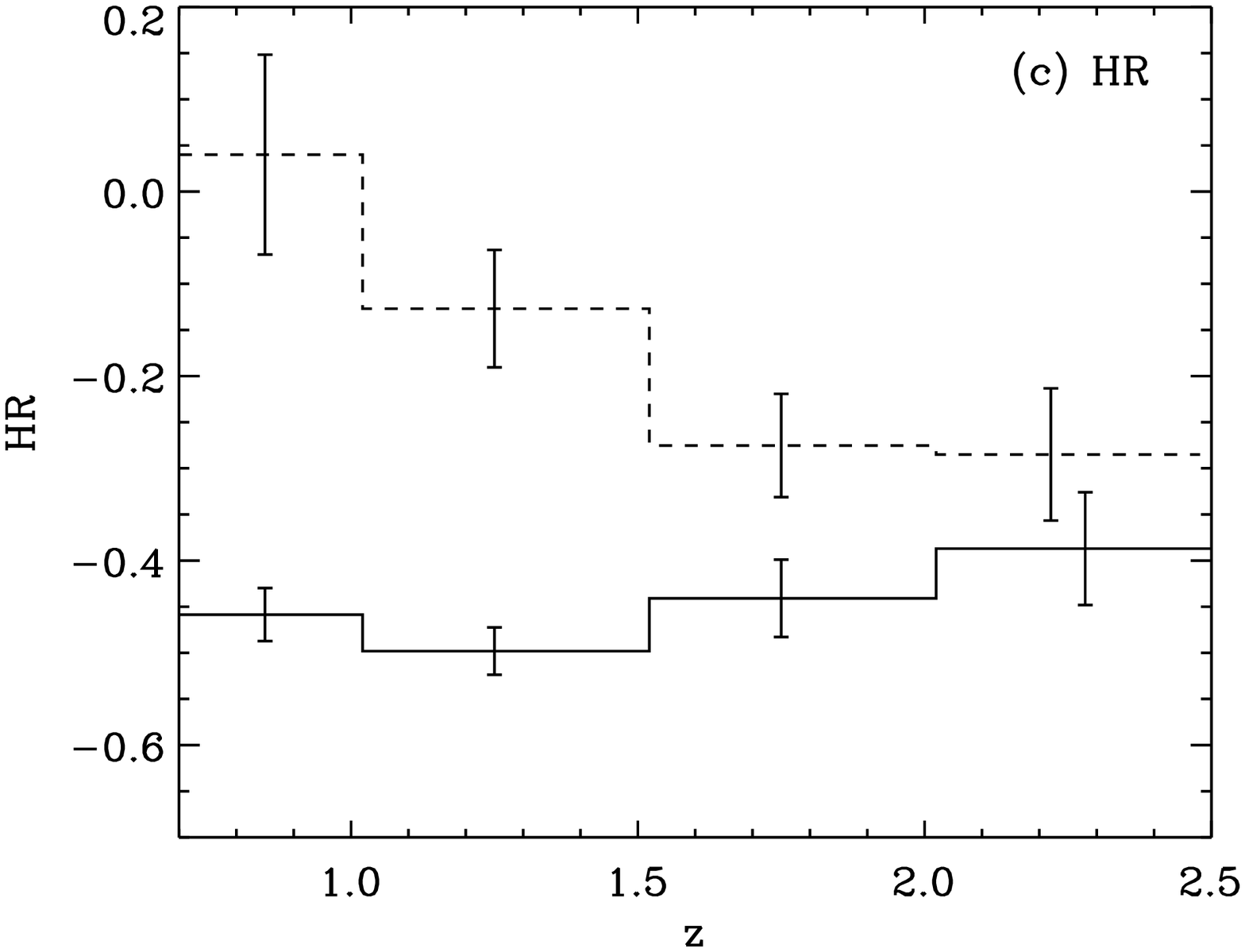}
\caption{Results from X-ray stacking analysis, in bins of redshift, for
  IRAGN 1s and 2s.  Here all sources are included, including those with X-ray detections.  Shown are the average $L_{\rm X}$ in the (a) 2--7
  keV, and (b) 0.5--2 keV bands, and (c) average hardness ratio.
  Note that the IRAGN 2s are consistently harder (larger HR) than
  the IRAGN 1s. \label{figlxhr}}
\end{figure}
 
\begin{figure}
\epsscale{1.2}
\plotone{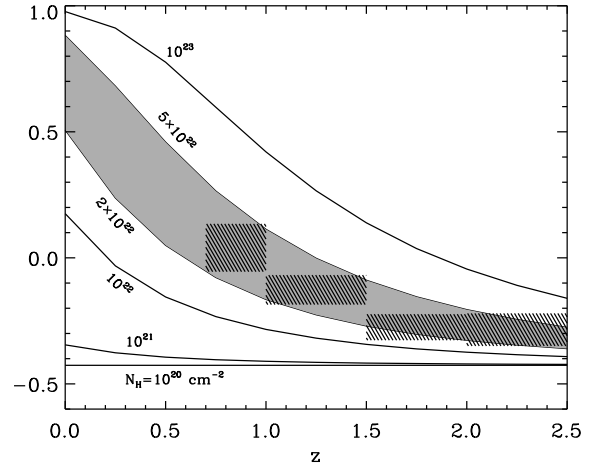}
\caption{X-ray hardness ratio HR vs.\ redshift for various \nh\
  (given in \cdens), given
  an intrinsic power-law photon index $\Gamma=1.8$ and the on-axis ACIS
  response function.  Hatched regions show the observed $1\sigma$ limits in
  HR for the IRAGN 2s in bins of redshift, as shown in Fig.\
  \ref{figlxhr}.  Assuming the intrinsic $\Gamma=1.8$, these HR values
  for IRAGN 2s are consistent with a constant $N_{\rm
  H}=$(2--5)$\times10^{22}$ \cdens, shown by the shaded
  region. \label{fignhz}}
\end{figure}

\begin{figure}
\epsscale{1.2}
\plotone{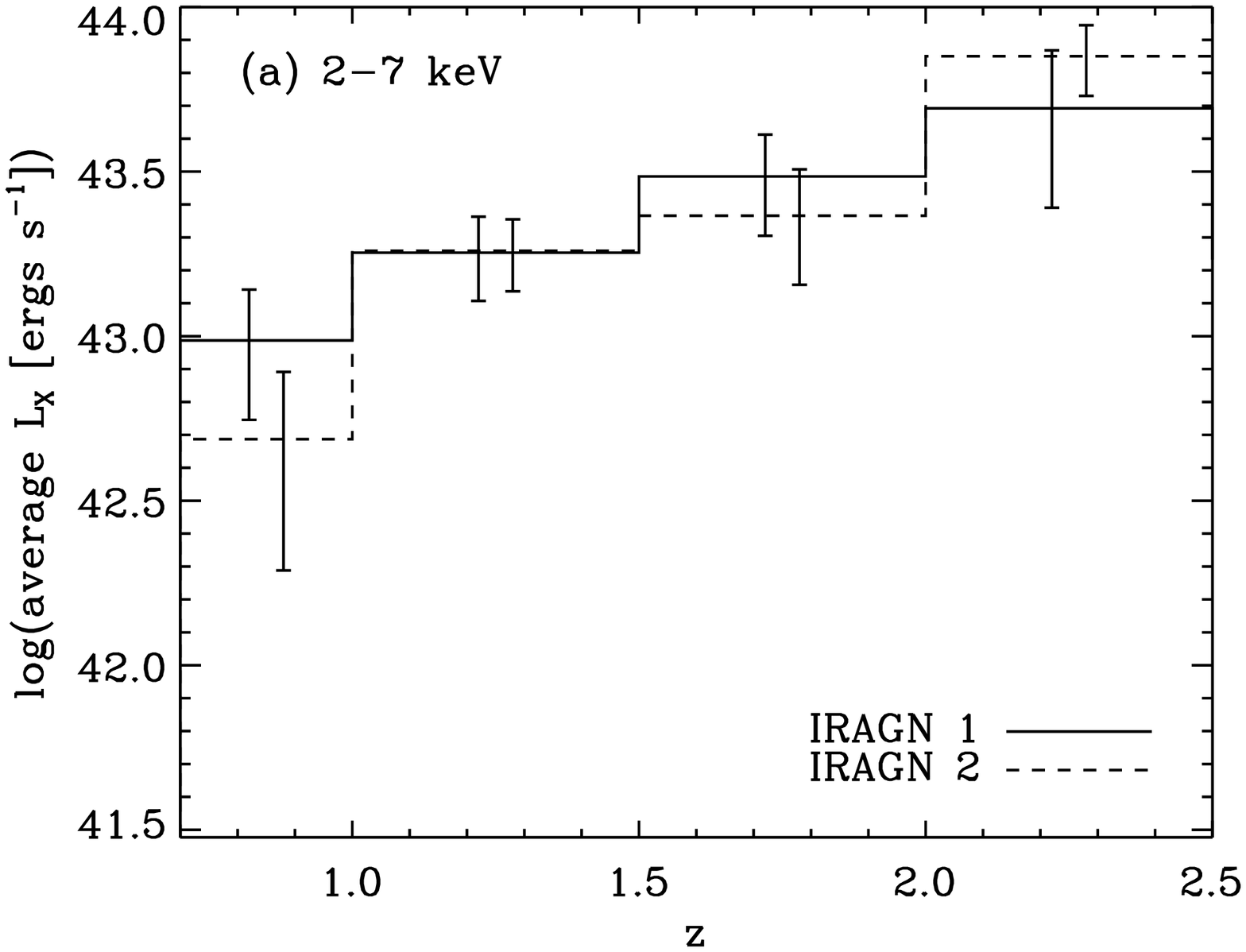}
\plotone{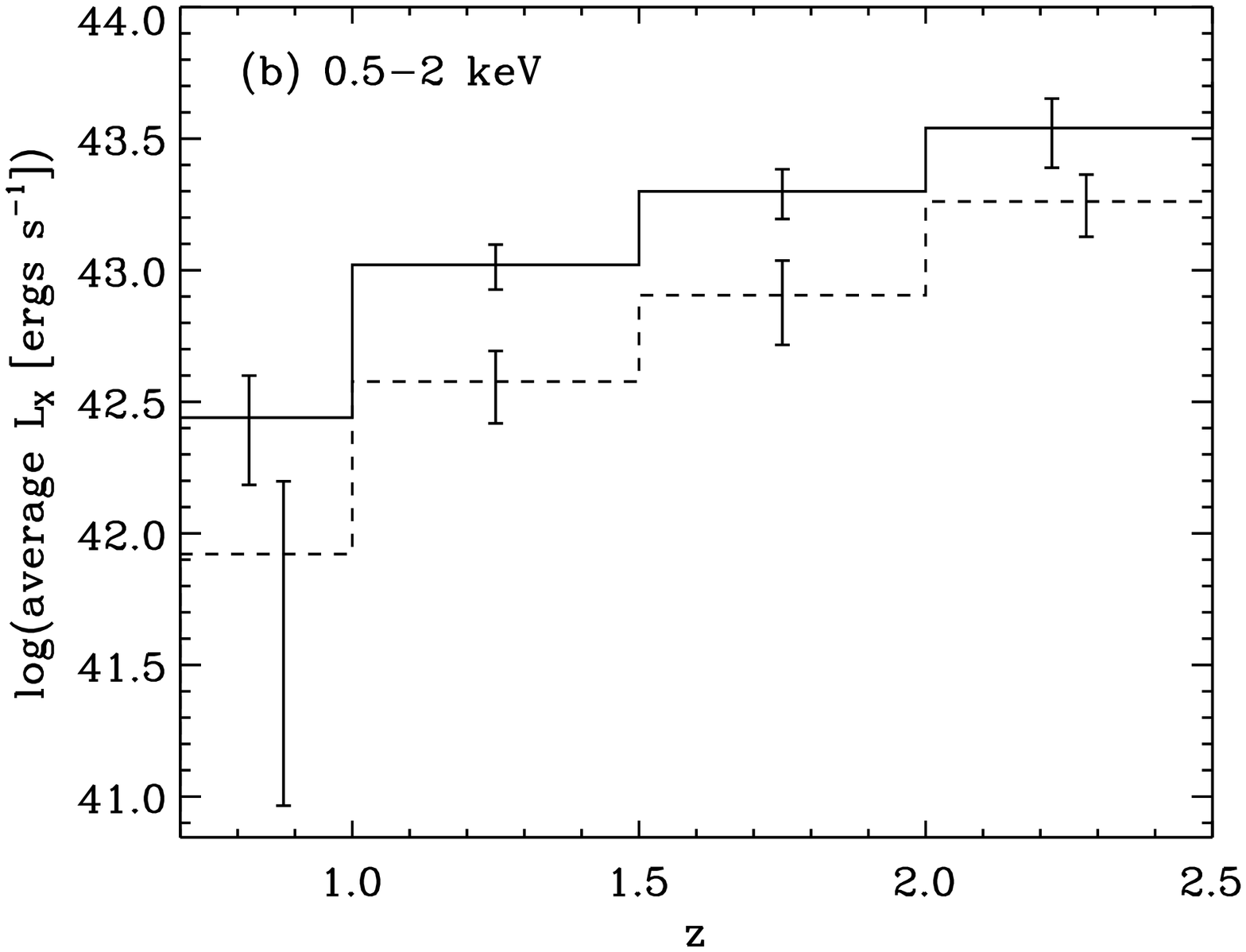}
\plotone{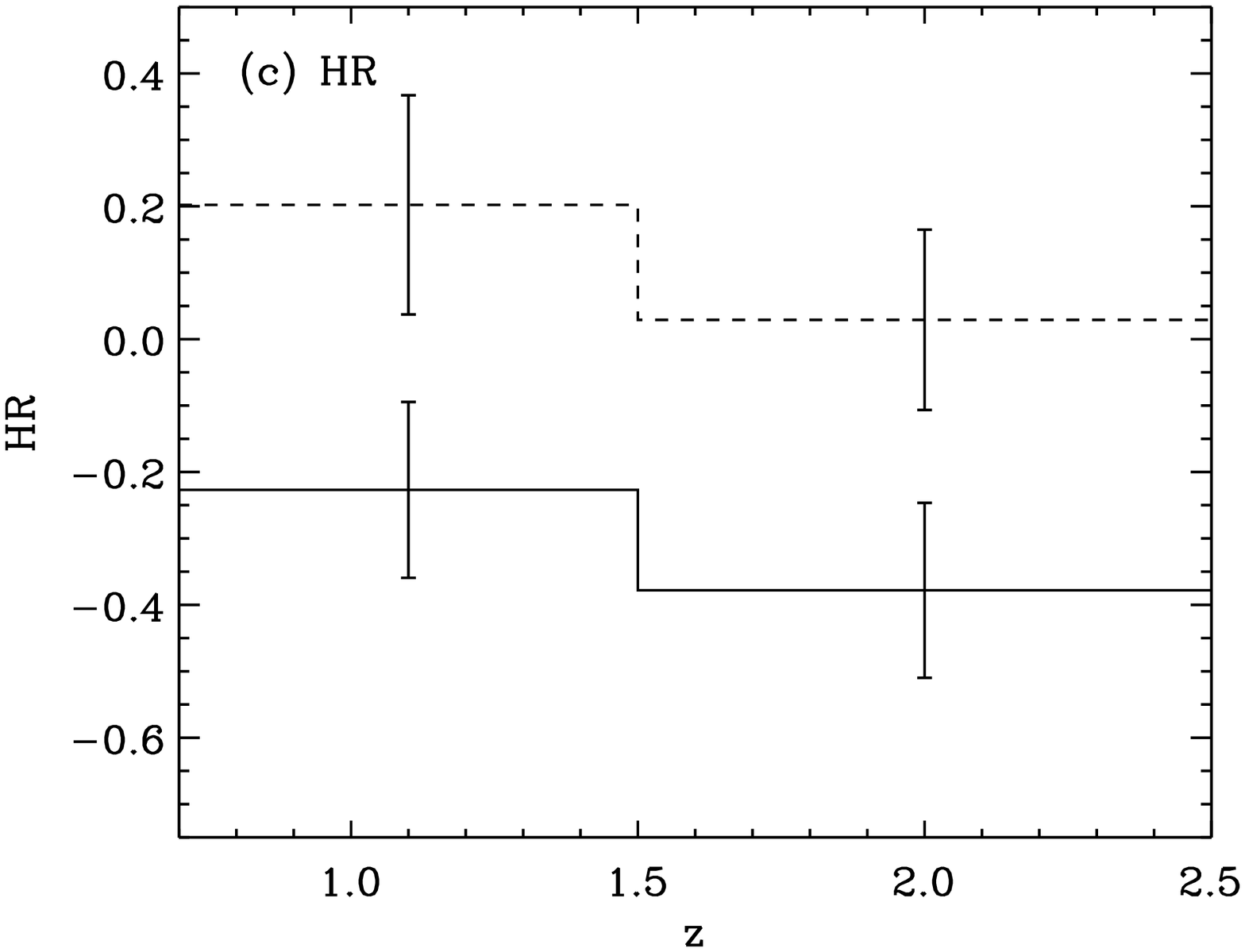}
\caption{Same as Fig. \ref{figlxhr}, only including sources with no
  detections in the X\bootes\ catalog \citep{kent05}. X-ray
  undetected IRAGNs have average 
  $L_{\rm 0.5-7\; keV}\sim10^{43}$ \ergs\ characteristic of luminous Seyfert
  galaxies.  The X-ray undetected IRAGN 2s have systematically larger average hardness
  ratios than the X-ray undetected IRAGN 1s. \label{figlxhr_max0}}
\end{figure}

\subsubsection{Calculation of \nh\ and \lx}
\label{xraycalc}
 Measuring \nh\ and \lx\ from X-ray fluxes requires an assumption for
the X-ray spectrum of the source, which for most AGNs can be modeled
by a simple power law with photon index $\Gamma$, such that the photon
flux density (in photons cm$^{-2}$ s$^{-1}$ keV$^{-1}$) $F\propto
E^{-{\Gamma}}$.  For all the AGNs in our sample, we assume an
intrinsic X-ray spectrum with $\Gamma=1.8$, typical for unabsorbed
AGNs \citep{tozz06}; as we show in \S\ \ref{xrayres}, this $\Gamma$
corresponds to the average spectral shape of the IRAGN 1s.  Although
X-ray AGNs do not all have the same intrinsic $\Gamma$, the typical
intrinsic spectrum does not vary significantly with luminosity or \nh\
\citep{tozz06}.  Therefore, it is reasonable to assume a constant
$\Gamma$ when estimating \nh\ and \lx\ for an ensemble of sources.

Given a constant intrinsic X-ray spectrum, \nh\ is directly related to
the observed hardness ratio.  Absorption by neutral gas preferentially
obscures lower energy X-rays, and so a larger \nh\ corresponds to a
larger (in this case, less negative) HR.  The conversion between
HR and \nh\ depends on the response function of the X-ray detector and the
redshift of the source.  With increasing $z$, the low-energy turnover
due to absorption by neutral hydrogen is increasingly redshifted out
of the 0.5--2 keV bandpass.  Therefore, for a power-law
spectrum attenuated by a fixed column density of gas, the observed
spectrum will become softer with increasing redshift, so that objects
with higher $z$, but equal HR, correspond to greater absorption.
Assuming $\Gamma=1.8$, we have calculated HR for a grid of
absorptions ($10^{20}\leq N_{\rm H}\leq10^{23}$ \cdens) and redshifts
($0\leq z\leq4$), and will use these to convert observed hardness ratios to
column densities.  Note that the Galactic $N_{\rm H}$ toward this
field is very small ($\sim$$10^{20}$ \cdens), so we neglect it in our
estimates of column density.

To derive the average \lx\ and \nh\ from stacking requires an estimate of redshift, so we perform the stacking in bins of $z$.  For each
bin we calculate the rms value of $d_{\rm L}$ for the
objects in the bin.  Using this distance, along with the average
fluxes from stacking, we calculate $\average{L_{\rm X}}$ in each bin.  We
also include a $K$-correction to $\average{L_{\rm X}}$, assuming
$\Gamma=1.8$, to account for the fact that the X-ray bands we observe
probe higher energies in the rest-frame spectrum.  This
$K$-correction varies from 0.9 at $z=0.7$ to 0.8 at $z=2$, so it has
a relatively small effect on $\average{L_{\rm X}}$.

\subsubsection{Average X-ray fluxes}
\label{xrayres}

As a first step in the stacking analysis, we compare the average X-ray
fluxes of different subsets of sources in the \bootes\ catalog.  We
divide the sources into (1) IRAGN 1s and (2) IRAGN 2s, defined by
\iiu\ as shown in Fig.\ \ref{figuvir},
and for comparison (3) objects with detections in all four IRAC bands
that are identified as optically normal galaxies in the optical from
their AGES spectra\footnotemark.  The IRAGNs are all selected to have
$z>0.7$, while the optically normal galaxies mainly lie at $z<0.7$.
In interpreting X-ray stacking results, it is a concern that the X-ray
brightest objects may dominate the average flux.  Therefore, we
perform the stacking analysis twice, first using all objects in each
subsample and then using only those objects that are not detected
with 4 or more counts in the X\bootes\ source catalog \citep{kent05}.

\footnotetext{For a detailed X-ray stacking analysis of normal
  galaxies in the AGES survey, see \citet{brand05}.}

Average fluxes in the soft and hard bands are given in Table
\ref{tblxray}.  Including all sources, the IRAGN 1s have a larger
average 2--7 keV flux than the IRAGN 2s, with $\average{F_{\rm 2-7\;
keV}}=10.9\times10^{-15}$ and $6.5\times10^{-15}$ \flux\ source$^{-1}$,
respectively.  However, when we exclude those objects that are detected in the
X\bootes\ catalog, the hard X-ray fluxes of the two subsets
closely agree.  This suggests that the IRAGN 1 sample contains more
bright X-ray sources than the IRAGN 2 sample, but for faint sources ($<$4
counts), the two IRAGN types have similar average fluxes.

While the hard X-ray fluxes are comparable between the IRAGN 1s and
2s, the soft X-ray fluxes are significantly smaller for the IRAGN 2s,
indicating that they are more absorbed.  The IRAGN 1s have an average
$HR=-0.47$, which is close to that expected for an unabsorbed AGN with
$\Gamma=1.8$.  In contrast, the IRAGN 2s have $HR=-0.20$.  The
optically normal galaxies have $HR=-0.19$, similar to the IRAGN 2s, but
with $\sim$5 times smaller average flux.

As mentioned in \S\ \ref{identify}, the X-ray data can be used to
verify our selection criterion for IRAGN 1s and 2s, by looking for
systematic differences in HR (and thus absorption) on either side of
our selection boundary.  We can therefore address concerns that the
selection may be biased by the AGES spectroscopic flux limits,
especially for $L_{\rm 4.5\mu m}<3\times10^{11}$ \lsun, where the
distribution in \iiu\ is not as clearly bimodal (Fig.\ \ref{figuvir}).
We perform the stacking analysis in bins of \iiu\ for the IRAGNs with
$L_{\rm 4.5 \mu m}<3\times10^{11}$ \lsun\ and plot HR versus \iiu\
in Fig.\ \ref{figstackxi}.  There is a significant increase in HR
across our IRAGN 1/IRAGN 2 boundary of $\log{L_R/L_{\rm 4.5\mu m
}}=-0.4$, verifying that this criterion effectively separates objects
with unabsorbed (IRAGN 1) and absorbed (IRAGN 2) X-ray emission and
is not significantly affected by the AGES flux limits.

Next, to confirm that the X-ray flux for these objects comes from
nuclear emission, we show that the average X-ray flux is significantly
larger than that expected for star formation.  The X-ray flux $F_X$
from star formation is related to the far-IR flux $F_{\rm FIR}$
\citep[see Eqn.~12 of][]{rana03}.  $F_{\rm FIR}$ is
defined as \citep{helo85}
\begin{equation}
F_{\rm  FIR}=1.26\times10^{-11}(2.58 S_{60 \mu {\rm m}} + S_{100 \mu {\rm
    m}}) {\rm \: ergs\: cm^{-2}\: s^{-1}},
\label{eqnfir}
\end{equation}
 where $S_{60 \mu {\rm m}}$ and $S_{100 \mu {\rm m}}$ are in Jy.  For
    the suite of starburst model SEDs given in \citet{sieb07}, we
    calculate the ratio of rest-frame $F_{\rm FIR}$ (in \flux) to the
    observed flux at 5.8 $\mu$m, $S_{\rm 5.8\mu m}$ (in $\mu$Jy).
    Excluding Arp 220 (which has an extreme star formation rate and
    is highly extincted in the optical, so it would not be detected
    in our survey), we find that $F_{\rm FIR}/S_{\rm 5.8\mu
    m}\sim(0.3-2)\times10^{-14}$ \flux\ $\mu {\rm Jy}^{-1}$ for the
    redshift range $0.7<z<3$.  Combining with the \citet{rana03}
    relation and converting from the rest-frame 2--10 keV band
    luminosity to our observed 2--7 keV band flux using a $\Gamma=1.8$
    power law spectrum (with the appropriate small $K$-correction), we
    have
\begin{equation}
\label{eqnf27}
F_{\rm 2-7\; keV}=S_{\rm 5.8\mu m }\times(0.5-4)\times10^{-18}{\rm \; ergs \; cm^{-2}\:
    s^{-1}\: \mu Jy^{-1}}
\end{equation}

Making the conservative assumption that the observed 5.8 \micron\ flux
for every IRAGN 2 is due entirely to stars and star formation, we can
put an upper limit on the \fhard\ we expect from star formation for
each object.  The distribution in these \fhard\ values, for every
IRAGN 2 that is not detected in X-rays, is shown in Fig.\
\ref{figsblx}.  By comparison, the average \fhard\ observed for these
objects is $>3$ times larger than that typically expected for star
formation, even for the largest typical ratio of X-ray to 5.8 $\mu$m flux.
This confirms that for most of the IRAGN 2s, the X-ray emission is not
powered by star formation, but nuclear accretion.

\subsubsection{Average \lx\ and \nh}
In order to use the X-ray stacking analysis to measure physical
parameters such as the accretion luminosity (\lx) or the gas
attenuation (\nh), we must include redshift information.  Therefore,
we have repeated the stacking for both types of IRAGN in bins of
redshift from $z=0.7$--2.5.  We do not include sources at $z>2.5$
because we do not have enough objects with best-fit $z>2.5$ to obtain
well-constrained fluxes.  We stress here that although there may be
significant uncertainties in photometric redshift estimates,
particularly for IRAGN 2s, there is no large bias in the photo-$z$'s,
as we show in \S\ \ref{photoz}.  Therefore, our stacking analysis
using large bins in redshift should not be strongly affected by
photo-$z$ uncertainties.

The stacking results as a function of $z$ are listed in Table
\ref{tblxray2}, and we plot $\average{L_{\rm X}}$ versus $z$ in
Fig.\ \ref{figlxhr}.  For both IRAGN types, $\average{L_{\rm X}}$ increases
by a factor of $\sim$2 between $z=0.5$ and $z=2.5$, due to the evolution
in the quasar luminosity function with redshift
\citep[e.g.,][]{ueda03,barg05,hasi05} and the IR and optical flux limits that
restrict us to selecting only the most luminous objects at high $z$.
The $\average{L_{\rm 2-7\; keV}}$ range
(0.3--3)$\times10^{44}$ \ergs\ is typical for Seyfert galaxies and
 quasars and much larger than the typical \lx\ of
starburst or normal galaxies.  Although the IRAGN 2s have
$\average{L_{\rm 0.5-2\; keV}}$ and $\average{L_{\rm 2-7\; keV}}$ that
are 3--5 and 2--3 times lower than the IRAGN 1s, respectively, these
\lx\ values are still typical of AGNs and not starburst galaxies.  

Plotting HR in redshift bins (Fig.\ \ref{figlxhr}), the IRAGN 2s are
significantly harder at all $z$.  The IRAGN 1s have $HR\simeq-0.45$
for all $z$, which corresponds to an intrinsic $\Gamma=1.8$ with no
absorption.  The IRAGN 2s are significantly harder, with
$HR\simeq-0.3$--0.1.  Assuming that these have the same intrinsic
$\Gamma$ as the IRAGN 1s, we estimate the corresponding \nh.  In Fig.\
\ref{fignhz} we plot HR versus $z$ for several values of \nh\
assuming $\Gamma=1.8$.  The hatched regions correspond to the
$1\sigma$ errors in HR for the IRAGN 2s in each of our redshift
bins.  For all redshifts, the IRAGN 2 HR values are consistent
with a column density of $N_{\rm H}=$(2--5)$\times10^{22}$ \cdens,
marked by the shaded region in Fig.\ \ref{fignhz}.

As mentioned in the previous section, it is a concern that the average
\lx\ and \nh\ we measure may be dominated by a few bright sources.  To
address this, we repeat the stacking as a function of $z$ but exclude
those objects that are detected in the X\bootes\ catalog.  The results
are shown in Fig.\ \ref{figlxhr_max0} and indicate that even those
objects that are not detected in X-rays have
$\average{L_{\rm X}}$ values consistent with AGNs.  In addition,
the IRAGN 2s have harder spectral shapes than the IRAGN 1s, even
at these fainter fluxes.

To summarize the X-ray stacking results, both IRAGN types have average
X-ray fluxes that are too large to be due to star formation and thus
strongly indicate AGN activity.  Performing the stacking as a function
of redshift, we find that both IRAGN 1s and 2s have average \lx\
values consistent with Seyferts and quasars, and the IRAGN 1s have
hardness ratios consistent with unabsorbed AGNs ($\Gamma=1.8$). The
IRAGN 2s, assuming the same intrinsic spectrum, correspond to absorbed
sources with $N_{\rm H}\sim 3\times10^{22}$ \cdens.

\subsection{Gas absorption and dust extinction}

\label{dust}

 In this section, we check that the dust extinction for the IRAGN 2s
 that we inferred from the optical/UV data is consistent with the \nh\
 we measure in X-rays, assuming that, in general, the X-ray-absorbing
 gas is coincident with the extincting dust.  Fig.~\ref{figav} shows
 that the SEDs of most of the IRAGN 2s are best fitted by templates with
 $0.7<A_V<7$.  The ratio of gas to dust in the Galaxy is such that
 $N_{\rm H}/A_V\simeq2\times10^{21} {\; \rm cm}^{-2}$, or $A_V\simeq15$
 for the observed average $N_{\rm H} = 3 \times10^{22}$ \cdens.  This
 extinction is more than enough to obscure the optical light from the
 nucleus, although it is somewhat larger than the typical $A_V$
 obtained from the SED fits. 

 However, there is evidence that AGNs have high gas-to-dust ratios,
 similar to or perhaps even greater than that of the SMC
 \citep[see][and references therein]{fall89,mart06}.  The SMC has
 $N_{\rm H}/A_V\simeq2\times10^{22} {\; \rm cm}^{-2}$, which corresponds
 to $A_V\simeq1.5$ for the observed average \nh\ and is close to the
 typical $A_V$ obtained by the SED fits to the IRAGN 2s.  Therefore,
 we conclude that the dust extinction implied by the optical and IR
 observations is generally consistent with the average \nh\ we derive
 from X-ray stacking.

The bimodality in the $A_V$ distribution from SED fits
(Fig.~\ref{figav}), as well as the clear separation of the two IRAGN
types in optical-IR color (Fig.~\ref{figuvir}), suggests that there is
a bimodal distribution in the dust column density to the IR-selected
AGNs.  There is no obvious selection effect that could produce this
bimodality, so we expect that it is real.
This is broadly consistent with previous results on the distribution
of \nh\ measured in X-rays.  These studies find many objects with
$N_{\rm H}<3\times10^{20}$ \cdens\ or $N_{\rm H}>3\times10^{21}$
\cdens, with relatively few at intermediate column densities
\citep[e.g.][]{trei05,tozz06}.  However, such a bimodal distribution
could simply be due to limitations of X-ray spectral fitting
techniques, with which it is difficult to measure $N_{\rm H}$ as low
as $\sim10^{21}$ \cdens, especially at high redshifts where X-ray
telescopes probe energies higher than the photoelectric cutoff at
$E\simeq 1$ keV \citep[e.g.,][]{akyl06}.  The colors we
observe in IR-selected AGNs suggests that such a bimodal obscuration
distribution does indeed exist, which has implications for models
of AGN obscuration, as we discuss in \S\ \ref{impl}.

\begin{figure} 
\epsscale{1.2} 
\plotone{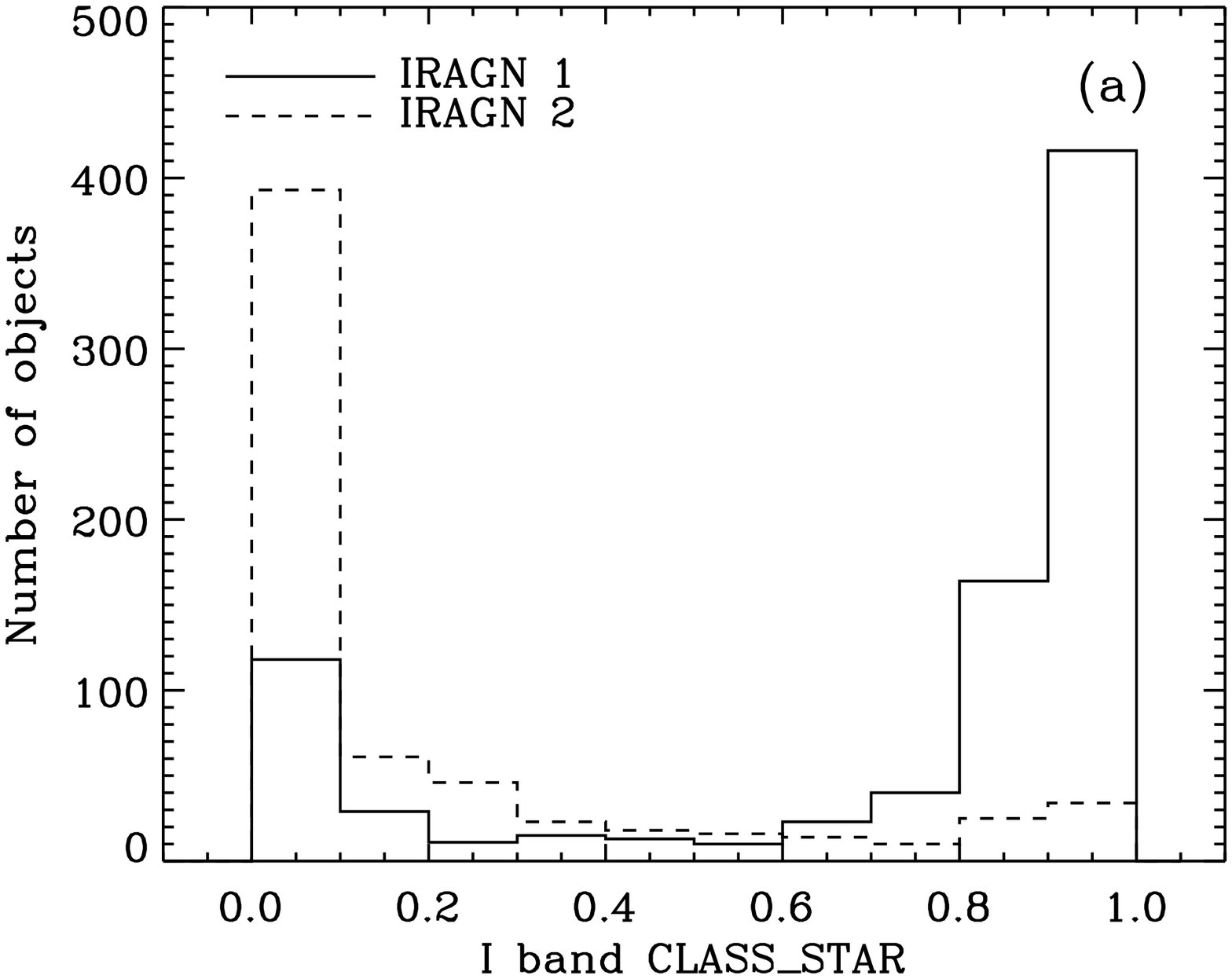}
\plotone{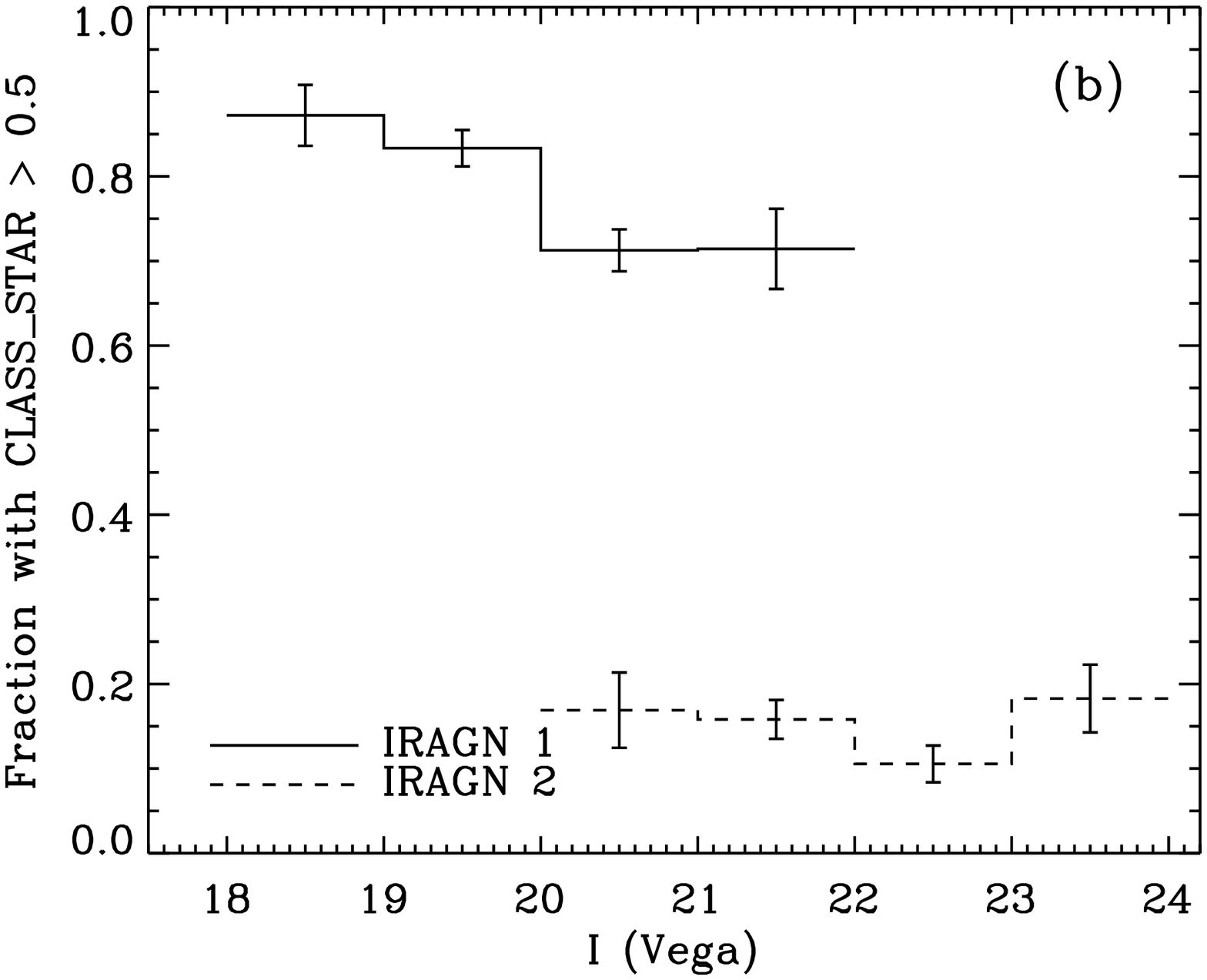}
\caption{Optical morphologies for the two types of IR-selected AGNs.
  (a) Histograms of the $I$ band stellarity parameter \cstar\ output
  by ${\tt SExtractor}$ (0 for extended sources, 1 for point sources).
  Note that the vast majority of IRAGN 1s are point-like, while the
  IRAGN 2s are extended, suggesting that their optical emission is
  dominated by the nucleus and by the host galaxy, respectively.  (b)
  Variation of $I$ band \cstar\ with $I$ magnitude, for the flux range
  in which the two IRAGN subsets overlap.  The $y$ axis shows the
  fraction of sources that have $I>0.5$ in each magnitude bin.  Errors
  shown are counting statistics, and only bins with $>$5 sources are
  shown.  The point-like fraction does not vary with $I$ and is
  significantly different between the two subsets.
  \label{figmorph} }
\vskip0.2cm
\end{figure}

\subsection{Optical morphologies and colors}
\label{galedd}

Since we expect the nuclear optical emission from the IRAGN 2s to be
extincted, their optical light should be dominated by their host
galaxies.  Normal galaxies differ from quasars in optical images in
two principal ways: (1) galaxies have extended morphologies, while
quasars are dominated by a small nucleus and so appear as point
sources; and (2) normal galaxies have redder colors, characteristic of
a composite stellar spectrum rather than a blue AGN continuum.  By
examining the optical morphologies and colors of our IRAGN sample, we
can confirm that optical emission is dominated by an AGN in IRAGN 1s
and by the host galaxy in IRAGN 2s.

To quantify morphologies, we use the \cstar\ parameter output by the ${\tt
SExtractor}$ photometry code \citep{bert96}.  \cstar\ is a measure of how well an
object can be approximated by a point source, with values ranging from
0 (extended) to 1 (point source).  In Fig.\ \ref{figmorph} (a), we
plot the distribution in \cstar\ in the $I$ band (which best
discriminates between the two IRAGN types) and find that 74\% of the
IRAGN 1s have \cstar\ $>0.7$, indicating that the emission is point-like,
while 85\% of the IRAGN 2s have \cstar\ $<0.5$, indicating mainly
extended emission.

However, for very faint objects, it is possible to obtain low \cstar\
values, even if the sources are point-like.  Therefore, we must confirm
that the lower \cstar\ values for IRAGN 2s are not simply a
result of their lower fluxes.  Fig.\ \ref{figmorph} (b) shows the
fraction of objects with ${\tt CLASS\_STAR}>0.5$ for each IRAGN subset
as a function of $I$ magnitude.  There is no clear trend in this
fraction with $I$ for the IRAGN 2s, and for the magnitudes in which
the subsets overlap, the IRAGN 2s have many fewer ``point-like''
morphologies than the IRAGN 1s.  We conclude that the IRAGN 2s do have
more extended morphologies than the IRAGN 1s, so that the color
selection described in \S\ \ref{identify} can effectively distinguish
between objects dominated by a nucleus and those dominated by extended
emission.

\begin{deluxetable}{lccc}
\tablewidth{3in}
\tabletypesize{\footnotesize}
\tablecaption{Best template fits to optical photometry \label{tblgalphot}}
\tablehead{
\colhead{} &
\multicolumn{3}{c}{Best-fit template}\\
\colhead{Subset} &
\colhead{Elliptical} &
\colhead{Sb} &
\colhead{Quasar}}
\startdata
optical BLAGN &          170 &          362 &          402 \\
optical NLAGN &           71 &           35 &        0 \\
optical galaxies &         4103 &         1167 &           47 \\
IRAGN 1 &          143 &          327 &          368 \\
IRAGN 2 &          452 &          168 &           11

\enddata
\tablecomments{Includes sources with detections in all three optical bands ($B_W$, $R$, and $I$) and all four IRAC bands.}
\end{deluxetable}

\begin{figure}
\epsscale{1.2}
\plotone{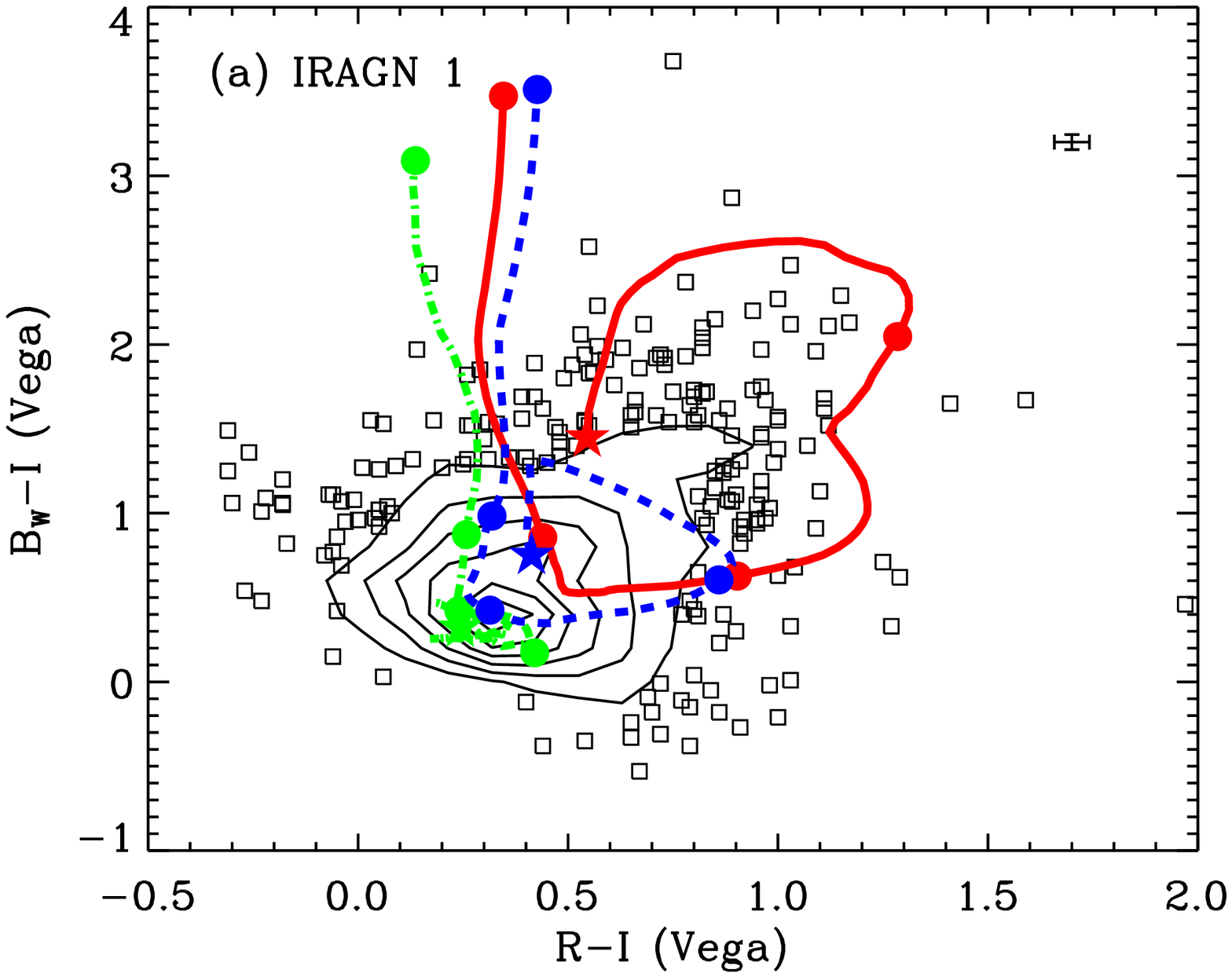}
\plotone{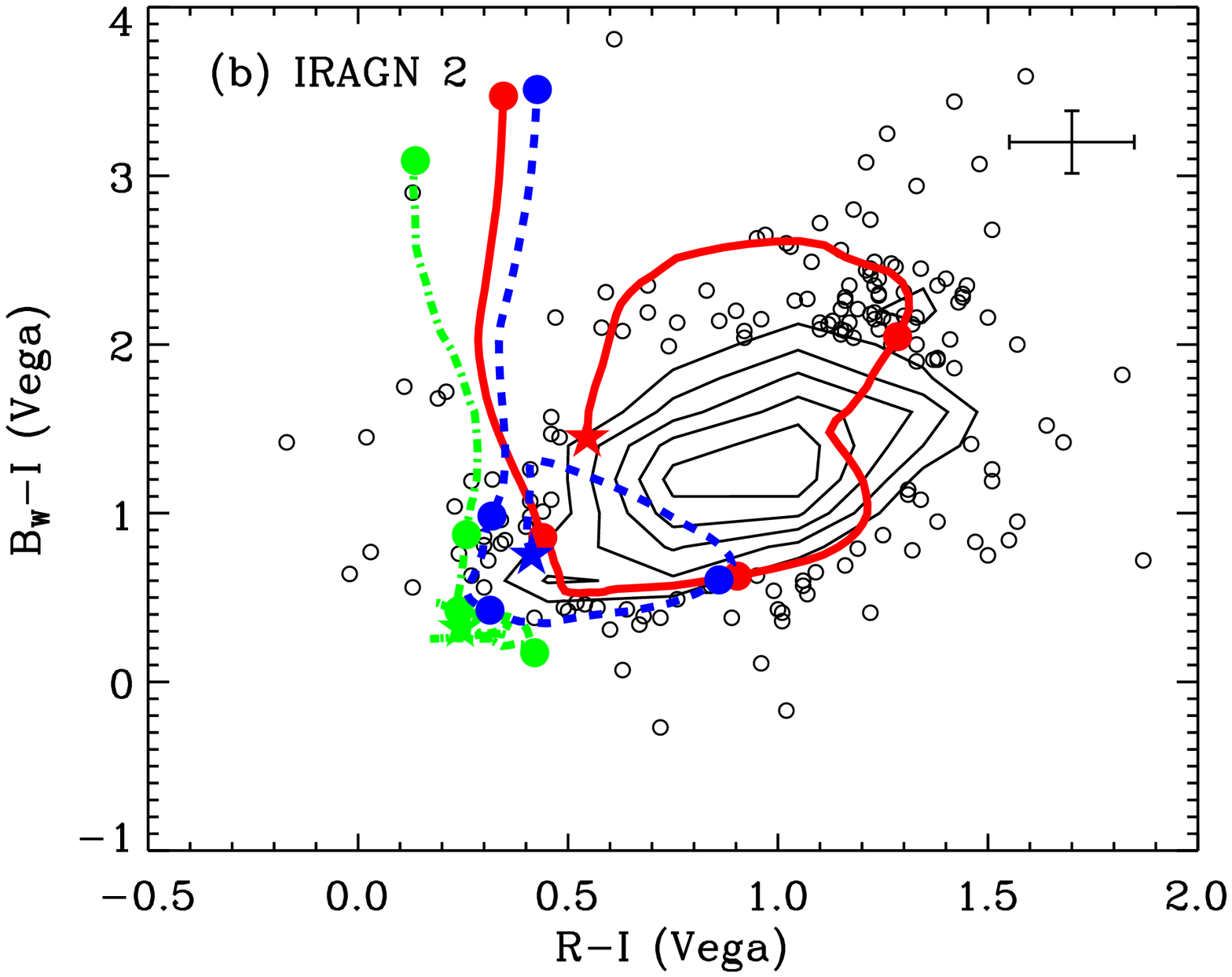}
\caption{Contours and points show observed $B_{W}-R$ and $R-I$ colors
  for (a) IRAGN 1s and (b) IRAGN 2s, compared to colors for elliptical
  galaxy (red solid line), Sb galaxy (blue dashed line), and quasar
  (green dot-dashed line) templates.  The model tracks run from
  $z=0-4$.  Stars show the color at $z=0$, and filled dots indicate
  $z=1$, 2, 3, and 4.  Error bars show the median uncertainty in the
  colors  for objects lacking
  spectroscopic redshifts.  \label{figcolcol}}
\vskip0.2cm
\end{figure}

\begin{figure*}
\epsscale{1.2}
\plotone{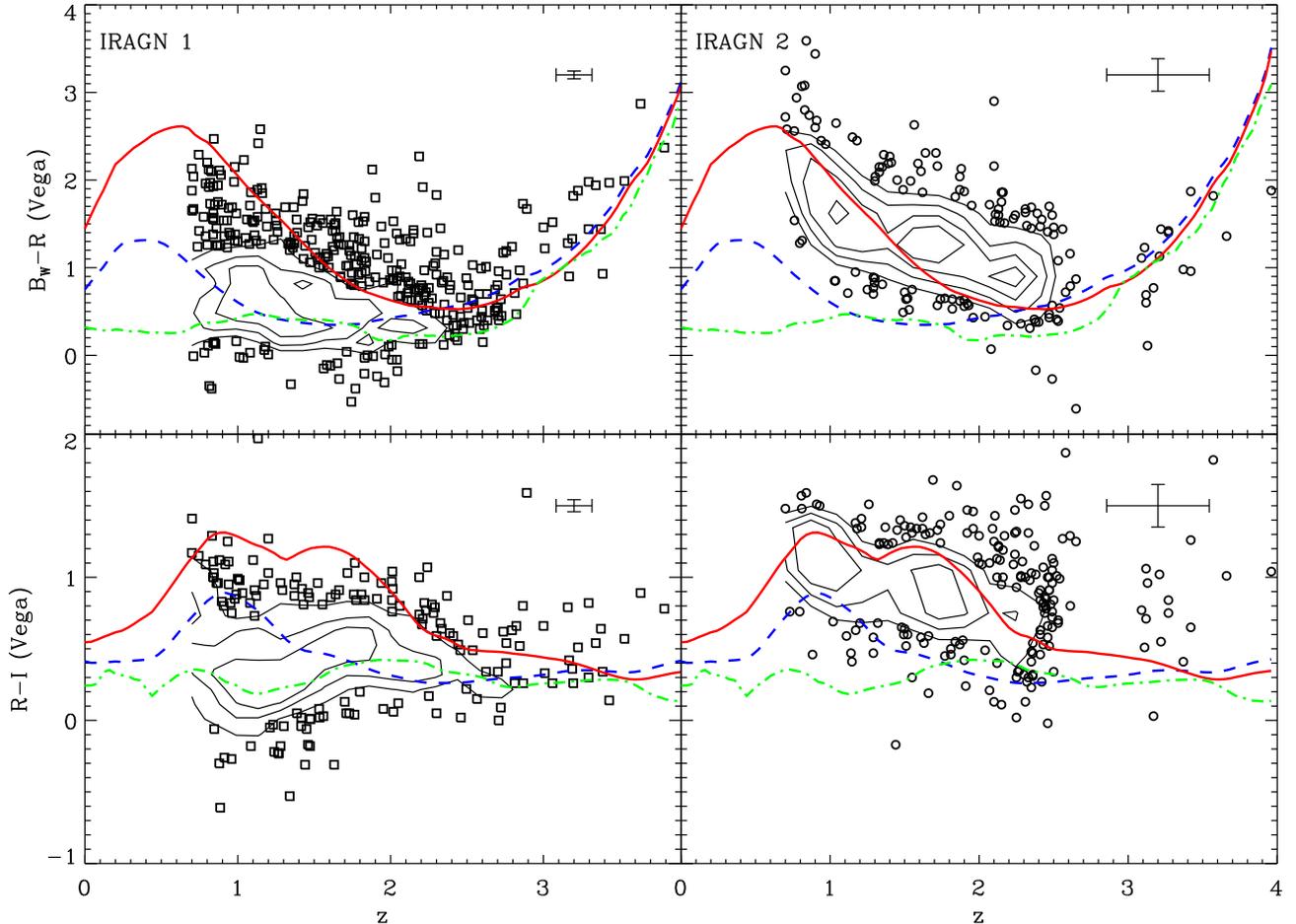}
\caption{Contours and points show observed $B_{W}-R$ and $R-I$ colors
  for IRAGN 1s (top) and IRAGN 2s (bottom) versus redshift, compared
  to model tracks as in Fig.~\ref{figcolcol}.  Error bars show the median uncertainty in the
  colors  for objects lacking
  spectroscopic redshifts. 
  \label{figgalagncol}}
\vskip0.2cm
\end{figure*}

We also examine the observed $B_{W}-R$ and $R-I$ colors of the two IRAGN
types and compare them to the colors of the galaxy and quasar templates
described in \S~\ref{lum}.  In Fig.\ \ref{figcolcol} we plot the
$B_{W}-R$ versus $R-I$ color tracks for the templates as a function of
redshift and overplot the observed colors for the two IRAGN types.
As expected, most IRAGN 1s have colors resembling a quasar spectrum,
although there are a few objects (even those with optical BLAGN
spectra) that have redder colors, owing to some optical extinction.  By comparison, the IRAGN 2s as a whole have colors that
are redder than those of the IRAGN 1s and lie between the elliptical
and spiral redshift tracks.  Fig.\ \ref{figgalagncol} shows a similar
plot, but as a function of redshift.  Again we see that the IRAGN 2s
have colors and trends with redshift that are closer to optically
normal galaxies than to quasars.

To quantify this further, we fit the quasar, elliptical, and Sb
templates (shown in Fig.\ \ref{figtemp}) to the $B_{W}$, $R$, and $I$
photometry for the 1469 IR-selected AGNs with detections in all three
optical bands.  For comparison, we perform the same fits for objects at all
redshifts that have four-band $5\sigma$ IRAC detections and AGES
classifications as BLAGN, NLAGN, and galaxies (as listed in the first
column of Table \ref{tblsample}).  We fix the redshift of the template
spectrum and leave only the normalization as a free parameter.  The
distribution of templates that fit the photometry with the lowest
$\chi^2$ is shown in Table \ref{tblgalphot}.  A total of 695 (83\%) of the IRAGN
1s are best fitted by the quasar or Sb templates (which have similar
colors for $z>1.5$), similar to the fits for optical BLAGNs.  By
contrast, 452 (71\%) of the IRAGN 2s are best fitted by the elliptical
template, with almost all the rest fit by the Sb template, similar to
the fits for optical NLAGNs and optically normal galaxies.  We
conclude that the IRAGN 2s, as a population, do indeed have optical
colors consistent with host galaxies and are markedly different from
the IRAGN 1s.

\section{Verification of photometric redshifts}
\label{photoz}

The selection criteria developed here for IRAGN 2s depend only on
observed color and so are independent of redshift.  However, our SED
fits and luminosity calculations depend on redshift, so it is
important to verify our redshift estimates.  Of the 1479 objects in
our IR-selected AGN sample, 751 have no spectroscopic redshift, so for
these we use photo-$z$'s calculated from IRAC and optical photometry.
As described in \citet{brod06}, photo-$z$'s using template-fitting
techniques generally fail for objects such as the IR-selected AGNs
that have featureless, power law SEDs.  To overcome this difficulty,
the technique of \citet{brod06} uses an artificial neural net to
estimate the photo-$z$'s for such objects, using those objects that
also have spectroscopic redshifts as a training set.

However, only 42 of the 640 IRAGN 2s have spectroscopic redshifts and
thus are included in the training set.  As shown in Fig.\ \ref{figr},
most of the IRAGN 2s are too faint to be spectroscopically targeted in
AGES.  Therefore, it is not immediately clear that
photo-$z$ estimates, which are calibrated against a training set of
optically brighter objects (many of them optical BLAGNs), will also be
valid for the IRAGN 2s that have significantly different mid-IR to
optical SEDs.  It is encouraging that the average X-ray hardness ratio
for the IRAGN 2s decreases with redshift as expected for a small range
in \nh\ (Fig.~\ref{fignhz}).  However, the sharp cutoff at
$z\simeq2.5$ in the redshift distribution of the IRAGN 2s (visible in
Fig.\ \ref{figgalagncol}) suggests a possible systematic bias in the
$z_{\rm phot}$.  It is important to verify that such errors do not
significantly affect our results.

\subsection{Comparison of spectroscopic and photometric redshifts}
\label{specphotz}
\begin{figure}
\epsscale{1.2}
\plotone{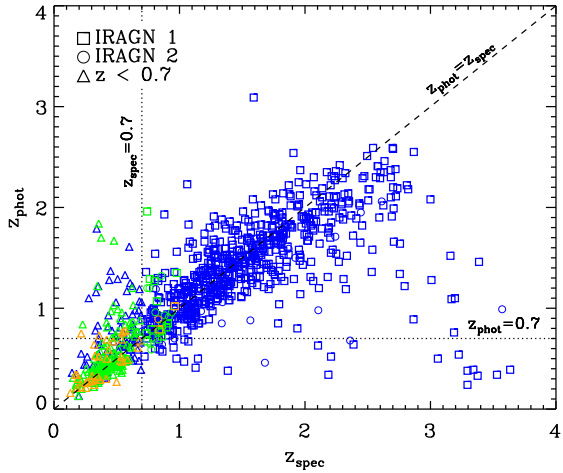}
\caption{Photometric redshifts ($z_{\rm phot}$) from the
  \citet{brod06} catalog versus spectroscopic redshifts ($z_{\rm
  spec}$), for those objects with AGES optical spectra.  Squares show
  IRAGN 1s, circles show IRAGN 2s.  BLAGNs are shown in blue, NLAGNs
  are shown in orange, and optically normal galaxies are shown in
  green.  The dashed line corresponds to $z_{\rm phot}=z_{\rm spec}$,
  while the dotted lines show $z_{\rm spec}=0.7$ and $z_{\rm
  phot}=0.7$.  Objects with no optical spectrum and with $z_{\rm
  phot}<0.7$ would not be included in the IRAGN sample.
\label{figz}}
\end{figure}

As a first check, we compare the photometric versus spectroscopic
redshifts for the IRAGNs, as shown in Fig.\ \ref{figz}.  For
completeness, this figure includes all objects selected by the
\citetalias{ster05} IRAC criteria (including those with $z<0.7$), but it
does not include 38 IRAGN 1s that have AGES spectroscopy but do not
have well-constrained photo-$z$'s from the \citet{brod06} catalog.  For
the IRAGNs, the distribution in $\delta z=(z_{\rm phot}-z_{\rm
spec})/(1+z_{\rm spec})$ is roughly Gaussian, with mean, dispersion,
and fraction of outliers (objects outside $2\sigma$ in the
distribution) of -0.03, 0.16, and 0.06, respectively.  These values
are (-0.03, 0.15, 0.05) for the 648 IRAGN 1s separately and (-0.06,
0.18, 0.10) for the 42 IRAGN 2s, indicating reasonably good agreement
for both IRAGN types.

The distribution in $\delta z$ is skewed somewhat by $\sim20$ quasars
at $z_{\rm spec}>2$ and $z_{\rm phot}<1$.  The presence of these
sources suggests that for some high-$z$ IRAGN, our reliance on
photo-$z$'s may give a large underestimate for the redshift (some such
objects would be eliminated from the sample by our requirement that
$z>0.7$).  Fig.~\ref{figz} also includes 73 sources with $z_{\rm
phot}>0.7$ and $z_{\rm spec}<0.7$, indicating that there could
be $\sim$10\% contamination from low-$z$ sources in the IRAGN sample.
Still, 54 of these 73 sources have $z_{\rm spec}>0.5$, so the
contamination from very low redshifts ($z<0.5$) is expected to be
$\lesssim 3$\%.  In addition, the sample includes 33 sources
($\sim$5\%) with $z_{\rm phot}<0.7$ and $z_{\rm spec}>0.7$ that would
not be included in the IRAGN sample.

Other redshift estimates in the \bootes\ field come from
\citet{houc05}, who obtained redshifts for 17 optically-faint sources
using the Infrared Spectrograph on \spitzer.  Of these, 5 have $5
\sigma$ detections in the IRAC bands, and 4 have IRAC colors inside
the \citetalias{ster05} selection region.  These four sources have 
photo-$z$ estimates from the \citet{brod06} catalog, although only two are in
our IRAGN 2 sample (the other two  have no detection in the
$R$ band).  Of the IRAGN 2s, one has $R=23.8$ and $(z_{\rm
spec}, z_{\rm phot})=(1.95, 2.35)$  while the other has $R=24.7$
and $(2.59, 3.96)$.  The two sources with no $R$ counterpart have
$(0.70, 0.99)$ and $(1.75, 1.01)$.  Based on only these four
objects it is difficult to make any conclusions about the whole
sample, except that photo-$z$'s are more uncertain for fainter sources.

\begin{figure}
\epsscale{1.2}
\plotone{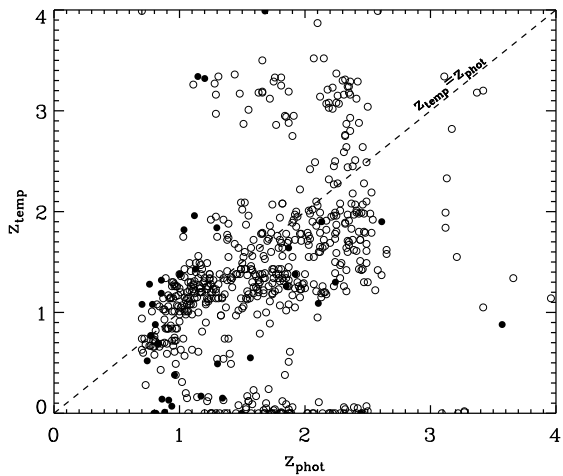}
\caption{Verification of \citet{brod06} photo-$z$'s ($z_{\rm phot}$)
for IRAGN 2s using three-band optical galaxy template redshifts
($z_{\rm temp}$).  The dashed line corresponds to $z_{\rm temp}=z_{\rm
phot}$.  Filled points have spectroscopic redshifts, while open points
have only photo-$z$'s; note that we plot $z_{\rm phot}$ here even for
objects with spectroscopic redshifts.
\label{figtempz}}
\end{figure}

\subsection{Comparison to optical template redshifts}

To test the photo-$z$'s for the entire IRAGN 2 sample, we note that
most IRAGN 2s have galaxy-like optical colors (\S\
\ref{galedd}). Therefore, for these sources we can perform a rough
template photo-$z$ estimate by using only the optical photometry,
fitting the $B_W$, $R$, and $I$ SED as in \S\ \ref{galedd}, but
allowing the redshift to vary.  

The accuracy of the template fits is limited by the fact that we have
only three optical photometric data points, so that the fits are
underdetermined if they include too many free parameters.  We have
tried fits using a wide range of galaxy and starburst templates with
varying ages and extinctions, and consistently find that if we include
two or more templates, the photo-$z$'s are poorly constrained.  We
therefore use a single, non-evolving template, of which the elliptical
galaxy model described in \S~\ref{lum} provides the best constraints
over the wide range in redshift ($0.7<z\lesssim3$) covered by our
sample.

The best-fit redshifts ($z_{\rm temp}$) from these template
fits are shown in Fig.\ \ref{figtempz}.  The $z_{\rm temp}$ estimates
follow the \citet{brod06} empirical photo-$z$'s reasonably well and
cover the same range in redshift, except for a group of 80 objects
that have very low $z_{\rm temp}<0.1$ (we note, however, that most of
these sources have a second minimum in the $\chi^2$ function that lies
within $\pm0.5$ of the $z_{\rm phot}$).   Excluding the sources with
$z_{\rm temp}<0.1$, 79\% of the IRAGN 2s have $\left| z_{\rm
temp}-z_{\rm phot} \right| < 0.25(1+z_{\rm phot})$, with a
bias toward lower redshifts at $1.7<z<2.2$.  Objects
with $z_{\rm phot}\sim2.5$ have a wide range in $z_{\rm temp}$, which
may indicate that the real redshift distribution of the IRAGN 2s
extends smoothly out to $z\gtrsim3$, similar to the IRAGN 1 sample.

We also obtain similar results with the ${\tt HyperZ}$ photometric
redshift package \citep{bolz00}, using the same fixed, non-evolving
template spectrum. These results give us confidence that the IRAGN 2s
lie at redshifts $0.7\lesssim z \lesssim 3$ and that the photo-$z$'s have no
systematic bias large enough to significantly affect the physical
interpretation of our results.

\section{Sample contamination and completeness}
\label{caveats}
In the previous section we showed that our mid-IR and optical
color classification for obscured AGNs is verified by the typical
X-ray, IR, and optical properties of these objects.  Therefore, we are
confident in the general technique of selecting obscured AGNs.
However, to make estimates of how our IRAGN 2 sample relates to the
total population of obscured AGNs, it is important to address issues
of contamination and completeness.  

\subsection{Photometric uncertainties and color selection}
\label{photoerr}

We first address the photometric uncertainty in the IRAC colors that
are used to select the IRAGN.  Photometric error will lead some
sources to move into or out of the \citetalias{ster05} selection
region, causing contamination or incompleteness, respectively.  These
will be dominated by the 5.8 and 8 \micron\ IRAC bands, which
are less sensitive than the shorter wavelength bands; the $1 \sigma$
uncertainty in the $[5.8]-[8.0]$ color is typically in the range
$0.1-0.4$, compared to $0.02-0.08$ for $[3.6]-[4.5]$.

\begin{figure}
\epsscale{1.2}
\plotone{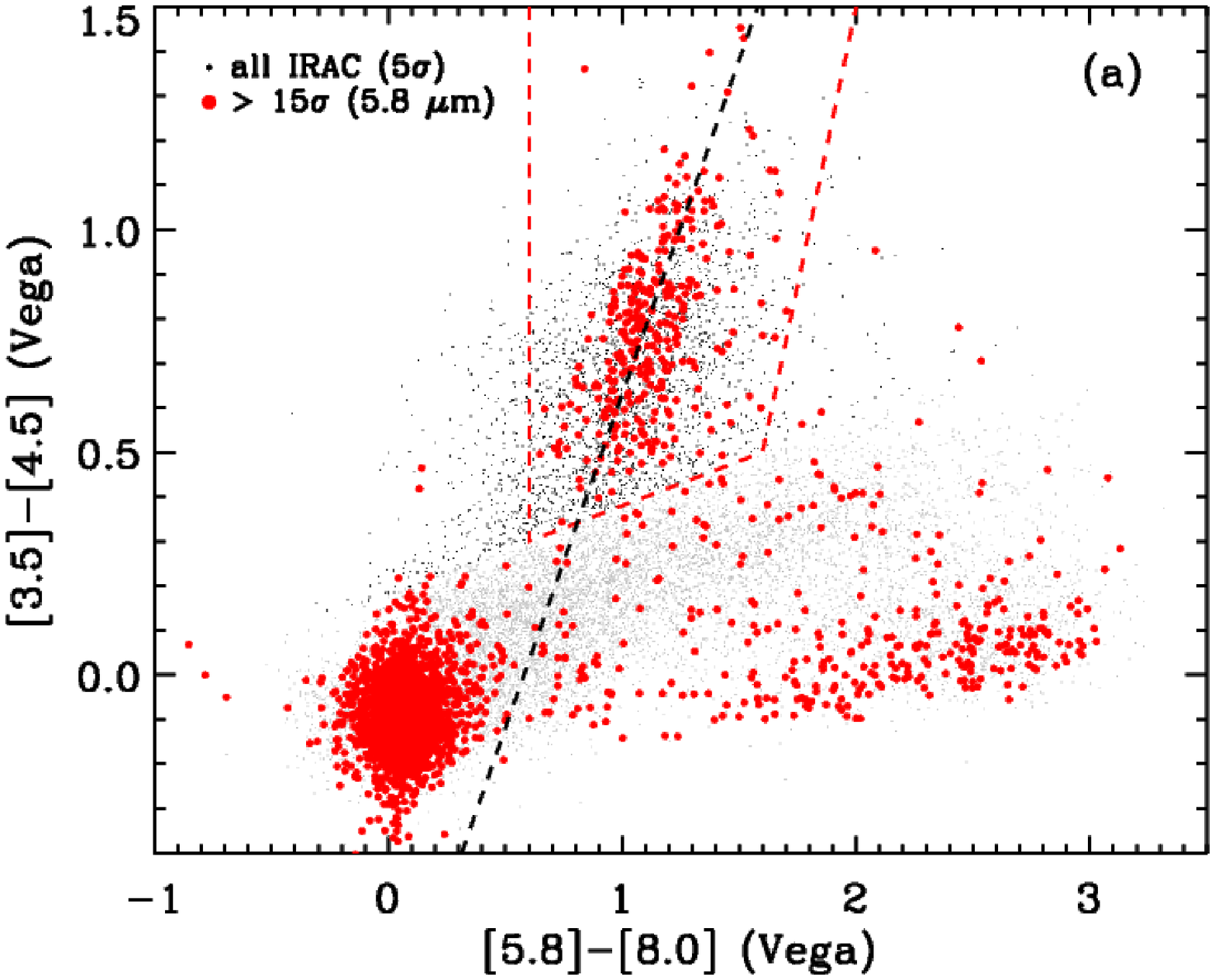}
\plotone{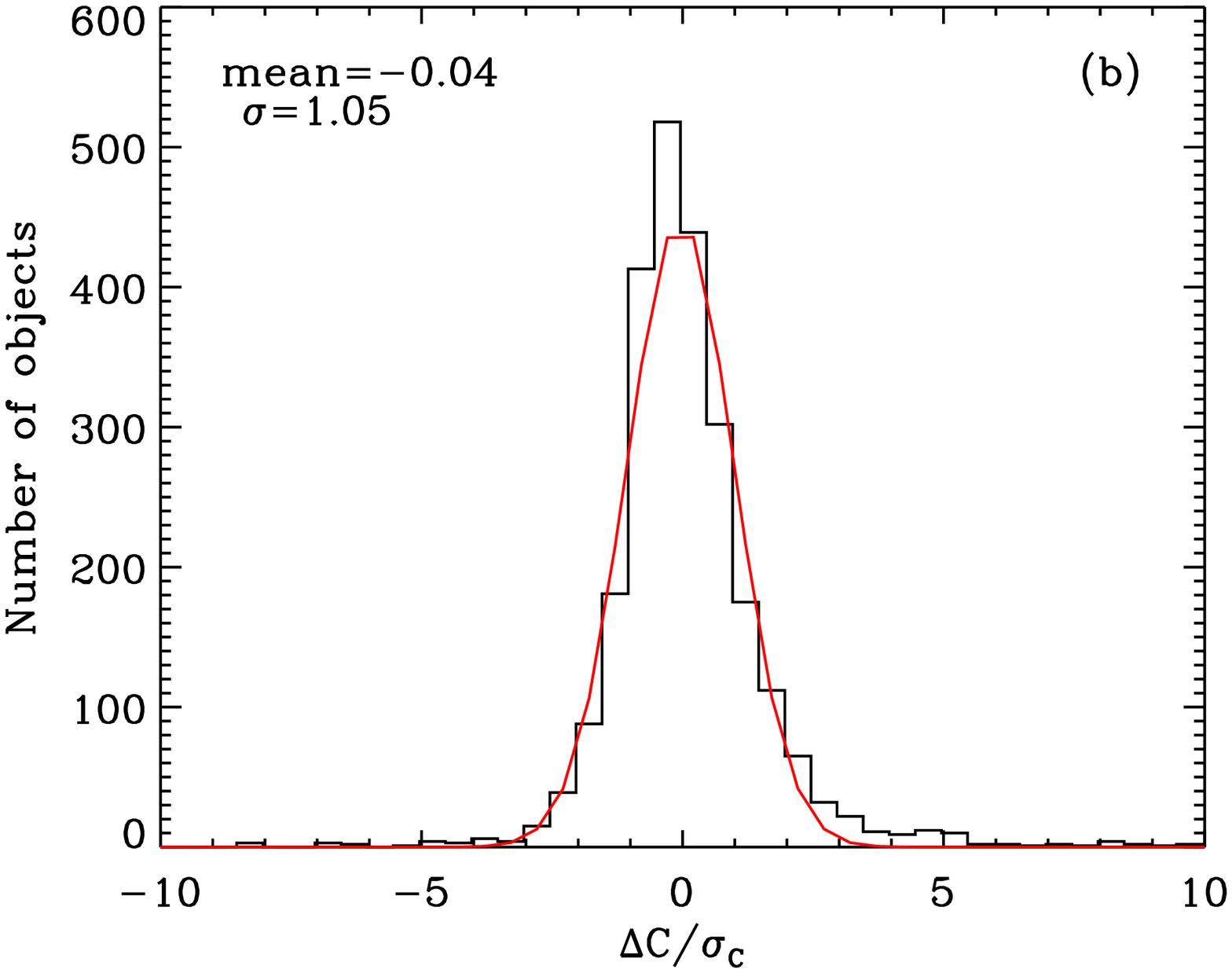}
\caption{(a) IRAC color-color distribution for the $\approx$15,500
  IRAC sources with four-band $5\sigma$ detections.  We have excluded
  from this figure 139 stars that are brighter than the saturation
  limits given in \citet{eise04}; all have $[5.8]-[8.0]<0.4$ and so do
  not lie in the \citetalias{ster05} AGN selection region (shown by
  the red dashed line).  Sources with $[3.6]-[4.5]$ greater than the
  lower boundary of the \citetalias{ster05} region are shown in black.
  Objects with $S_{5.8}/\sigma_{5.8}>15$ are shown in red, and the
  locus defined by these points (Eqn.~\ref{eqstern}) is shown by the
  diagonal line.  (b) Distribution of the deviation  ($\Delta C$) of $[5.8]-[8.0]$ colors from the
  diagonal line for the black points in (a), in units of
  the $1 \sigma$ uncertainty in the $[5.8]-[8.0]$ color.  A Gaussian fit to the
  distribution is shown in red, and is consistent with most points
  having intrinsic colors defined by the line shown in (a).
\label{figcol_stern_errors}}
\vskip0.4cm
\end{figure}

The color-color distribution indicates that incompleteness is a greater
problem than contamination.  Fig.~\ref{figcol_stern_errors}(a) shows the
IRAC color-color distribution, highlighting those objects with
$S_{5.8}/\sigma_{5.8}>15$, where $\sigma_{5.8}$ is the error in the
5.8 \micron\ band flux.  This shows that the bright sources in the \citetalias{ster05} AGN region occupy a small locus in color-color space around a line defined by 
\begin{equation}
\label{eqstern}
[3.6]-[4.5]=0.2([5.8]-[8.0])+1.8.
\end{equation}
The spread of points about this line is consistent with the
photometric uncertainties.  For all sources lying above the lower
boundary in the \citetalias{ster05} criteria (shown as black points in Fig.~\ref{figcol_stern_errors}(a), we derive the difference
$\Delta C$ between the observed $[5.8]-[8.0]$ and the line defined
above.  The distribution in $\Delta C/\sigma_C$, where $\sigma_C$ is
the $1 \sigma$ uncertainty in the color, is shown in
Fig.~\ref{figcol_stern_errors}(b), and is well fitted by a Gaussian
 with mean $-0.04$ and $\sigma=1.05$.  This indicates that
most of the objects with high $[3.6]-[4.5]$ can be associated with the
\citetalias{ster05} region and in fact may occupy a remarkably tight
locus in color-color space.  However, photometric errors cause
$\sim$10\% to be observed outside the AGN selection region.
Conversely, we only expect $\sim$100 sources to be scattered into this
region, indicating that contamination due to photometric errors is
$\lesssim$5\%.

\subsection{Reliability of obscured AGN selection}
It is important to estimate the reliability of our classification of
IRAGNs based on IR-optical colors; that is, how many IRAGN 1s are
actually obscured, and how many IRAGN 2s are unobscured?  In the
sample of 839 IRAGN 1s, 719 have BLAGN spectra from the AGES data set,
or have point-like optical morphologies (${\tt CLASS\_STAR}>0.7$) and
are best fitted by blue (quasar or Sb) optical templates.  These are
strong indicators that a source is an unobscured, type 1 AGN, so the
color selection is at least 85\% reliable for IRAGN 1s.  Of the 640
IRAGN 2s, 517 have galaxy-like optical colors and ${\tt
CLASS\_STAR}\leq0.5$, and do not have BLAGN optical spectra (only 29,
or 3\%, of the IRAGNs with BLAGN spectra are classified as IRAGN 2s).  These criteria
only indicate that an object is dominated by the host galaxy in the
optical; however, as discussed in \S\ \ref{identify}, the high IRAC
luminosities of these sources would suggest dominant nuclear emission
in the optical, were they unobscured.  We conclude that our selection
of obscured AGNs based on optical-IR color is at least 80\% reliable.

\subsection{Contamination from starburst and normal galaxies}
\label{contamination}

We expect almost all of the IRAGN 1s to be AGNs rather than starburst
or normal galaxies.  Most of these objects were targeted by AGES, so
we can verify their classification with optical spectra.  Of the 686
IRAGN 1s that have AGES spectroscopy (or 82\% of the IRAGN 1 sample),
668 (97\%) are BLAGNs, while 3 are NLAGNs and 15 are optically normal
galaxies.  Keeping in mind that many AGES targets were selected to be
X-ray sources and thus are biased toward bright AGNs, we consider
separately the 233 IRAGN 1s that have optical spectra but no X-ray
counterpart.  These should be a representative sample of the IRAGN 1s
that are not detected in X-rays, and of these 226 (97\%) are BLAGNs, one
is a NLAGN, and five are optically normal galaxies.  We conclude that
there is little ($<$5\%) contamination in the IRAGN 1 sample.

It is more difficult to estimate contamination in the IRAGN 2s.  Only
42 IRAGN 2s have AGES spectra (29 BLAGNs, 1 NLAGN, 12 galaxies),
because most IRAGN 2s are fainter than the AGES flux limits (Fig.\
\ref{figr}).  A total of 155 of the IRAGN 2s have X-ray detections and
thus \lx\ values that imply that they must be powered by accretion.
Of the remaining 485 objects, some at high redshifts might not be AGNs
but instead luminous starburst galaxies with IRAC colors that lie
inside the \citetalias{ster05} AGN color-color selection region.  As
mentioned in \S~\ref{lum}, heavily extincted starbursts (i.e., Arp
220) can have very red IRAC colors.  However, the \citet{sieb07} Arp
220 template is also very red in the optical ($B_W-I>3$ at $z>0.7$),
which is much redder than observed for the IRAGN 2s
(Fig.~\ref{figgalagncol}).  Still, it is possible that some high-$z$
starbursts have similar IRAC colors to Arp 220 but are bluer in the
optical, and these could contaminate the IRAGN 2 sample.  In addition,
at $z\gtrsim 3$, the colors of less obscured starbursts (e.g., M82)
would also lie in the \citetalias{ster05} region \citep[see Fig.~6
of][]{barm06}.  However, to be detected to our IRAC flux limits at
$z>3$, a source must have a very high $L_{\rm 5.8 \mu m}>10^{12}$
\lsun\ (where $L_{\rm 5.8 \mu m}$ is the observed $\nu L_\nu$ in the
5.8 \micron\ band).  For a typical ratio of rest-frame far-infrared
(FIR) to observed 5.8 \micron\ fluxes for starburst galaxies
(\S~\ref{xrayres}), this implies $L_{\rm FIR}\gtrsim10^{13}$ \lsun.
In most such ``hyperluminous infrared galaxies'', a significant (and
often dominant) contribution to the IR emission comes from an AGN
\citep[e.g.,][]{farr02}.  Also considering that our IRAGN sample
contains only 27 objects at $z>3$ that do not have BLAGN optical
classifications, contamination from such high-$z$ starbursts should be
small.

One empirical constraint on contamination comes from the X-ray
stacking results, due to the fact that starburst galaxies tend to be
significantly fainter in the X-rays than AGNs.  If we exclude
sources that have X-ray counterparts, the IRAGN 1s and 2s have similar
average X-ray fluxes in the 2--7 keV band of $0.47\pm0.06$ and
$0.46\pm0.05$ counts source$^{-1}$, respectively (Table
\ref{tblxray}).  Because there is little contamination in the IRAGN 1
sample, 0.47 counts source$^{-1}$ should be typical for IR-selected
AGNs that are fainter than the X\bootes\ detection limit.  We thus
consider the possibility that the AGNs among the X-ray--undetected
IRAGN 2s have the same average flux, but the sample is 40\%
contaminated by starburst galaxies, which have 0.5--7 keV fluxes that
are 5 times smaller.  The observed average flux from stacking would
then be 68\% of that for the IRAGN 1s, or 0.32 counts source$^{-1}$,
which is $\simeq3\sigma$ below the observed value.  We are therefore
confident that $<40\%$ of the 485 X-ray--undetected IRAGN 2s are
contaminating starbursts, implying a $3 \sigma$ upper limit of
$\sim$30\% contamination for the total sample of IRAGN 2s.

Results from deeper surveys can help put more concrete limits on
contamination.  \citet{alon06} examined a population of objects in the
Chandra Deep Field-South (CDF-S) selected using the
\citetalias{ster05} IRAC color-color criteria.  Based on X-ray
luminosities and spectral shapes for the individual sources,
\citet{alon06} find that at least 70\% of the IR-selected objects are
AGNs.  We conclude that while it is difficult to accurately estimate
the contamination by normal galaxies of the IRAGN 2 sample, we expect
it to be no larger than $\sim$30\%.  In addition, a further $\sim$10\%
contamination of the IRAGN 2 sample could come from objects at $z<0.7$
(as discussed in \S~\ref{specphotz}), although many of these would
likely be AGNs rather than galaxies.

\subsection{Sample completeness}
\label{incompleteness}

We next estimate our selection completeness; that is, of the AGNs
brighter than the flux limits of the survey, how many are included in
the IRAGN sample?  For AGNs with broad-line optical spectra at
$z>0.7$, the IRAC color-color selection is highly complete to the IRAC
flux limits.  The AGES sample contains 1306 BLAGNs at $z>0.7$ in the
area observed by IRAC, of which 784 (60\%) have $5\sigma$ detections
in all four IRAC bands.  Of these, 697 (89\%) have IRAC colors in the
\citetalias{ster05} selection region.  Of four NLAGNs with $z>0.7$, all
have four-band IRAC detections and are selected by the
\citetalias{ster05} criteria.

For optically faint or obscured AGNs, however, the completeness is more
difficult to estimate.  Of the 1298 X\bootes\ sources with four-band 
$5\sigma$ IRAC counterparts (almost all of which are AGNs), 879 (68\%)
are selected by the \citetalias{ster05} criteria.  Likewise, in the
much deeper \spitzer\ and \chandra\ data from the EGS, \citet{barm06}
find that only $\sim$50\% of X-ray AGNs are selected by the
\citetalias{ster05} criteria.

This incompleteness can be caused by either obscuration or dilution.
Heavy obscuration can absorb even mid-IR emission.  We consider an AGN
with $N_{\rm H}=6\times10^{23}$ \cdens\ (roughly 20 times higher than
the typical column for the IRAGN 2s), for which an SMC gas-to-dust ratio of
$N_{\rm H}/A_V\simeq2\times10^{22}$ \cdens\  implies $A_V=30$.
This corresponds to a rest-frame extinction at 2 \micron\ of 3.6
mag (this is largely independent of the choice of extinction
curve, which are very similar redward of the $V$ band); for a smaller
Galactic dust-to-gas ratio, the IR extinction would be even higher.
Therefore, high column densities can obscure the nucleus such that
either the IRAC fluxes drop below our detection limits or the IRAC
color-color selection criteria would not select such an object as an
AGN (note that Fig.~\ref{figcol} shows that at $z\gtrsim0.7$, sources
with $A_V\gtrsim30$ move out of the \citetalias{ster05} AGN color
selection).  For these reasons, we expect our IRAGN sample to include
very few highly absorbed objects ($N_{\rm H}\gg10^{23}$ \cdens\ in the
X-ray).

AGNs can also be missed if their IR emission is diluted by starburst
activity.  The luminosities of the IRAGNs in our sample, $L_{\rm 4.5
\mu m}\sim(0.3-3)\times10^{12}$ \lsun, are comparable to that of
luminous starburst galaxies \citep[e.g.,][]{rowa05}.  There is
compelling evidence that starburst activity and AGN activity are often
linked \citep{ho05,king05, farr03}, so we expect some sources with an
AGN also to have a powerful starburst that dominates the mid-IR
luminosity.  Such an object would have a starburst-like SED
(corresponding to a low $f_{\rm AGN}$ as in Fig.~\ref{figcol}) and
would not be selected using the AGN color-color technique.

One way to estimate this incompleteness is to examine bright radio
sources with relatively faint IR counterparts.  These are likely to be
AGNs and not starbursts, and the radio emission will not be strongly
affected by extinction.  The VLA FIRST 20 cm radio survey
\citep{beck95} detects 301 radio sources that are brighter than 5 mJy
in the area covered by IRAC.  Of these, 24 are matched to the
four-band IRAC catalog with 3.6 $\mu$m magnitude fainter than 15.  The
\citetalias{ster05} color-color criteria select 14 of these objects as
AGNs, of which 11 are in our IRAGN sample at $z>0.7$.  Of these 11
sources, 7 are IRAGN 1s and 4 are IRAGN 2s.  These results suggest
that the completeness of our IRAC color-color selection may be as low
as $\sim$60\% for AGNs at our IRAC flux limits.  However, radio-loud
AGNs may be different from the more numerous radio-quiet objects, so
it is difficult to draw conclusions about the total AGN population,
except to say that incompleteness effects may be significant.

\begin{figure}
\epsscale{1.2}
\plotone{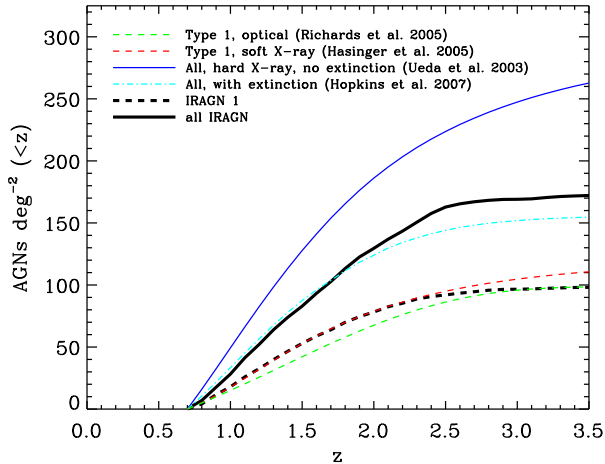}
\caption{Cumulative redshift distribution of IRAGNs, compared to
  predictions using optical and X-ray luminosity functions, and considering the
  \bootes\ flux limits.  See text in \S~\ref{compmodel} for discussion.
  \label{figmodel}}
\end{figure}

\subsection{Comparison to predictions from luminosity functions}
\label{compmodel}
Another check on the completeness of our selection is to compare
the number of IRAGNs in our sample with the number that are predicted by
optical, IR, and X-ray luminosity functions, accounting for the
\bootes\ flux limits.  Our AGN detection is usually limited by the 5.8
$\mu$m band (the 8 $\mu$m band has a similar flux limit, but since the
quasars have red SEDs, they are usually brighter at 8 $\mu$m than at
5.8 $\mu$m).  To convert X-ray and optical luminosities to IRAC fluxes,
we use the AGN model (SDSS optical spectrum plus mid-IR power law)
described in \S\ \ref{lum}.  We include a distribution in IR slopes,
modeled by a Gaussian with $\average{\alpha_\nu}=-1$ and
$\sigma_{\alpha_\nu}=0.5$.  This distribution approximately reproduces
the observed dispersion in the optical to IR colors for type 1 AGNs
from \citet{rich06}, as shown in Fig.~\ref{figtemp}.

We first compare the IRAGN 1 sample to the predictions of the
broad-line quasar luminosity function of \citet{rich05} from the 2QZ
survey. Unlike more recent luminosity functions derived from SDSS
data, this data set includes objects below the ``knee'' of the
luminosity function.  This model SED, convolved with the
\citet{rich05} luminosity function, predicts 840 type 1 quasars
brighter than our flux limits in the 8.5 deg$^2$ field covered by IRAC
(see the the green dashed line in Fig.~\ref{figmodel}).  We detect 839
IRAGN 1s, very close to this total.  This indicates that the IRAC
color selection is highly complete for broad-line quasars (although
some highly-reddened IRAGN 1s might not be included in the
\citet{rich05} sample).

We also evaluate predictions for X-ray luminosity functions.  We use
the same model UV/IR spectrum described above, and take the
relationship between UV luminosity and the UV/X-ray spectral slope
$\alpha_{\rm ox}$ from \citet{stef06}.  For simplicity, we assume a
constant unabsorbed X-ray spectrum with $\Gamma=2$.  The 0.5--2 keV
luminosity function for unabsorbed (type 1) quasars of \citet{hasi05}
predicts 970 AGNs above our flux limits at $z>0.7$ (see the red
dashed curve in Fig.~\ref{figmodel}), or 16\% more than
the number of IRAGN 1s.  

In contrast, the 2--10 keV luminosity function of \citet{ueda03}
predicts $\sim$2300 total AGNs, 55\% more than we detect.
However, although the \citet{ueda03} X-ray sample includes many AGNs
that are X-ray absorbed, our model AGN SED used here does not
include corresponding dust extinction.  The solid blue line in
Fig.~\ref{figmodel} therefore represents the number of total AGNs that would
be observed at the \bootes\ flux limits, in the absence of dust
extinction.

The effects of dust extinction were included by \citet{hopk07qlf}, who
determined a parametrization of the {\it bolometric} luminosity
function by fitting observed X-ray, optical, and IR luminosity
functions.  This work used the distribution in \nh\ observed by
\citet{ueda03}, and a typical Galactic gas-to-dust ratio, in
predicting the numbers of observed AGNs.  The total number of AGNs
for the \bootes\ 4.5 $\mu$m flux limit is shown by the dot-dashed cyan line in
Fig.~\ref{figmodel}.  The model predicts 1320 detectable AGNs at
$z>0.7$, 10\% fewer than the total number of IRAGNs we observe.

We note that the \citet{hopk07qlf} predictions may provide only a
lower limit on the number of detectable AGNs.  If AGNs typically have
a gas-to-dust ratio that is higher than the Galactic value (see
\S~\ref{dust}), this would tend to decrease the dust extinction for a
given \nh\ distribution, and so would increase the number of
detectable AGNs at a given flux limit.  Still, the agreement between
the observed number of IRAGN 1s and 2s and the predictions from
luminosity functions indicates that our selection is reasonably
complete {\it to the flux limits of the survey}.  However, we are
likely missing half or more of the total obscured AGN population
because dust extinction causes them to fall below our IRAC flux limits
or out of the \citetalias{ster05} AGN color selection.

\section{Discussion}
\label{discussion}

\subsection{Comparison to other obscured AGN samples}

\begin{figure}
\epsscale{1.2}
\plotone{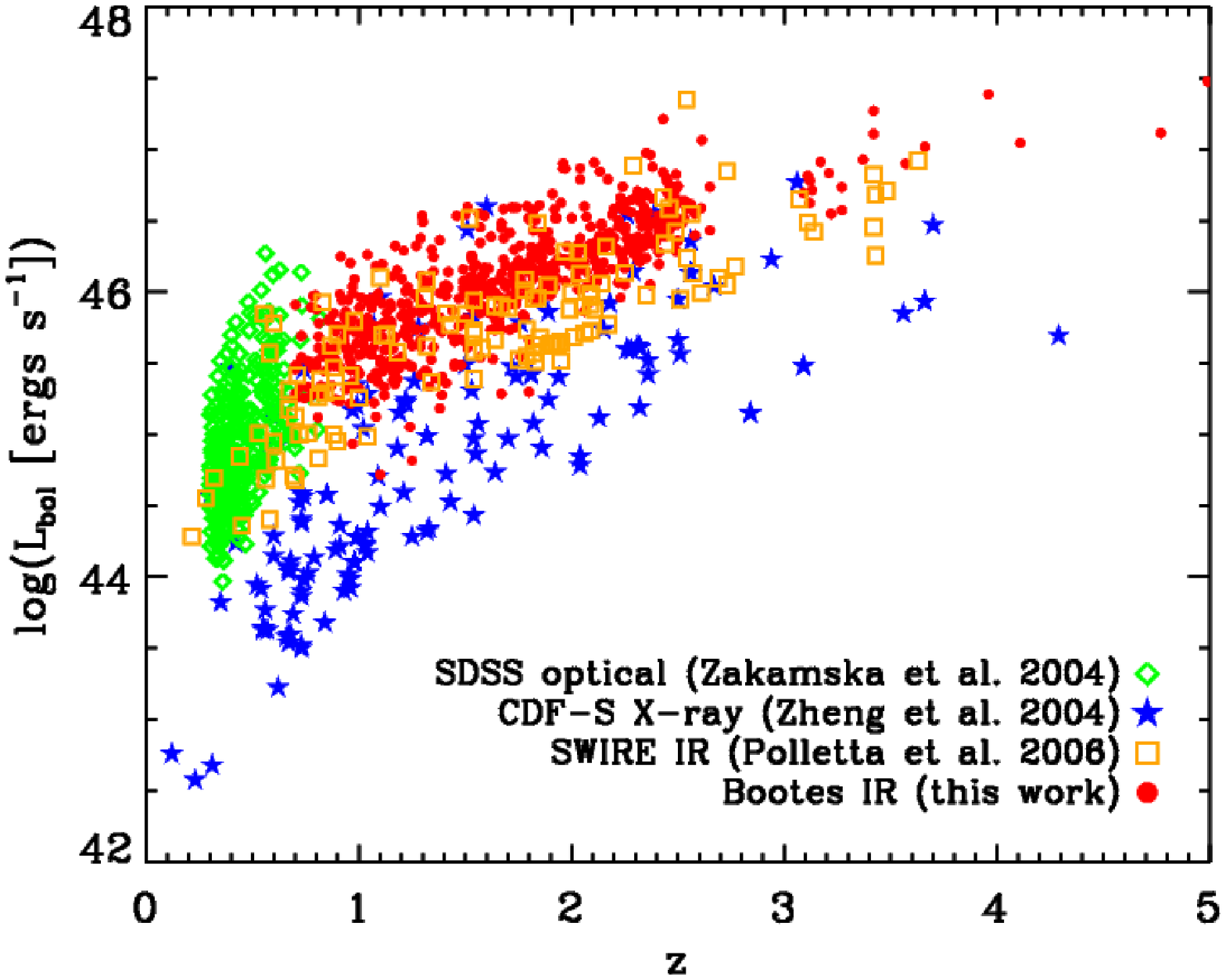}
\caption{$L_{\rm bol}$ versus redshift for four large samples
  of obscured AGNs: optically selected objects from SDSS
  \citep[green triangles]{zaka03}, X-ray selected objects from the CDF-S
  \citep[blue stars]{zhen04}, IR-selected AGNs from the \chandra/SWIRE
  survey \citep[orange squares]{poll06}, and IRAGNs from this work
  (red circles).  $L_{\rm bol}$ is estimated using bolometric corrections
  from \citet{hopk07qlf}, as described in \S\ \ref{othersamples}.
  \label{figz_lum}}
\vskip0.5cm
\end{figure}

\label{othersamples}
In order to perform a complete census of obscured accretion in the
Universe, it is important to place the IRAGN 2 sample described
in this paper in the context of obscured AGNs found in other surveys.
We compare the distribution in $z$ and \lbol\ of our sample with three
of the largest ($>$100 objects) samples of obscured AGNs with redshift
estimates.

The first sample consists of 291 optically-selected, luminous type 2
AGNs at $0.3<z<0.83$ from the SDSS \citep{zaka03}.  To estimate
intrinsic \lbol\ from the observed spectroscopic properties, we assume
that the unobscured SED of the type 2 quasars in this sample is
similar to that for type 1 optical AGNs.  In this case we use Equation
(8) of \citet{zaka03} to convert the observed [\ion{O}{3}]
$\lambda$5007 luminosity to the intrinsic, unobscured $L_B$.  We then use the bolometric corrections from the model of \citet{hopk07qlf}, for which $BC_B=9-12$.

The second sample consists of 145 X-ray sources in the CDF-S that are
classified as type 2 AGNs on the basis of their X-ray spectrum and/or
optical properties \citep{zhen04}.  Of these objects, 68 have
spectroscopic redshifts, while 78 have photo-$z$ estimates.  For these
objects we estimate an unabsorbed $L_{\rm 0.5-8\; keV}$ on the basis
of the redshift, observed flux, and HR, assuming an intrinsic
power-law spectrum with $\Gamma=1.8$.  We then convert these values to
$L_{\rm bol}$, again using the model of \citet{hopk07qlf}, for which
$BC_{\rm 0.5-8\; keV}\simeq20-80$.

The third sample consists of the 120 obscured AGNs selected from
optical and IR SEDs in the \chandra/SWIRE survey by \citet{poll06}.
Of these objects, 11 have spectroscopic redshifts, and the remainder
have photo-$z$'s.  We calculate \lbol\ from the mid-IR luminosity
\lirac, as described in \S\ \ref{bolometric}.  As a further check of
the robustness of these $L_{\rm bol}$ estimates, we estimate $L_{\rm
bol}$ for the 41 objects in the survey with X-ray detections, using
the unabsorbed X-ray luminosities calculated by \citet{poll06} and the
\citet{hopk07qlf} SED model.  We find that while there is significant
scatter between the IR and X-ray estimates of $L_{\rm bol}$, on
average they agree to within a factor of $\lesssim 2$.  Nevertheless,
these discrepancies between $L_{\rm bol}$ estimates require that our
comparison between samples selected at different wavelengths is at best
qualitative.

These samples of obscured AGNs selected in the optical \citep{zaka03},
X-ray \citep{zhen04}, and IR \citep[this work]{poll06} are shown
in Fig.\ \ref{figz_lum}.  The \bootes\ IRAGN 2s are more luminous than
the X-ray selected AGNs from the CDF-S and at higher redshifts than
those in the SDSS sample.  The SWIRE obscured AGNs and \bootes\ IRAGN
2s have similar distributions in $z$ and \lbol, but because of the
much larger area in the \bootes\ field (8.5 deg$^2$ vs. 0.6
deg$^2$), the \bootes\ IRAGN 2 sample contains $\sim$6 times more
objects.  We note that even accounting for possible 30\% contamination (\S\
\ref{contamination}), the \bootes\ IRAGN 2s contain the largest sample
to date of luminous, moderately obscured AGNs at high redshift.

\subsection{Contribution to the cosmic X-ray background}
\label{cxb}
Synthesis models require a population of obscured AGNs to produce the
intensity and spectral shape of the CXB.  From our stacking analysis,
we estimate the contribution to the total CXB of the two IRAGN types
defined in this paper.  The extragalactic component of the CXB is
well modeled by a power law with $\Gamma=1.4$ and normalization $10.9\pm0.5$
photons cm$^{-1}$ s$^{-1}$ keV$^{-1}$ sr$^{-1}$ at 1 keV
\citep[e.g.,][]{hick06a,delu04}.  This gives CXB intensities of
$7.6\times10^{-12}$ \intens\ in the 0.5--2 keV band and
$1.5\times10^{-11}$ \intens\ in the 2--7 keV band.

Over the 2.9 deg$^2$ for which we perform the X-ray stacking
analysis, the total fluxes listed in Table \ref{tblxray} correspond to
intensities for the IRAGN 1s of $1.2\times10^{-12}$ \intens\ in the
0.5--2 keV band and $1.3\times10^{-12}$ \intens\ in the 2--7 keV
band, which represent 16\% and 9\%, respectively, of the total CXB.
For the IRAGN 2s the intensities are $3.0\times10^{-13}$ \intens\ in
the 0.5--2 keV band and $6.0\times10^{-13}$ \intens\ in the 2--7 keV
band, or 4\% and 4\% of the CXB, respectively.  This indicates that
selecting IR AGNs in shallow exposures and at $z>0.7$ only captures a
small fraction of the sources that produce the X-ray background.  

\subsection{Implications of a bimodal distribution in obscuration}
\label{impl}
The bimodal distribution in \iiu\ (and accordingly $A_V$) observed in
the IRAGNs may give clues to the distribution of material that
obscures the central engine.  A detailed comparison of the
distribution of dust extinction to AGN obscuration models is beyond
the scope of this paper, but we qualitatively consider two
explanations for the bimodal extinction: a hard-edged torus, or
obscuration as an evolutionary phase.

In the unified model, the obscuring material is in an extended
distribution that surrounds the nucleus on scales of $\lesssim100$ pc,
possibly in the shape of a torus, such that the level of obscuration
depends on the observer's line of sight. These obscuring structures
are well-established for local Seyfert galaxies \citep[see ][and
references therein]{anto93}, and there is evidence that they also
exist in more distant, luminous quasars, for example from the
detection of broad emission lines in polarized light that is scattered
from the nucleus \citep{zaka05}.  In some models of the torus
\citep[e.g.,][]{trei04, ibar07}, the obscuring medium is not
homogeneous but varies in density with radius and angle from the axis
of symmetry, so that there is a slow increase in obscuring column as
the torus is seen more edge-on.  However, this slow increase is
inconsistent with the $A_V$ distribution we observe.  Instead, a
bimodal $A_V$ distribution could indicate an abrupt edge to the
obscuring material rather than a smooth distribution and so provides a
constraint on the obscuring geometry.

Alternatively, the obscuring material could be in the form of
irregular clouds that surround the nucleus on scales as large as
kiloparsecs.  This material can be driven to the center of the galaxy
by major galaxy mergers and can feed the AGNs (as well as nuclear
starbursts) while also obscuring the central engine
\citep[e.g.,][]{sand88,hopk06merge}.  In time, AGN winds may blow this
material away from the nucleus, leading to an unobscured phase of AGN
activity \citep[e.g.,][]{silk98, spri05,hopk06apjs}.  In this picture,
obscured accretion is an evolutionary stage in the life of the quasar;
as long as the blowout phase is short-lived, quasars will be seen with
either significant or very little obscuration.  Therefore, the bimodal
distribution we observe may place constraints on the timescale for AGN
feedback.
 
\vspace{1cm}

\section{Summary}
\label{summary}

In this paper we analyze a sample of 1479 AGNs at $0.7<z\lesssim3$
from the wide-field multiwavelength \bootes\ survey, selected on the
basis of their IRAC colors.   This work has two key elements
that together make it unique among studies of IR-selected AGN: (1) the
wide area and deep optical photometry in the \bootes\ field allow us
to identify a large number of obscured sources, and (2) the contiguous
X-ray coverage allows us to verify independently that the IRAGN 2s are
obscured AGNs, and to measure their neutral gas column densities. 

Key results of this paper are as follows:
\begin{enumerate}

\im The optical-IR color distribution of the IR-selected AGNs is
bimodal, with a boundary of $R-[4.5]=6.1$ (Vega) between the two
subsets.  Based on this color criterion, we divide our sample into 640
obscured (IRAGN 2) and 839 unobscured (IRAGN 1) AGNs.  The optical-IR
color distribution can be interpreted in terms of dust extinction of
the nuclear optical emission for the IRAGN 2s.  The obscured AGN color
selection is valid for AGNs at $z>0.7$ and with mid-IR luminosities
$\gtrsim10^{11}$ \lsun.

\im X-ray and optical data confirm our selection of obscured AGNs.
X-ray stacking shows that both subsets of IRAGNs have average X-ray
luminosities characteristic of luminous AGNs.  The IRAGN 1s have
average X-ray hardness ratios typical of unobscured sources, while the
IRAGN 2s have harder X-ray spectra, corresponding to absorption with
typical $N_{\rm H}\sim3\times10^{22}$ \cdens.  The optical colors and
morphologies are typical of galaxies for most IRAGN 2s and of quasars
for most IRAGN 1s, consistent with the optical emission from the IRAGN
2s being extincted.

\im For a typical range of AGN gas-to-dust ratios, the \nh\ for the
IRAGN 2s (derived from X-ray stacking) corresponds to $1\lesssim A_V
\lesssim10$, consistent with the $A_V$ values derived from optical-IR
SED fits, and sufficient to completely extinct the nuclear optical
emission.  This indicates that, on average, absorption by neutral gas
and extinction by dust are correlated in these luminous AGNs.

\im The \citetalias{ster05} IRAC color-color AGN selection is
reasonably complete to our survey  flux limits. The numbers
of IRAGN 1s and 2s are within $\sim$15\% of predictions from
optical and X-ray luminosity functions.  We expect the optical/IR
color selection to be at least 80\% reliable in distinguishing between
unobscured and obscured AGNs, while contamination from starburst
galaxies in the IRAGN 2 sample should be at most $\sim$30\% and is likely
much lower.

\im The bimodal distribution in optical-IR color for IRAGNs suggests
that these objects have either low ($A_V\lesssim0.1$) or significant
($A_V\gtrsim0.7$) extinction.  This distribution may have implications
for models of AGN obscuration.  In the context of the unified model,
this may imply a hard edge to distribution of obscuring material.
Alternatively, obscuration may be an evolutionary phase that is
followed by rapid blowout of the obscuring dust, leading to a bimodal
distribution in $A_V$.

\im The IRAGN 2s comprise the largest sample to date of AGNs with high
redshifts ($0.7<z<3$), high bolometric luminosities
($10^{45}\lesssim L_{\rm bol}\lesssim 10^{47}$ \ergs), and moderate absorption
($10^{22}\lesssim N_{\rm H}\lesssim 10^{23}$ \cdens), even after accounting
for possible sample contamination of at most $\sim$30\%.   This work
shows that IRAC and
optical selection is a powerful tool for identifying  large numbers
of luminous,
obscured AGNs for follow-up study.

\end{enumerate}

\acknowledgements We thank our colleagues on the AGES, IRAC Shallow
Survey, NDWFS, and X\bootes\ teams, and Ramesh Narayan, Michael Pahre,
Pauline Barmby, and Kamson Lai for productive discussions.  We are
grateful to the referee for suggestions that significantly
strengthened the paper.  This paper would not have been possible
without the efforts of the \chandra, \spitzer, KPNO, and MMT support
staffs.  This work is based in part on observations made by the {\it
Spitzer Space Telescope}, which is operated by the Jet Propulsion
Laboratory, California Institute of Technology under a contract with
NASA.  This research was supported by the National Optical Astronomy
Observatory, which is operated by the Association of Universities for
Research in Astronomy (AURA), Inc., under a cooperative agreement with
the National Science Foundation.  Optical spectroscopy discussed in
this paper was obtained at the MMT Observatory, a joint facility of
the Smithsonian Institution and the University of Arizona.  R.C.H. was
supported by a NASA GSRP Fellowship and a Harvard Merit Fellowship.

\end{document}